\documentclass[preprint,prd,aps,floatfix,preprintnumbers,superscriptaddress,bibnotes,nofootinbib]{revtex4-1}

\usepackage{epsfig}
\usepackage{amsmath}
\usepackage{latexsym}
\usepackage[psamsfonts]{amssymb}
\usepackage{graphicx}
\usepackage{ulem}
\usepackage{longtable}
\usepackage{epstopdf}
\usepackage{bm}
\usepackage {tablefootnote} 
\usepackage{color}

\newcommand{\be}{\begin{equation}}
\newcommand{\ee}{\end{equation}}
\newcommand{\bea}{\begin{eqnarray}}
\newcommand{\eea}{\end{eqnarray}}
\newcommand{\bi}{\begin{itemize}}
\newcommand{\ei}{\end{itemize}}

\begin{document}
\preprint{UTHEP-802, UTCCS-P-165, HUPD-2503, KEK-TH-2715}

\title{
Method for high-precision determination of the nucleon axial structure 
using lattice QCD: \\
Removing $\pi N$-state contamination
}

%
\author{Yasumichi Aoki}
\affiliation{RIKEN Center for Computational Science, Kobe 650-0047, Japan}
\author{Ken-Ichi~Ishikawa}
\affiliation{Core of Research for the Energetic Universe, Graduate School of Advanced Science and Engineering, Hiroshima University, Higashi-Hiroshima 739-8526, Japan}
\affiliation{Graduate School of Advanced Science and Engineering,
Hiroshima University, Higashi-Hiroshima 739-8526, Japan}
\author{Yoshinobu~Kuramashi}
\affiliation{Center for Computational Sciences, University of Tsukuba, Tsukuba, Ibaraki 305-8577, Japan}
\author{Shoichi~Sasaki}
\email[E-mail: ]{ssasaki@nucl.phys.tohoku.ac.jp}
\affiliation{Department of Physics, Tohoku University, Sendai 980-8578, Japan}
\author{Kohei~Sato}
\affiliation{
Office of the President, Seikei University, 3-3-1 Kichijoji-Kitamachi,
Musashino-shi, Tokyo 180-8633, Japan
}
%
\author{Eigo~Shintani}
\affiliation{Center for Computational Sciences, University of Tsukuba, Tsukuba, Ibaraki 305-8577, Japan}
\affiliation{Graduate School of Engineering, University of Tokyo, Hongo 7-3-1, Bunkyo, Tokyo, Japan}

\author{\\Ryutaro Tsuji}
\email[E-mail: ]{rtsuji@post.kek.jp}
\affiliation{High Energy Accelerator Research Organization (KEK), Ibaraki 305-0801, Japan}
\author{Hiromasa~Watanabe}
\affiliation{
Department of Physics, and Research and Education Center for Natural Sciences, Keio University, 4-1-1 Hiyoshi, Yokohama, Kanagawa 223-8521, Japan
}
%
\author{Takeshi~Yamazaki}
\affiliation{Institute of Pure and Applied Sciences, University of Tsukuba, Tsukuba, Ibaraki, 305-8571, Japan}
\affiliation{Center for Computational Sciences, University of Tsukuba, Tsukuba, Ibaraki 305-8577, Japan}
\collaboration{PACS Collaboration}

\date{\today}
\begin{abstract}

We perform a precise calculation of physical quantities related to the axial structure of the nucleon using 2+1 flavor lattice QCD gauge configuration (PACS10 configuration) generated at the physical point with lattice volume larger than $(10\;{\mathrm{fm}})^4$ by the PACS Collaboration. The nucleon matrix element of the axial-vector current has two types of the nucleon form factors, the axial-vector ($F_A$) form factor and the induced pseudoscalar ($F_P$) form factor. 
Recently lattice QCD simulations have succeeded in reproducing the experimental value of the axial-vector coupling, $g_A$, determined from $F_A(q^2)$ at zero momentum transfer $q^2=0$, at a percent level of statistical accuracy.
However, the $F_P$ form factor so far has not reproduced the experimental values well due to strong $\pi N$ excited-state contamination. Therefore, we propose a simple subtraction method for removing the so-called leading $\pi N$-state contribution. This method succeeds in reproducing the values obtained by two experiments of muon capture on the proton and pion electro-production for $F_P(q^2)$. The novel approach can also be 
applied to determine the pseudoscalar ($G_P$) form factor from the nucleon pseudoscalar matrix element
with the help of the axial Ward-Takahashi identity. The resulting form factors, $F_P(q^2)$ and $G_P(q^2)$, are in good agreement with the prediction of the pion-pole dominance model. In the new analysis, the induced pseudoscalar coupling $g_P^\ast$ and the pion-nucleon coupling $g_{\pi NN}$ can be evaluated with a few percent accuracy including systematic uncertainties
using existing data calculated at two lattice spacings.

\end{abstract}

\maketitle

 
\section{Introduction}
\label{sec:INTRO}

The axial structure of the nucleon plays an important role 
to understand the weak interaction between nucleons. 
Experimentally, weak processes mediated by the weak charged current like neutron beta decay $n\rightarrow p+e^{-}+\bar{\nu}_e$, muon capture on the proton $\mu^{-}+p\rightarrow \nu_\mu +n$, or quasi-elastic neutrino scattering $\bar{\nu}_\mu+p\rightarrow \mu^{+}+n$
are mainly exploited for studying the weak nucleon form factors.
In recent years, the physics of neutrino oscillations, which inevitably involves the effects of nucleons and nuclear structure, has required accurate knowledge of the axial structure of the nucleon~\cite{{Kronfeld:2019nfb},{Meyer:2022mix},{Tomalak:2023pdi},{Petti:2023abz}}. This is because the vector part of weak processes can be constrained by knowledge of the isovector combination between the electromagnetic form factors of the proton and the neutron under the assumption of the conserved-vector-current (CVC) hypothesis using isospin relations~\cite{Thomas:2001kw}, while the axial-vector part described by the axial-vector ($F_A$) and induced pseudoscalar ($F_P$) form factors is less known except for the axial-vector coupling $g_A=F_A(q^2=0)$ that is most accurately measured by neutron beta decay.

From a theoretical perspective, the axial structure of the nucleon is highly connected with the physics
of chiral symmetry and its spontaneous breaking, which ensures the presence of pseudo-Nambu-Goldstone particles such as the pion.
This is empirically known as the partially conserved axial-vector current (PCAC) hypothesis, where the divergence of the axial-vector current
is proportional to the pion field. Applying this idea to the isovector axial-vector matrix element of the nucleon given by
%
%
\begin{align}
\label{eq:NME_AV}
\langle N(p^\prime)|A_\alpha(x)|N(p)\rangle = \overline{u}_N(p^\prime)
\left[
\gamma_\alpha\gamma_5 F_A(q^2) +i q_\alpha \gamma_5 F_P(q^2)
\right]u_N(p)e^{iq\cdot x}
\end{align}
with $q=p -p^\prime$, a specific relation, known as the Goldberger-Treiman (GT) relation~\cite{Goldberger:1958vp}, is
derived between the axial-vector coupling $g_A$ and the residue of the pion-pole structure in the $F_P$ form factor
together with the nucleon mass $M_N$.
Instead of the PCAC hypothesis, the axial Ward-Takahashi identity, $\partial_{\alpha} A_{\alpha}(x) = 2m P(x)$
with a degenerate up and down quark mass $m=m_u=m_d$, leads to the generalized GT relation~\cite{{Weisberger:1966ip},{Bernard:1994wn}, {Sasaki:2007gw}}: 
%
%
\begin{align}
\label{eq:GGT}
2M_N F_A(q^2)-q^2 F_P(q^2)= 2m G_P(q^2),
\end{align}
which is satisfied among the three nucleon form factors including the pseudoscalar ($G_P$) form factor defined in the isovector pseudoscalar matrix element of the nucleon as
%
%
\begin{align}
\label{eq:NME_PS}
\langle N(p^\prime)|P(x)|N(p)\rangle = \overline{u}_N(p^\prime)
\left[
\gamma_5 G_P(q^2)
\right]u_N(p)e^{iq\cdot x}.
\end{align}
In addition, the following pion-pole dominance (PPD) ans\"atz~\cite{{Bernard:1994wn},{Sasaki:2007gw},{Nambu:1960xd}}
for $F_P(q^2)$ and $G_P(q^2)$ at low $q^2$, 
%
%
\begin{align}
\label{eq:PPD_FP}
F_P^{\mathrm{PPD}}(q^2)=\frac{2M_NF_A(q^2)}{q^2+M_\pi^2}
\end{align}
with the pion mass $M_\pi$ and
%
%
\begin{align}
\label{eq:PPD_GP}
2m G_P^{\mathrm{PPD}}(q^2)=2M_NF_A(q^2)\frac{M_\pi^2}{q^2+M_\pi^2}, 
\end{align}
satisfy the generalized GT relation~(\ref{eq:GGT}). For details on the derivation of Eqs.~(\ref{eq:GGT}), (\ref{eq:PPD_FP}) and (\ref{eq:PPD_GP}), see Appendix B in Ref.~\cite{Sasaki:2007gw}. 

We have extensively studied the axial structure of the nucleon in lattice QCD at the physical point using two sets of the PACS10 gauge configurations~\footnote{In the PACS10 project, the PACS Collaboration have generated three sets of the PACS10 gauge configurations at the physical point with lattice volume larger than $(10\;{\rm fm})^4$ and three different lattice spacings~\cite{Yamazaki:2024otm}.
Our study using the third PACS10 gauge configurations is underway~\cite{Tsuji:2024scy}.} in our previous works~\cite{{Shintani:2018ozy},{Tsuji:2023llh},{Ishikawa:2021eut},{Tsuji:2022ric}}, which perform lattice QCD simulations as listed in Table~\ref{tab:measurement_details}.
In our strategy, the smearing parameters of the nucleon interpolation operator were highly optimized to eliminate as much as possible the contribution of excited states in the nucleon two-point function. 
This strategy was quite successful in calculations of the $F_A$ form factor, while the $F_P$ and $G_P$ form factors remained strongly affected by residual contamination of $\pi N$-state contribution. 
As a result, both the $F_P$ and $G_P$ form factors are significantly underestimated in the low-$q^2$ region compared to the PPD model, and
they fail to satisfy the generalized GT relation~\cite{{Shintani:2018ozy},{Tsuji:2023llh}}.

In this work, we propose a simple method to remove the $\pi N$-state contamination from the $F_P$ and $G_P$ form factors given by
the standard ratio method in lattice QCD~\cite{Sasaki:2025qro}.
The application of the novel method
will enable the precise determination of the induced pseudoscalar coupling ($g_P^\ast$) and the pion-nucleon coupling ($g_{\pi NN}$) from existing data calculated at two lattice spacings, 
since these quantities are defined using the $F_P$ form factor~\cite{Thomas:2001kw} as 
%
%
\begin{align}
g_P^\ast &=m_\mu F_P(0.88m_\mu^2), \nonumber\\
g_{\pi NN}&=\lim_{q^2\rightarrow \infty}(q^2+M_\pi^2)\frac{F_P(q^2)}{2F_\pi}, \nonumber
\end{align}
with the muon mass $m_\mu$, and 
can still be compared to respective experimental values.


This paper is organized as follows: in Sec.~\ref{Sec:Method}, after a brief introduction to the standard ratio method for calculating the nucleon form factors in lattice QCD simulations, we describe our proposal to remove $\pi N$-state contamination from 
two specific nucleon form factors $F_P(q^2)$ and $G_P(q^2)$
(hereafter referred to as the leading $\pi N$ subtraction method).
Section~\ref{Sec:Num_results} presents the numerical results of 
the $F_P$ and $G_P$ form factors obtained by the leading $\pi N$ subtraction method with the datasets generated in Refs.~\cite{{Shintani:2018ozy},{Tsuji:2023llh},{Ishikawa:2021eut}}. We then determine the values of $g_P^{\ast}$ and
$g_{\pi NN}$. Finally, we close with a summary in Sec.~\ref{sec:summary}.

In this paper, the nucleon matrix elements are given in the Euclidean metric convention. $\gamma_5$ is defined by $\gamma_5\equiv \gamma_1\gamma_2\gamma_3\gamma_4=-\gamma_5^{M}$, which has the opposite sign relative to that in the Minkowski convention ($\vec{\gamma}^{M}=i\vec{\gamma}$ and $\gamma^{M}_0=\gamma_4$) adopted in the Particle Data Group.
The sign of all the form factors is chosen to be positive. 
The Euclidean four-momentum squared $q^2$ corresponds to the spacelike momentum squared as $q_M^2=-q^2<0$ in Minkowski space.

%
%
\begin{table}[h]
\caption{
Details of gauge ensembles: the $\beta$ value, lattice volume ($L^3\times T$), spatial extent ($La$), 
the simulated pion mass ($M_\pi$), the number of configurations ($N_{\mathrm{conf}}$) and the total number of the measurements ($N_{\mathrm{meas}})$ using on the all-mode-averaging technique~\cite{{Blum:2012uh},{Shintani:2014vja},{vonHippel:2016wid}}
to be analyzed 
for studying the isovector nucleon form factors in our previous works~\cite{{Shintani:2018ozy},{Tsuji:2023llh},{Ishikawa:2021eut},{Tsuji:2022ric}} and also this study.
\label{tab:measurement_details}}
\begin{ruledtabular}
\begin{tabular}{ccccccc}
 $\beta$ & Label  & $L^3\times T$ & $La$ [fm] & $M_\pi$
 [GeV]
 & $N_{\mathrm{conf}}$  & $N_{\mathrm{meas}}$ \\ \hline
 1.82 & PACS10/L128 & $128^3\times 128$ & 10.8 & 0.135 & 20 & $O(2500)$-$O(10000)$\\
 & PACS5/L64 & $64^3\times 64$ & 5.9 & 0.138 & 100 & $O(25000)$-$O(200000)$\\
 2.00& PACS10/L160 &  $160^3\times 160$ & 10.1 & 0.138 & 19 & $O(5000)$-$O(58000)$\\
\end{tabular}
\end{ruledtabular}
\end{table}

\section{Method}
\label{Sec:Method}

\subsection{Standard ratio method}
\label{Sec:Std_method}
First of all, the nucleon operator with a three-dimensional momentum $\bm{p}$ 
is given for the proton state by
%
%
\begin{align}
\label{eq:NucOP}
N(t, \bm{p})=\sum_{\bm{x}}e^{-i\bm{p}\cdot\bm{x}}\varepsilon_{abc}\left[
u_a^T(t,\bm{x})C\gamma_5 d_b(t,\bm{x})
\right]u_c(t,\bm{x})
\end{align}
with the charge conjugation matrix, $C=\gamma_4\gamma_2$. The superscript $T$ denotes
a transposition, while the indices $a$, $b$, $c$ and $u$, $d$ label the color and the flavor, respectively. 
The nucleon two-point function from the source-time position (denoted as $t_{\mathrm{src}}$) to the sink-time position 
(denoted as $t_{\mathrm{sink}}$) is defined as 
%
%
\begin{align}
\label{eq:2ptC}
C_N(t_{\mathrm{sink}}-t_{\mathrm{src}};\bm{p})=\frac{1}{4}{\rm Tr}\left\{
{\cal P}_{+}\langle N(t_{\mathrm{sink}};\bm{p})
\overline{N}(t_{\mathrm{src}};-\bm{p})\rangle\right\}.
\end{align}
On the other hand, when the momentum $\bm{p}$ and $\bm{p}^{\prime}$ are given to the initial and final states, respectively, the nucleon three-point function with fixed source and sink separation is given by
%
%
\begin{align}
\label{eq:3ptC}
&C_{J}^{5z}(t; \bm{p}^{\prime}, \bm{p})=\frac{1}{4}{\rm Tr}\left\{
{\cal P}^{5z}\langle N(t_{\mathrm{sink}};\bm{p}^\prime)\widetilde{J}(t;\bm{q})
\overline{N}(t_{\mathrm{src}};-\bm{p})
\right\},
\end{align}
where the local current $\widetilde{J}$~\footnote{
Hereafter, the currents and the form factors with and without tildes indicate
bare and renormalized ones.} located at the time slice $t$ denotes the axial-vector current ($J=A_\alpha$) or  
the pseudoscalar current ($J=P$), carrying the momentum transfer $\bm{q}=\bm{p}-\bm{p}^\prime$. 
The projection operator ${\cal P}_{+}=\frac{1+\gamma_4}{2}$ appearing in Eq.~(\ref{eq:2ptC}) 
can eliminate contributions from the
opposite-parity state for the case of $|\bm{p}|=0$~\cite{{Sasaki:2001nf},{Sasaki:2005ug}}, while the projection operator ${\cal P}^{5z}={\cal P}_{+}\gamma_5\gamma_3$ appearing in Eq.~(\ref{eq:3ptC}) means that the $z$ direction is chosen as the polarization direction.

In a conventional way to extract the nucleon form factors, the following ratios are constructed by appropriate combinations of two-point and three-point functions 
with the source-sink separation ($t_{\mathrm{sep}}\equiv t_{\mathrm{sink}}-t_{\mathrm{src}}$)~\cite{{Hagler:2003jd},{Gockeler:2003ay}} as 
%
%
\begin{align}
\mathcal{R}_{J}^{5z}(t; \bm{p}^{\prime}, \bm{p}) 
=\frac{C_{J}^{5z}(t; \bm{p}^{\prime}, \bm{p})}{C_2(t, t_{\rm sep}; \bm{p}^\prime, \bm{p})},
\label{eq:ratio_3pt_2pt}
\end{align}
where $C_2(t, t_{\rm sep}; \bm{p}^\prime, \bm{p})\propto e^{-E_N(\bm{p})(t-t_{\mathrm{src}})}e^{-E_N(\bm{p}^\prime)(t_{\mathrm{sink}}-t)}$ is given by the following combination:
%
%
\begin{align}
&C_2(t, t_{\rm sep}; \bm{p}^\prime, \bm{p})\cr
&\equiv
\frac{C_{N}(t_{\mathrm{sink}}-t_{\mathrm{src}}; \bm{p}^{\prime})}{
\sqrt{\frac{C_{N}(t_{\mathrm{sink}}-t; \bm{p}) C_{N}(t-t_{\mathrm{src}}; \bm{p}^{\prime}) C_{N}(t_{\mathrm{sink}}-t_{\mathrm{src}}; \bm{p}^{\prime})}{C_{N}(t_{\mathrm{sink}}-t; \bm{p}^{\prime}) C_{N}(t-t_{\mathrm{src}}; \bm{p}) C_{N}(t_{\mathrm{sink}}-t_{\mathrm{src}}; \boldsymbol{p})}}
},
\end{align}
%
which is representative of $t$ dependence of the nucleon three-point function due to the contribution of the nucleon ground state.
All $t$ dependence due to the contribution of the nucleon ground state can be eliminated 
in the ratio $\mathcal{R}_{J}^{5z}(t; \bm{p}^{\prime}, \bm{p})$,
while it remains for the excited-state contributions.
Therefore, if the condition $t_{\mathrm{sep}}/a\gg
(t-t_{\mathrm{src}})/a\gg 1$ is satisfied, 
the target quantity
can be read off from an asymptotic plateau of the ratio $\mathcal{R}_{J}^{5z}(t; \bm{p}^{\prime}, \bm{p})$,
being independent of the choice of $t_{\mathrm{sep}}$.

For a simplicity, we consider only the rest frame of the final state with $\bm{p}^\prime=\bm{0}$,
which leads to the condition of 
$\bm{q}=\bm{p}-\bm{p}^{\prime}=\bm{p}$. 
Therefore, 
the squared four-momentum transfer is given by $q^2=2M_N(E_N(\bm{q})-M_N)$ where $M_N$ and $E_N(\bm{q})$ represent the nucleon mass and energy with the momentum $\bm{q}$. 
In this kinematics, $\mathcal{R}_{J}^{5z}(t; \bm{p}^{\prime},
\bm{p})$ and $C_{J}^{5z}(t; \bm{p}^{\prime}, \bm{p})$ are rewritten by a simple notation $\mathcal{R}_{J}^{5z}(t; \bm{q})$ and $C_{J}^{5z}(t; \bm{q})$.

%
%
\begin{figure*}[t]
\centering
\includegraphics[width=0.48\textwidth,bb=0 0 864 720,clip]{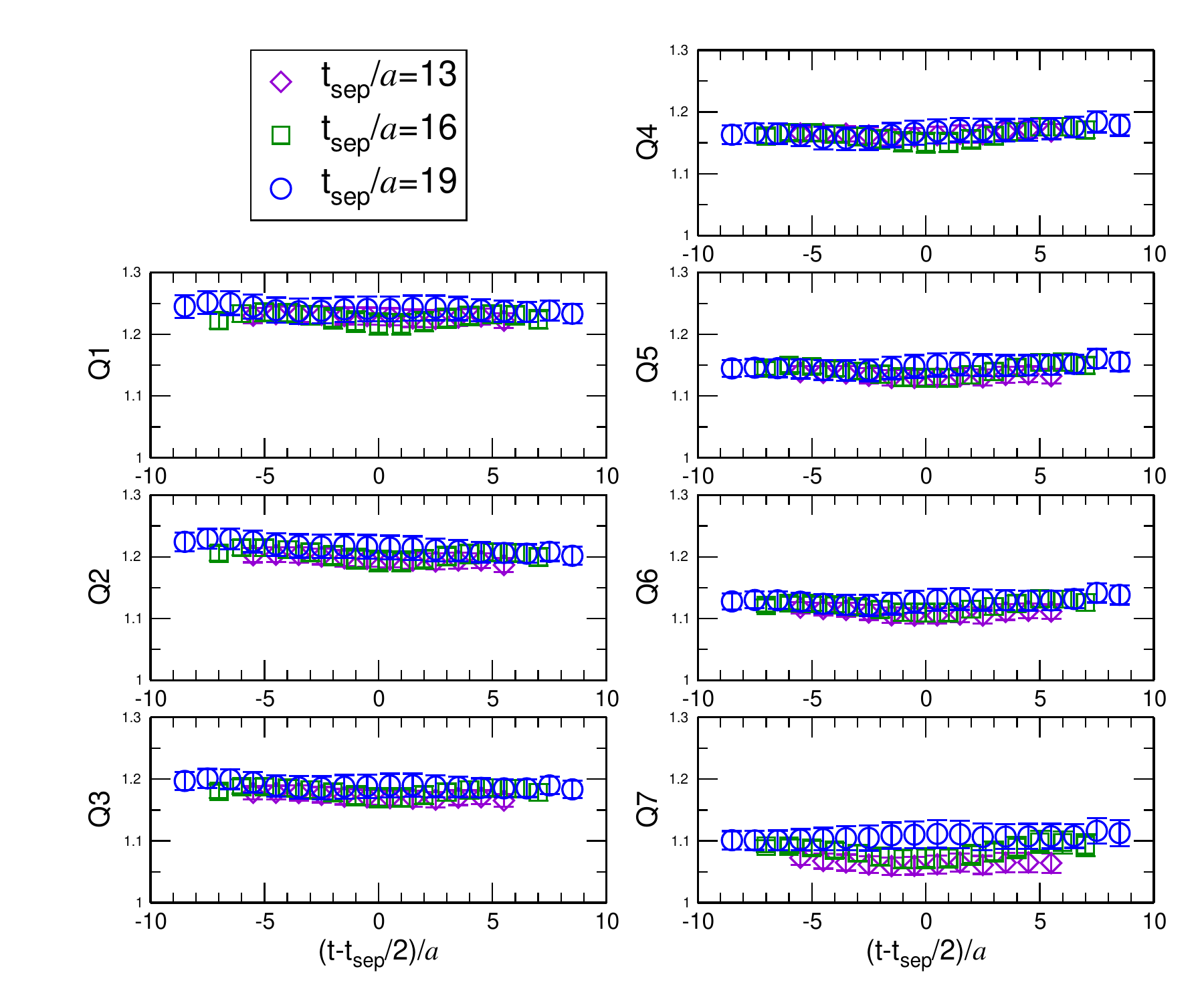}
\includegraphics[width=0.48\textwidth,bb=0 0 864 720,clip]{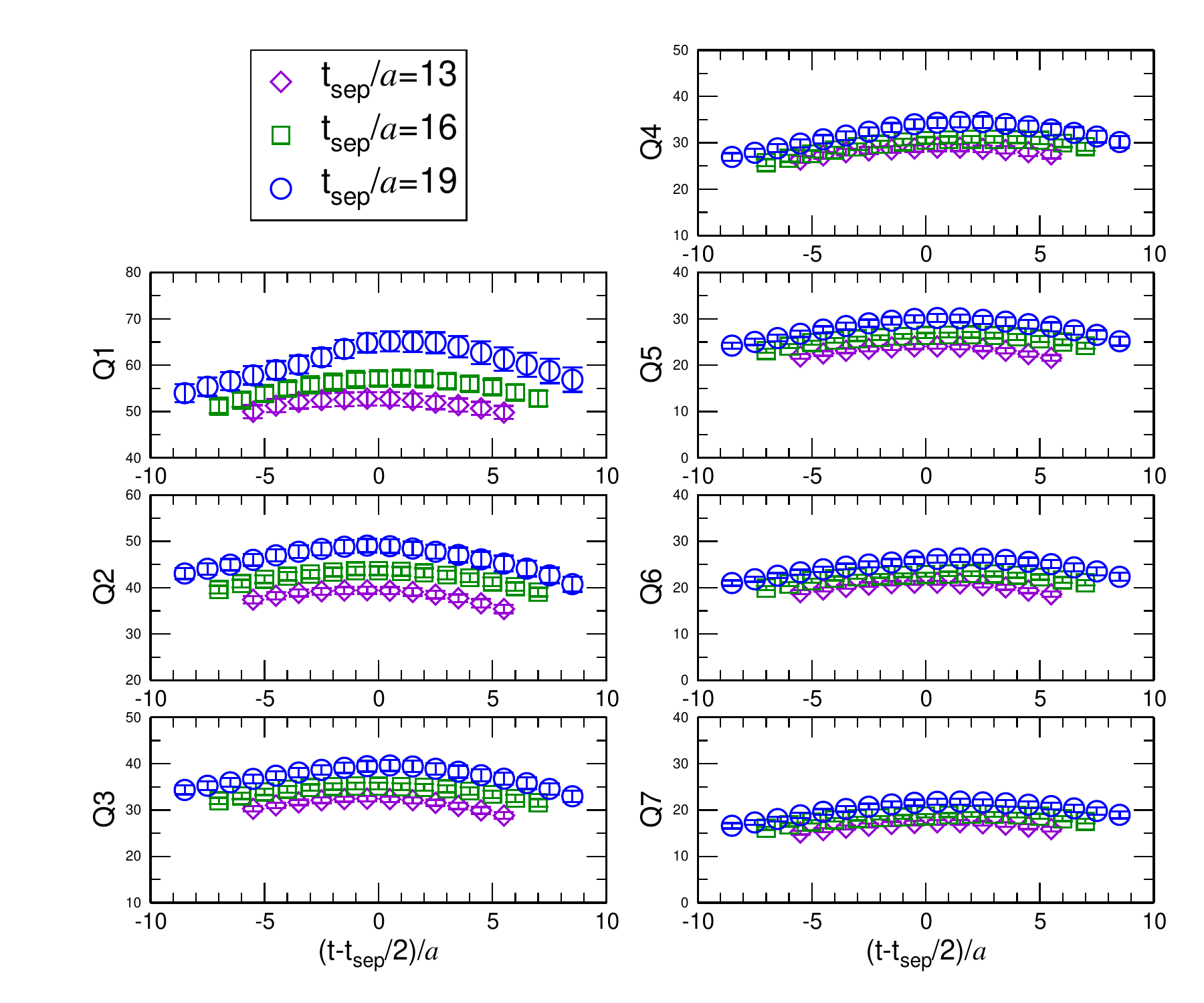}
\includegraphics[width=0.48\textwidth,bb=0 0 864 720,clip]{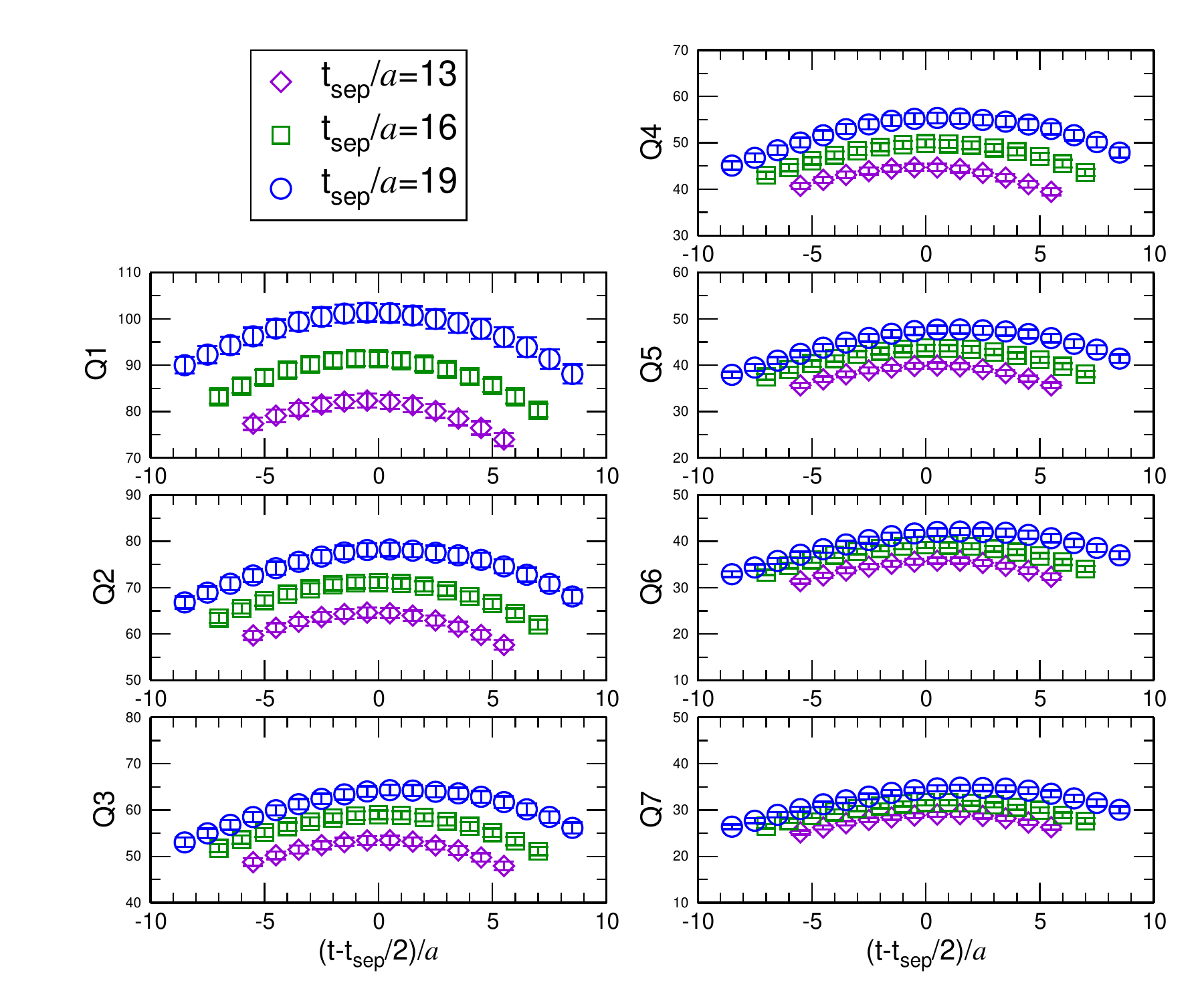}
\caption{
Using the standard ratio method, the values of $F_A$ (upper-left panel), $2M_N F_P$ (upper-right panel) and $\widetilde{G}_P$ (lower panel) calculated from PACS10/L160 with $t_{\mathrm{sep}}/a=13$ (diamonds), $16$ (squares) and $19$ (circles) for seven nonzero momentum transfers (labeled from Q1 to Q7) as functions of the current insertion time slice $t$. 
}
\label{fig:Three_FF_plateau_160}
\end{figure*}

The ratio $\mathcal{R}_{J}^{5z}(t; \bm{q})$ 
gives the following asymptotic values including the respective bare form factors in the asymptotic region~\cite{Sasaki:2007gw}, as follows:
%
%
\begin{widetext}
\begin{align}
    \label{eq:fa_def}
    \mathcal{R}^{5z}_{A_i}(t;\bm{q})
    &=
    K^{-1}
    \left[        (E_N(\bm{q})+M_N)\widetilde{F}_A(q^2)\delta_{i3}-q_iq_3\widetilde{F}_P(q^2) 
    \right] + \cdot\cdot\cdot,\\
    \label{eq:fa4_def}
    \mathcal{R}^{5z}_{A_4}(t;\bm{q})
    &=iq_3 K^{-1}
    \left[
        \widetilde{F}_A(q^2)-(E_N(\bm{q})-M_N)\widetilde{F}_P(q^2)
    \right]+ \cdot\cdot\cdot, \\
    \label{eq:gp_def}
    \mathcal{R}_{P}^{5z}(t; \boldsymbol{q})
    & =iq_3 K^{-1}
    \widetilde{G}_{P}\left(q^{2}\right)+ \cdot\cdot\cdot
\end{align}
\end{widetext}
with $K=\sqrt{2E_N(\bm{q})(E_N(\bm{q})+M_N)}$. The ellipsis denotes excited state contributions, 
which are supposed to be ignored in the case of $t_{\mathrm{sep}}/a\gg (t-t_{\mathrm{src}})/a \gg 1$.
Three target quantities: the axial ($\widetilde{F}_A$), induced pseudoscalar ($\widetilde{F}_P$) and pseudoscalar ($\widetilde{G}_P$) form factors can be read off from an asymptotic plateau
of the ratio $\mathcal{R}_{J}^{5z}(t; \bm{q})$,
being independent of the choice of $t_{\mathrm{sep}}$. This approach is hereafter referred to as the 
standard ratio method.

In our previous works~\cite{{Ishikawa:2018rew},{Shintani:2018ozy},{Ishikawa:2021eut},{Tsuji:2023llh}} 
the correlation functions involving the $A_4$ current are
not taken into account for the calculation of the $\widetilde{F}_A(q^2)$ and $\widetilde{F}_P(q^2)$
form factors. 
Instead of using Eq.~(\ref{eq:fa4_def}), kinematically different types of the three-point functions~(\ref{eq:fa_def}) are 
used to extract the two independent form factors $\widetilde{F}_A(q^2)$ and $\widetilde{F}_P(q^2)$, since 
the $z$ direction is chosen as the polarization direction by the projection operator ${\cal P}^{5z}$~\cite{Sasaki:2007gw}. 
Indeed, the longitudinal momentum ($q_3$) dependence explicitly appears in Eq.~(\ref{eq:fa_def}). 
Therefore, in the transverse case ($i=1$ or 2), ${\cal R}^{5z}_{A_i}(t,{\bm q})$ only contains 
the $\widetilde{F}_P(q^2)$ form factor, while ${\cal R}^{5z}_{A_3}(t,{\bm q})$
contains both $\widetilde{F}_A(q^2)$ and $\widetilde{F}_P(q^2)$ contributions as shown in Eq.~(\ref{eq:fa_def}). 
Even in the longitudinal case ($i=3$), the $\widetilde{F}_A(q^2)$ and $\widetilde{F}_P(q^2)$ can be 
separately determined due to the fact that there are two types of kinematics: $q_3= 0$ or $q_3\neq 0$. 
For the former case, ${\cal R}^{5z}_{A_3}(t,{\bm q})$ only contains the $\widetilde{F}_A(q^2)$. 
Therefore, the subtracted correlator ratio $\overline{\cal R}^{5z}_{A_i}(t,{\bm q})$ that
only includes the $\widetilde{F}_P(q^2)$ contribution can be defined as
%
%
\begin{align}
\overline{\cal R}^{5z}_{A_i}(t,{\bm q})
&\equiv
{\cal R}^{5z}_{A_i}(t,{\bm q})-\delta_{i3}{\cal R}^{5z}_{A_3}(t,{\bm q}_0)\cr
&=-q_iq_3K^{-1}\widetilde{F}_P(q^2)+\cdot\cdot\cdot,
\end{align}
where the second term is evaluated with ${\bm q}_0=(q_1, q_2, 0)$ satisfying $|{\bm q}_0|=|{\bm q}|$.

Finally, we recall that the quark local currents on the lattice receive finite renormalizations relative to their continuum counterparts in general.
The renormalized values of the form factors require the renormalization factors $Z_O\ (O=A,P)$, as follows:
%
%
\begin{align}
  F_A(q^2) &= Z_A\widetilde{F}_A(q^2),\\
  F_P(q^2) &= Z_A\widetilde{F}_P(q^2),\\
  G_P(q^2) &= Z_P\widetilde{G}_P(q^2),
\end{align}
where the renormalization factors are defined through the renormalization of the quark local currents,
$A_\alpha=Z_A \widetilde{A}_\alpha$
and $P=Z_P \widetilde{P}$.
The renormalization factor $Z_A$ is scale independent, while $Z_P$ depends on the renormalization scale. 
To make a comparison with the experimental values, 
two form factors, $F_A$ and $F_P$, will be properly renormalized with $Z_A$,
which is determined by the Schr\"odinger functional method.
On the other hand, the pseudoscalar form factor is provided only as a bare quantity as indicated by $\widetilde{G}_P$ in this study.

In our previous works~\cite{{Ishikawa:2018rew},{Shintani:2018ozy},{Ishikawa:2021eut},{Tsuji:2023llh}}, 
we adopted the strategy that the smearing parameters of the nucleon interpolating operator are highly optimized to eliminate as much as possible the contribution of excited states in the nucleon two-point function.
Figure~\ref{fig:Three_FF_plateau_160} shows
the values of $F_A$ (upper-left panel), $2M_N F_P$ (upper-right panel) and $\widetilde{G}_P$ (lower panel) computed on the $160^4$ lattice (PACS10/L160) with $t_{\mathrm{sep}}/a=13$ (diamonds), $16$ (squares) and $19$ (circles) for 
seven nonzero momentum transfers (labeled from Q1 to Q7) as functions of the current insertion time slice $t$. Although $F_A(q^2)$ shows a clear plateau and no $t_{\mathrm{sep}}$ dependence, suggesting that the ground-state dominance is manifest thanks to the elaborate tuning of the sink and source smearing functions, the excited-state contributions are not completely eliminated in $F_P(q^2)$ and $\widetilde{G}_P(q^2)$.

\subsection{Leading $\pi N$ subtraction method}
\label{sec:New_method}
\subsubsection{Induced pseudoscalar form factor $\widetilde{F}_P(q^2)$}

\begin{figure}[h]
\centering
\includegraphics[width=0.80\textwidth,bb=0 0 864 720,clip]{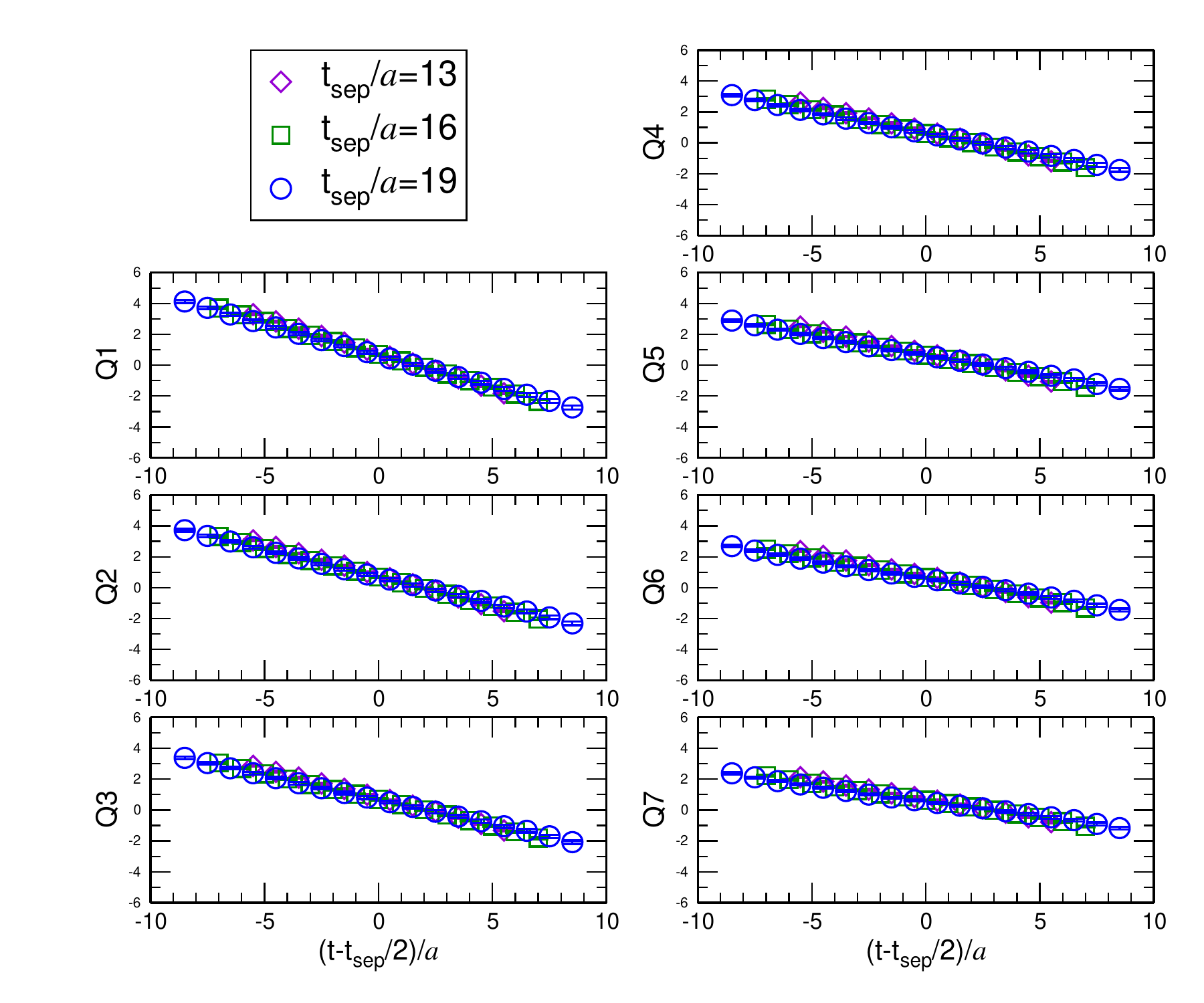}
\caption{
Results of the ratio $\mathcal{R}^{5z}_{A_4}(t;\bm{q})/(iq_3 K^{-1})$
calculated from PACS10/L160 with $t_{\mathrm{sep}}/a=13$ (diamonds), $16$ (squares) and $19$ (circles) for seven nonzero momentum transfers (labeled from Q1 to Q7) as functions of the current insertion time slice $t$. 
The nucleon matrix element of $A_4$ exposes the presence of 
$\pi N$-state contribution described by $\Delta_-(t)$ defined in Eq.~(\ref{Eq:Delta_ex}).
}
\label{fig:RA4_correlator}
\end{figure}

In our previous works~\cite{{Ishikawa:2018rew},{Shintani:2018ozy},{Ishikawa:2021eut},{Tsuji:2023llh}}, the correlation functions involving the $A_4$ current are
not taken into account for the calculation of the $\widetilde{F}_A(q^2)$ and $\widetilde{F}_P(q^2)$
form factors.
This is simply because, to the best of our knowledge, the $A_4$ correlator was found to be statistically very noisy in Ref.~\cite{Ishikawa:2018rew}, where the time-reversal averaging was performed using both forward and backward propagations in time for all three-point correlation functions.
However, as pointed out for the first time in Ref.~\cite{Bali:2018qus}, 
the correlator ratio ${\cal R}^{5z}_{A_4}(t,{\bm q})$ does not show a plateau, but 
rather a peculiar behavior that depends linearly on the current insertion time $t$
with a steep negative slope under their kinematic setup.
When no time-reversal averaging is applied in our data as shown in
Fig.~\ref{fig:RA4_correlator}, an almost linear $t$ dependence is indeed confirmed, giving the same slope, although the direction is reversed according to the respective kinematics.

%
%

\begin{figure*}[t]
\centering
\includegraphics[width=0.3\textwidth,bb=0 0 485 392,clip]{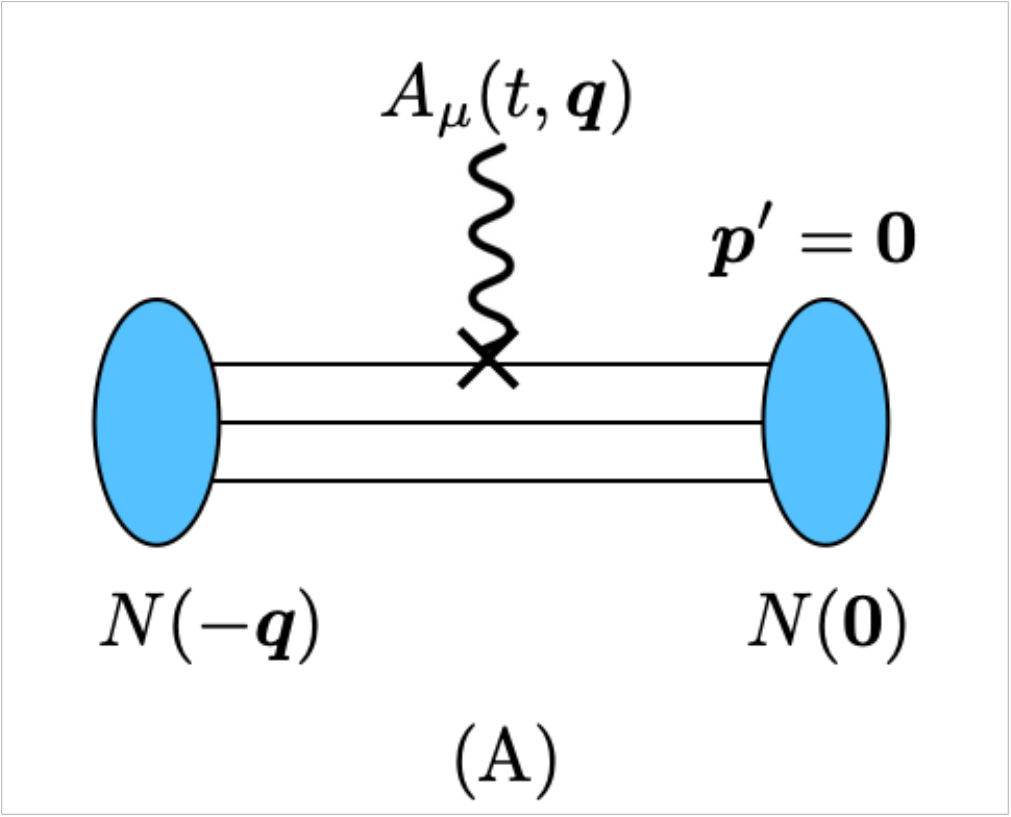}
\includegraphics[width=0.3\textwidth,bb=0 0 485 392,clip]{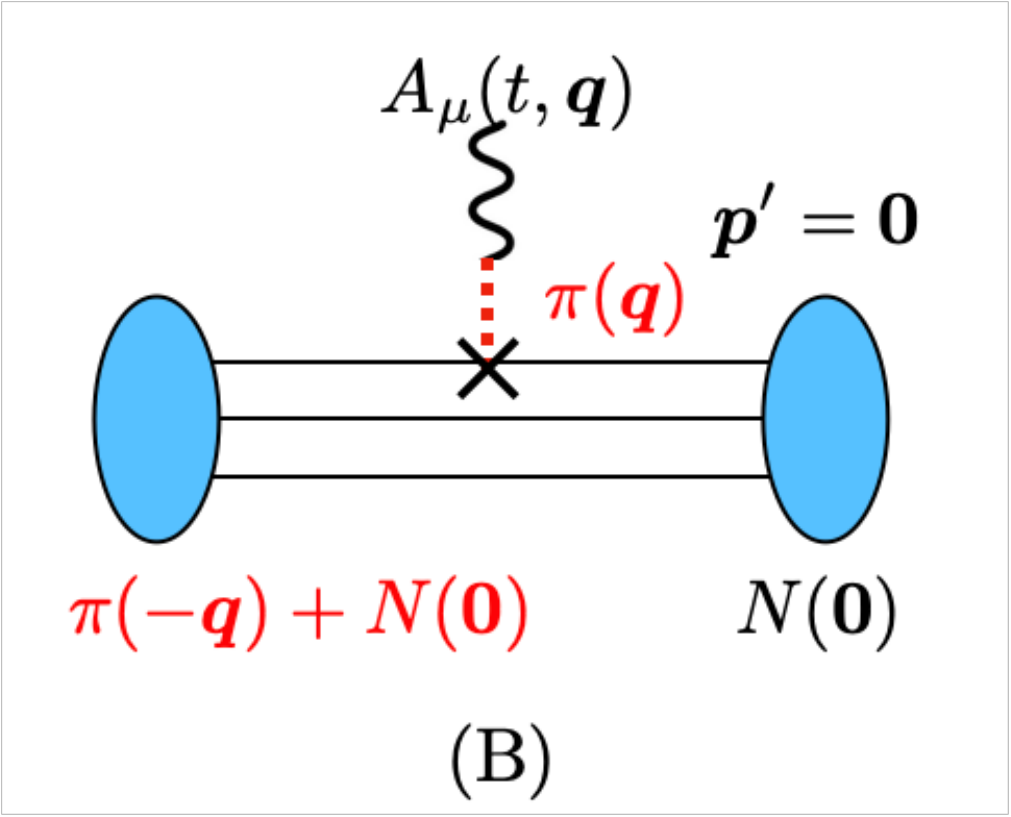}
\includegraphics[width=0.3\textwidth,bb=0 0 485 392,clip]{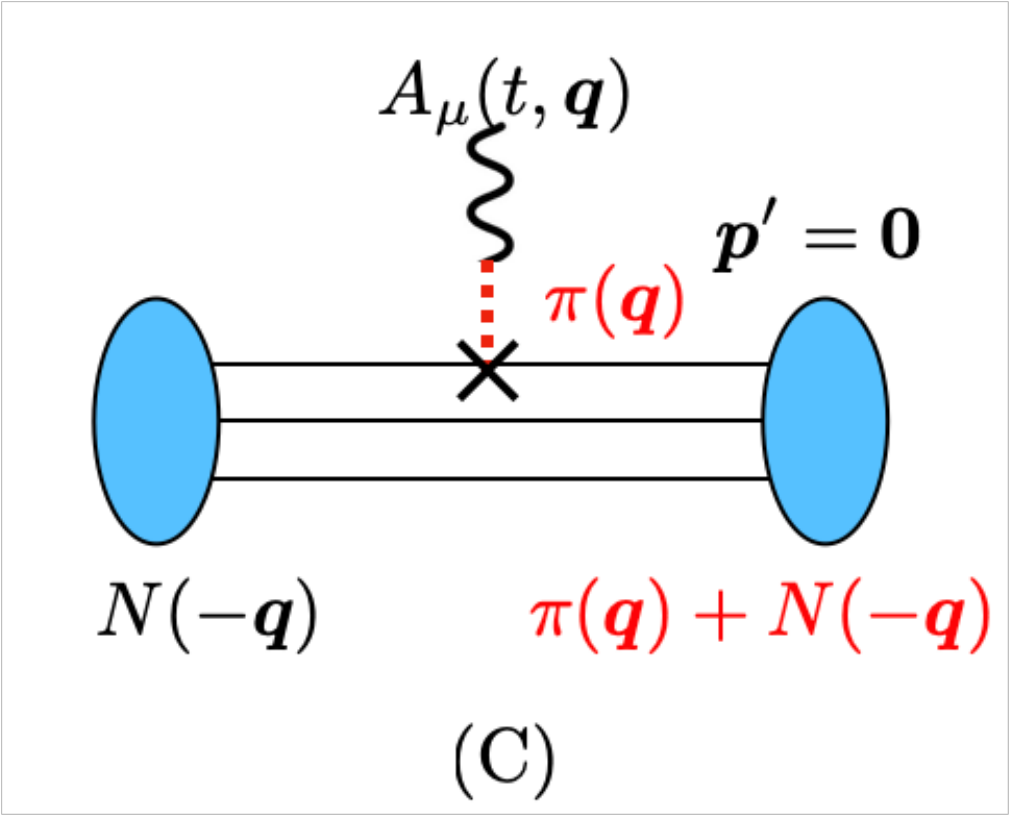}
\caption{Schematic view of the ground-state contribution \rm{(A)}
and two types of the leading $\pi N$ contributions \rm{(B)} and \rm{(C)} for the axial-vector matrix element. 
}
\label{fig:quarkline_diag}
\end{figure*}

As discussed in Refs.~\cite{{Bar:2018xyi},{Bar:2019gfx}}, such peculiar time dependence is understood as 
the leading contribution from the $\pi N$ state in ${\cal R}^{5z}_{A_4}(t,{\bm q})$, 
arising in the tree diagram of the baryon chiral perturbation theory (ChPT). 
Importantly, the momentum $\bm{q}$ injected by the axial-vector current is entirely inherited 
by the pion state, since the pion in such $\pi N$ state remains in the on-mass shell. 
As illustrated in Fig.~\ref{fig:quarkline_diag}, 
the kinematics of the leading 
correction to
the ground-state contribution $N(-\bm{q})\rightarrow N(\bm{0})$ [depicted in Fig.~\ref{fig:quarkline_diag}(A)]
is thus constrained to two special cases~\cite{{Meyer:2018twz},{RQCD:2019jai}}:
%
%
%
\begin{itemize}
\item $\pi(-\bm{q})N(\bm{0}) \rightarrow N(\bm{0})$
[depicted in Fig.~\ref{fig:quarkline_diag}(B)]
\item $N(-\bm{q}) \rightarrow \pi(\bm{q})N(-\bm{q})$
[depicted in Fig.~\ref{fig:quarkline_diag}(C)]
\end{itemize}
where the momentum ${\bm q}$ is carried by the axial-vector current operator.
It is important to note that neither case is always the lowest energy of the possible $\pi N$ 
states as excited states. However, in Ref.~\cite{Jang:2019vkm}, a detailed analysis was carried out 
to identify the main contributions to excited-state contamination in ${\cal R}^{5z}_{A_4}(t,{\bm q})$
using multi-state fits of the two-point function and the $A_4$ correlator ${\cal C}^{5z}_{A_4}(t,{\bm q})$, 
and indeed the particular $\pi N$ state described in Fig.~\ref{fig:quarkline_diag} was confirmed to 
be a major contribution as the leading $\pi N$ contribution in ${\cal R}^{5z}_{A_4}(t,{\bm q})$.

Similar contributions from the same $\pi N$ states may appear even in ${\cal R}^{5z}_{A_i}(t,{\bm q})$. 
Indeed, Ref.~\cite{RQCD:2019jai} proved it by providing a more general discussion at tree level in baryon ChPT 
compared to Refs.~\cite{{Bar:2018xyi},{Bar:2019gfx}}, and showed that the relative signs of the two $\pi N$ contributions are reversed between the $A_4$ and $A_{i}$ correlators.

%
%

As described earlier at the end of Sec.~\ref{Sec:Std_method}, 
in our previous works~\cite{{Ishikawa:2018rew},{Shintani:2018ozy},{Ishikawa:2021eut},{Tsuji:2023llh}}, 
the $F_P$ form factor obtained from $\overline{\cal R}^{5z}_{A_i}(t,{\bm q})$ 
was indeed significantly affected by the excited-state contamination, though no such effect was observed for 
the $F_A$ form factor. Recent studies solving the generalized eigenvalue problem including $\pi N$ operators also show that the $\pi N$ contributions are strong in $\widetilde{F}_P(q^2)$ and $\widetilde{G}_P(q^2)$, 
but not in $\widetilde{F}_A(q^2)$~\cite{{Barca:2022uhi},{Alexandrou:2024tin}}.
Therefore, we assume that the contributions from the leading $\pi N$ state
for ${\cal R}^{5z}_{A_\mu}(t,{\bm q})$ can be described as follows
%
%
\begin{align}
\overline{\cal R}^{5z}_{A_i}(t,{\bm q})
&\equiv{\cal R}^{5z}_{A_i}(t,{\bm q})-\delta_{i3}{\cal R}^{5z}_{A_3}(t,{\bm q}_0)
\cr
&=
- q_3q_iK^{-1}\left[\widetilde{F}_P(q^2)-\Delta_{+}(t, t_{\mathrm{sep}}; {\bm q}) \right], 
\label{Eq:A_space} \\
%
%
{\cal R}^{5z}_{A_4}(t,{\bm q})
&=iq_3K^{-1}\left[
\left(\widetilde{F}_A(q^2)-(E_N({\bm q})-M_N)\widetilde{F}_P(q^2)\right) \right.\cr
&\left.+E_\pi({\bm q}) \Delta_{-}(t, t_{\mathrm{sep}}; {\bm q}) \right],
\label{Eq:A_time}
\end{align}
where the functions $\Delta_{\pm}(t, t_{\mathrm{sep}}; {\bm q})$ encode the leading $\pi N$ contributions, which
provide the residual $t$ dependence with a given $t_{\mathrm{sep}}$. See Appendix~\ref{app:MOS}, for detailed derivations based on an argument given by Meyer-Ottnad-Schulz~\cite{Meyer:2018twz}. 

For the case when the axial-vector current operator carries the momentum ${\bm q}$, 
the $\pi N$ contribution can be expressed by the following form with $t$-independent coefficients $B(\bm{q})$ and $C(\bm{q})$:
%
%
\begin{align}
\Delta_{\pm}(t, t_{\mathrm{sep}}; {\bm q})=
B(\bm{q}) e^{-\Delta E({\bm q},-{\bm q})t} \pm C(\bm{q}) e^{-\Delta E({\bm 0},{\bm q})(t_{\mathrm{sep}}-t)},
\label{Eq:Delta_ex}
\end{align}
where the non-interacting estimate $\Delta E({\bm q},{\bm k})=E_{\pi}({\bm k})+E_{N}({\bm q}+{\bm k})-E_{N}({\bm q})$ 
may be used. Therefore, the time derivative of the $\pi N$ contribution $\Delta_{\pm}(t, t_{\mathrm{sep}}; {\bm q})$ may have 
the following property:
%
%
\begin{align}
\partial_4 \Delta_{\pm}(t, t_{\mathrm{sep}}; {\bm q})=-E_\pi({\bm q}) \Delta_{\mp}(t, t_{\mathrm{sep}}; {\bm q})+(E_N({\bm q})-M_N)\Delta_{\pm}(t, t_{\mathrm{sep}}; {\bm q}),
\label{eq:Time_Derivative_Delta}
\end{align}
which offers us to separate the $\pi N$ contribution $\Delta_{\pm}(t, t_{\mathrm{sep}}; {\bm q})$ from the $\widetilde{F}_P$ form factor using two types of the time derivative of the correlator ratio, $\partial_4\overline{\cal R}^{5z}_{A_i}(t,{\bm q})$ and
$\partial_4 {\cal R}^{5z}_{A_4}(t,{\bm q})$.~\footnote{The time derivative $\partial_4$ is defined by $\partial_4 f(t)=\frac{1}{2}\left(f(t+1)-f(t-1)\right)$ so that Eq.~(\ref{eq:Time_Derivative_Delta}) is valid up to ${\cal O}(a^2)$.}
Hereafter the nucleon energy $E_N(\bm{q})$ and the pion energy $E_\pi(\bm{q})$ are
simply abbreviated by shorthand notations $E_N$ and $E_\pi$, respectively.

The leading $\pi N$ subtraction method for determining $\widetilde{F}_P(q^2)$, including the time derivatives of 
the $A_4$ and $A_i$ correlators, is given by
%
%
\begin{widetext}
\begin{align}
\label{eq:new_FP}
\widetilde{F}_P(q^2)
&=-K\frac{\overline{\cal R}^{5z}_{A_i}(t, {\bm q})}{q_iq_3}
+\frac{K}{(\Delta E_N)^2-E_\pi^2}
\left[
\Delta E_N\frac{\partial_4 \overline{\cal R}^{5z}_{A_i}(t, {\bm q})}{q_iq_3}
+\frac{\partial_4 {\cal R}^{5z}_{A_4}(t, {\bm q})}{iq_3}
\right] 
\end{align}
\end{widetext}
with $\Delta E_N\equiv E_N-M_N$ and $K=\sqrt{2E_N(E_N+M_N)}$.
The first term corresponds to $\widetilde{F}_P^{\mathrm{std}}(q^2)$ in the standard ratio method. The leading $\pi N$ contributions represented in terms of $\Delta_{+}(t, t_{\mathrm{sep}}; {\bm q})$ 
and $\Delta_{-}(t, t_{\mathrm{sep}}; {\bm q})$ can be completely eliminated by adding the second term in Eq.~(\ref{eq:new_FP}). For the ground-state contribution, namely $\widetilde{F}_P(q^2)$, 
Eq.~(\ref{eq:new_FP}) is just a harmless linear combination exploiting the redundancy in the determination of
 $\widetilde{F}_P(q^2)$ from both of $C_{A_4}^{5z}(t; \bm{q})$ and ${C}_{A_i}^{5z}(t; \bm{q})$. Furthermore, 
 from Eqs.~(\ref{Eq:A_space}) and (\ref{Eq:A_time}), it is clear that the terms in the parentheses in Eq.~(\ref{eq:new_FP}) do not have the ground-state contribution.
Therefore, if Eq.~(\ref{eq:new_FP}) successfully shows good plateau behavior, independent of the choice of $t_{\mathrm{sep}}$, it guarantees that the 
ground-state contribution can be read accurately without excited-state contamination.

\subsubsection{Pseudoscalar form factor $\widetilde{G}_P(q^2)$}

The leading $\pi N$ contributions are due to the pion-pole contribution of the target form factor, as discussed in Appendix~\ref{app:MOS}. Therefore, the $P$ correlator $C_P^{5z}(t; \bm{q})$ also suffers from the contamination of the particular $\pi N$ state contributions. Indeed, the $\widetilde{G}_P(q^2)$ was observed to be strongly contaminated from the excited state, similar to the $\widetilde{F}_P(q^2)$ in our previous works~\cite{{Ishikawa:2018rew},{Shintani:2018ozy},{Ishikawa:2021eut},{Tsuji:2023llh}}.
We simply express that  
%
%
\begin{align}
\label{Eq:P_space}
{\cal R}^{5z}_{P}(t,{\bm q})
&=iq_3K^{-1}
\left[\widetilde{G}_P(q^2)-\Delta_{P}(t, t_{\mathrm{sep}}; {\bm q}) \right],
\end{align}
where $\Delta_{P}(t, t_{\mathrm{sep}}; {\bm q})$ encodes the leading $\pi N$-state contributions that
cause a residual $t$ dependence in ${\cal R}_P^{5z}(t; \bm{q})$ as shown in the lower panel of Fig.~\ref{fig:Three_FF_plateau_160}.
Unlike in the case of the axial-vector currents, only a single correlator cannot remove the $\pi N$ 
contribution $\Delta_{P}(t, t_{\mathrm{sep}}; {\bm q})$.

It is important to recall here that in our previous study that the axial Ward-Takahashi identity is well satisfied 
in terms of the three-point functions of the nucleon, as follows:
%
%
\begin{align}
\label{eq:AWTI}
Z_A [
\partial_\alpha C_{A_{\alpha}}^{5z}(t; \bm{q})
]=2 m_{\rm PCAC} C_{P}^{5z}(t; \bm{q}),
\end{align}
where $m_{\rm PCAC}$ corresponds to the bare quark mass which coincides with
the value determined from the pion two-point correlation functions~\cite{Tsuji:2023llh}. 
It should be emphasized that Eq.~(\ref{eq:AWTI}) is satisfied without isolating the ground-state
contribution from the excited-state contributions, as first pointed out in Ref.~\cite{Bali:2018qus}.
Thus, Eq.~(\ref{eq:AWTI}) leads to the following PCAC relation for the leading $\pi N$ contributions involved in $\overline{\cal R}_{A_i}^{5z}(t; \bm{q})$ and 
${\cal R}_P^{5z}(t; \bm{q})$:
%
%
\begin{align}
\label{eq:piN_AWTI}
\Delta_{P}(t, t_{\mathrm{sep}}; {\bm q})=Z_A \frac{M_\pi^2}{2m_{\rm PCAC}}\Delta_{+}(t, t_{\mathrm{sep}}; {\bm q}),
\end{align}
which also offers a simple subtraction method for determining the $\widetilde{G}_P(q^2)$, as follows:
%
%
\begin{widetext}
\begin{align}
\label{eq:new_GP}
\widetilde{G}_P(q^2)
&=K\frac{{\cal R}^{5z}_{P}(t, {\bm q})}{iq_3}
+\frac{Z_A B_0 K}{(\Delta E_N)^2-E_\pi^2}
\left[
\Delta E_N\frac{
\partial_4 \overline{\cal R}^{5z}_{A_i}(t, {\bm q})
}{q_iq_3}
+\frac{\partial_4 {\cal R}^{5z}_{A_4}(t, {\bm q})}{iq_3}
\right] 
\end{align}
\end{widetext}
with the bare low-energy constant $B_0=\frac{M_\pi^2}{2m_{\mathrm{PCAC}}}$.
The first term corresponds to $\widetilde{G}_P^{\mathrm{std}}(q^2)$ in the standard ratio method.
See Appendix~\ref{app:derivation_eq_25} for details on the derivation of Eq.~(\ref{eq:piN_AWTI}).

\section{Numerical results}
\label{Sec:Num_results}
In this study, we reanalyze the datasets generated in Refs.~\cite{{Shintani:2018ozy},{Tsuji:2023llh},{Ishikawa:2021eut}} for the induced pseudoscalar
form factor $\widetilde{F}_{P}(q^2)$ and pseudoscalar form factor $\widetilde{G}_{P}(q^2)$ using the leading $\pi N$ subtraction method described in Sec.~\ref{sec:New_method}.
The two datasets are computed with the first and second PACS10 ensembles, which are two sets of gauge configurations generated in a large volume of over $(10\;\mathrm{fm})^4$ 
by the PACS Collaboration with the six stout-smeared 
${\cal O}(a)$ improved Wilson-clover quark action and Iwasaki gauge action~\cite{Iwasaki:1983iya} at $\beta=1.82$ and 2.00 corresponding
to the lattice spacings of 0.09 fm (coarse) and 0.06 fm (fine)~\cite{{Ishikawa:2018jee},{PACS:2019ofv},{Shintani:2019wai}}, respectively.
At the coarse lattice spacing, we have two lattice volumes (linear spatial extents of 10.9 fm
and 5.5 fm)~\cite{{Ishikawa:2018jee},{PACS:2019ofv}} to examine the finite volume effect. A brief summary of the simulation parameters is given in Table~\ref{tab:simulation_details}.
The simulated pion masses on the coarse $128^4$
and $64^4$ lattices and the fine $160^4$ lattice are almost at the physical point as listed in Table~\ref{tab:measurement_details}.

In the calculation of nucleon two-point and three-point functions, the all-mode-averaging (AMA) technique~\cite{{Blum:2012uh},{Shintani:2014vja},{vonHippel:2016wid}} is used to significantly reduce the computational cost of multiple measurements and to achieve a much higher statistical accuracy.
We compute the combination of the correlation function with the quark propagator in the
high-precision and low-precision calculations. 
For the low-precision calculations,
the position of the source operator is
changed with respect to the translational
symmetry $G$ on the lattice and the temporal direction of the configuration is also changed using the hypercubic symmetry of our lattice setup ($L=T$). If the number of the low-precision calculations is given by $N_{G}$,
the total number of measurements
(denoted by $N_{\mathrm{meas}}=N_G \times N_{\mathrm{conf}}$) is $O(10^4\mbox{-}10^5)$
for the $64^4$ lattice or $O(10^3\mbox{-}10^4)$ for the $128^4$ and $160^4$ lattices. See
Refs.~\cite{{Tsuji:2022ric},{Tsuji:2023llh}}
for further details.

The nucleon interpolating operator defined in Eq.~(\ref{eq:NucOP})
is highly optimized to eliminate as much as possible the contribution of excited states in the nucleon two-point function 
using exponentially smeared source (sink) with Coulomb gauge fixing. Details of our choice of smearing parameters are
described in Refs.~\cite{{Shintani:2018ozy},{Tsuji:2023llh}}. 

As for the three-point functions, the sequential source method
with a fixed source-sink separation~\cite{{Martinelli:1988rr},{Sasaki:2003jh}} is employed and calculated with $t_{\mathrm{sep}}/a=\{10, 12, 14, 16\}$ ($t_{\mathrm{sep}}/a=\{12, 14, 16\}$) for the coarse $128^4$ ($64^4$) lattice ensemble and $t_{\mathrm{sep}}/a=\{13, 16, 19\}$ for the fine $160^4$ lattice ensemble. 
Seven nonzero spatial momenta ${\bm q}=\frac{2\pi}{aL}$ with ${\bm n}=(1,0,0)$, $(1,1,0)$, $(1,1,1)$, $(2,0,0)$, $(2,1,0)$, $(2,1,1)$, $(2,2,0)$ (labeled from Q1 to Q7) are used to determine the nucleon form factors at $q^2\neq 0$. In addition, the renormalization factor $Z_A$ is determined by the Schr\"odinger functional (SF) method~\cite{
{PACS:2019ofv},{Tsuji:2023llh}} as tabulated in Table~\ref{tab:PCAC_GGT_details}.

%
%
\begin{table}[h]
\caption{
Summary of simulation parameters in 2+1 flavor PACS10 ensembles with two different
lattice spacings: gauge coupling ($\beta=6/g^2$), hopping parameters for light ($\kappa_{ud}$) and strange ($\kappa_s$) quarks, clover coefficient ($c_{\mathrm{SW}}$), lattice cutoff ($a^{-1}$) and lattice spacing ($a$). 
See Refs.~\cite{{Ishikawa:2018jee},{PACS:2019ofv},{Shintani:2019wai}} for further details.
\label{tab:simulation_details}}
\begin{ruledtabular}
\begin{tabular}{cccccc}
 $\beta$ & $\kappa_{ud}$ & $\kappa_{s}$ & $c_{\mathrm{SW}}$ & $a^{-1}$ [GeV] &  $a$ [fm] \cr
\hline
1.82&0.126117 & 0.124902 & 1.11&2.3162(44)& 0.08520(16)  \\
2.00& 0.125814& 0.124925 & 1.02 &3.1108(70) & 0.06333(14) \\
\end{tabular}
\end{ruledtabular}
\end{table}

%
%
\begin{table}[h]
\caption{
Summary of results for basic physical quantities in 2+1 flavor PACS10 ensembles with two different
lattice spacings: the $\beta$ value, the PCAC quark mass ($\hat{m}_{\mathrm{PCAC}}$)~\cite{{PACS:2019ofv},{Tsuji:2023llh}}, the pion decay constant ($F_\pi$)~\cite{Ishikawa:2022ulx} and the nucleon mass ($M_N$)~\cite{{Shintani:2018ozy},{Tsuji:2023llh}} in lattice units, 
and the renormalization constant for the axial-vector current determined by the SF scheme ($Z_A^{\mathrm{SF}}$)~\cite{{PACS:2019ofv},{Tsuji:2023llh}}.
\label{tab:PCAC_GGT_details}}
\begin{ruledtabular}
\begin{tabular}{ccccc}
 $\beta$ &  $a\hat{m}_{\rm PCAC}$~\footnotemark[1]\footnotetext[1]{The bare quark mass $m_{\mathrm{PCAC}}$ that appears in Eq.~(\ref{eq:AWTI}) is defined by
$m_{\mathrm{PCAC}}=Z_A^{\mathrm{SF}}\hat{m}_{\mathrm{PCAC}}$.}  
 & $aF_\pi$~\footnotemark[2]\footnotetext[2]{
We use a traditional convention as $F_\pi=f_\pi/\sqrt{2}\sim 93$ MeV, while $f_\pi$ is quoted in Ref.~\cite{Ishikawa:2022ulx}.}
 & $aM_N$  & $Z_A^{\mathrm{SF}}$ \\
\hline
1.82& 0.0013663(143)
& 
0.040244(62)
& 0.4041(47)
& 0.9650(68) \\
2.00 & 0.0010511(038)
& 0.030387(22)
& 0.3045(8)
&0.9783(21)\\
\end{tabular}
\end{ruledtabular}
\end{table}

\subsection{Induced pseudoscalar form factor ${F}_{P}(q^2)$}

\begin{figure}[h]
\centering
\includegraphics[width=0.80\textwidth,bb=0 0 864 720,clip]{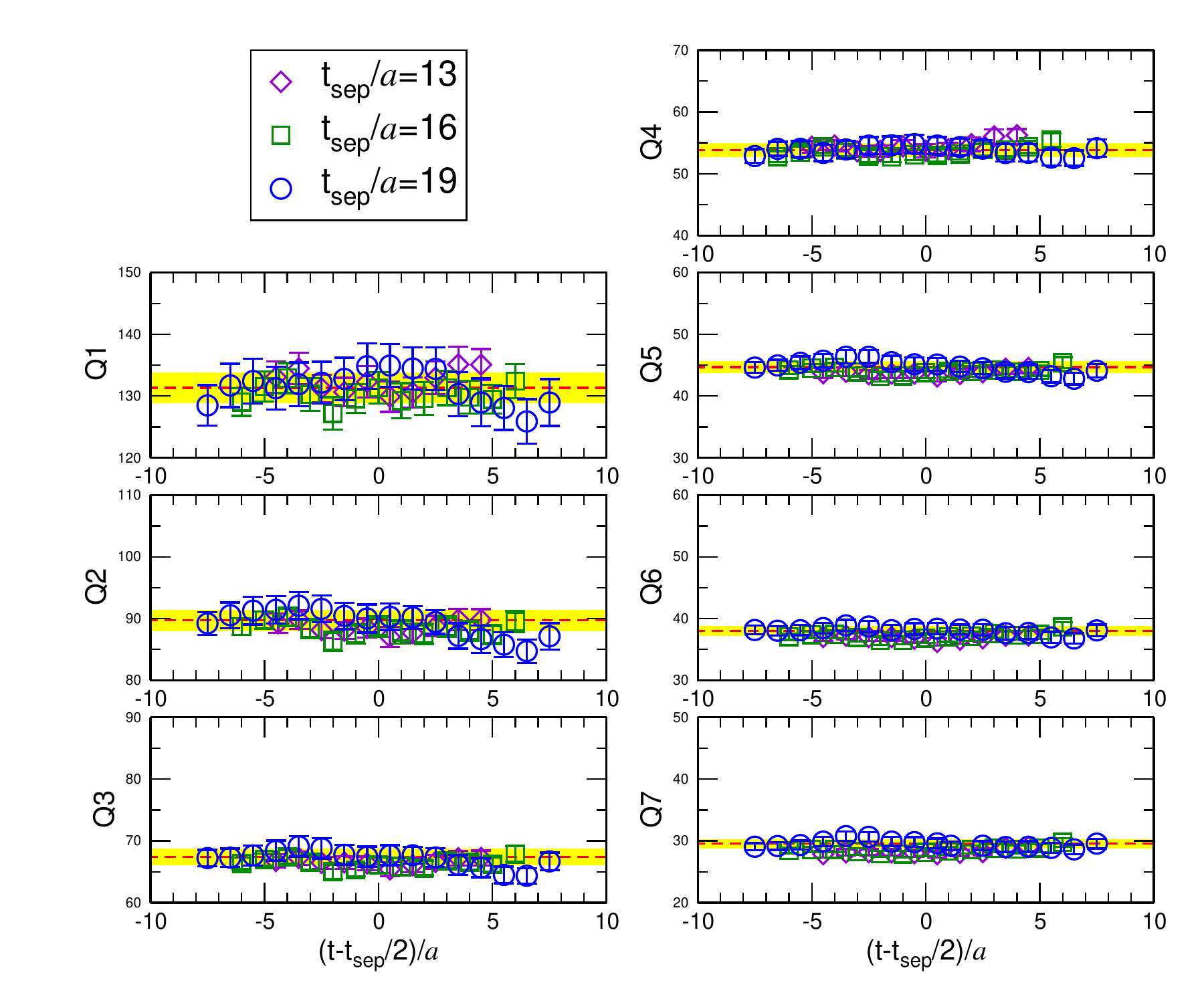}
\caption{
Using the leading $\pi N$ subtraction method as defined in Eq.~(\ref{eq:new_FP}),
the values of the $F_P$ form factor multiplied by $2M_N$ calculated from  PACS10/L160
with $t_{\mathrm{sep}}/a=13$ (diamonds), $16$ (squares) and $19$ (circles) are shown for seven nonzero momentum transfers (labeled from Q1 to Q7) as functions of the current insertion time slice $t$. The horizontal dashed line together with yellow band in each panel is calculated from the PPD model ($2M_N F_P^{\rm PPD}(q^2)$) given in Eq.(\ref{eq:PPD_FP})}.
\label{fig:F_P_New_plateau}
\end{figure}

The $\widetilde{F}_P$ form factor is extracted from Eq.~(\ref{eq:new_FP}) as a function of the current insertion time $t$.
Figure~\ref{fig:F_P_New_plateau} shows the $t$ dependence
and $t_{\mathrm{sep}}$ dependence of $F_P(q^2)=Z_A^{\mathrm{SF}} \widetilde{F}_P(q^2)$ obtained by the leading $\pi N$ subtraction method for the $160^4$ lattice ensemble (PACS10/L160) as typical examples. The leading $\pi N$ subtraction method is really effective in obtaining an asymptotic plateau in all choices of $t_{\mathrm{sep}}/a=\{13,16,19\}$ for all variations of $q^2$. 
Indeed, the good plateaus show that the $t$ dependence is eliminated and the $t_{\mathrm{sep}}$ dependence is not visible either. 
Furthermore, the plateau values are consistent with the PPD model estimates given in Eq.~(\ref{eq:PPD_FP}) within the calculated range of $q^2$.

%
%

\begin{figure}[h]
\centering
\includegraphics[width=0.48\textwidth,bb=0 0 864 720,clip]{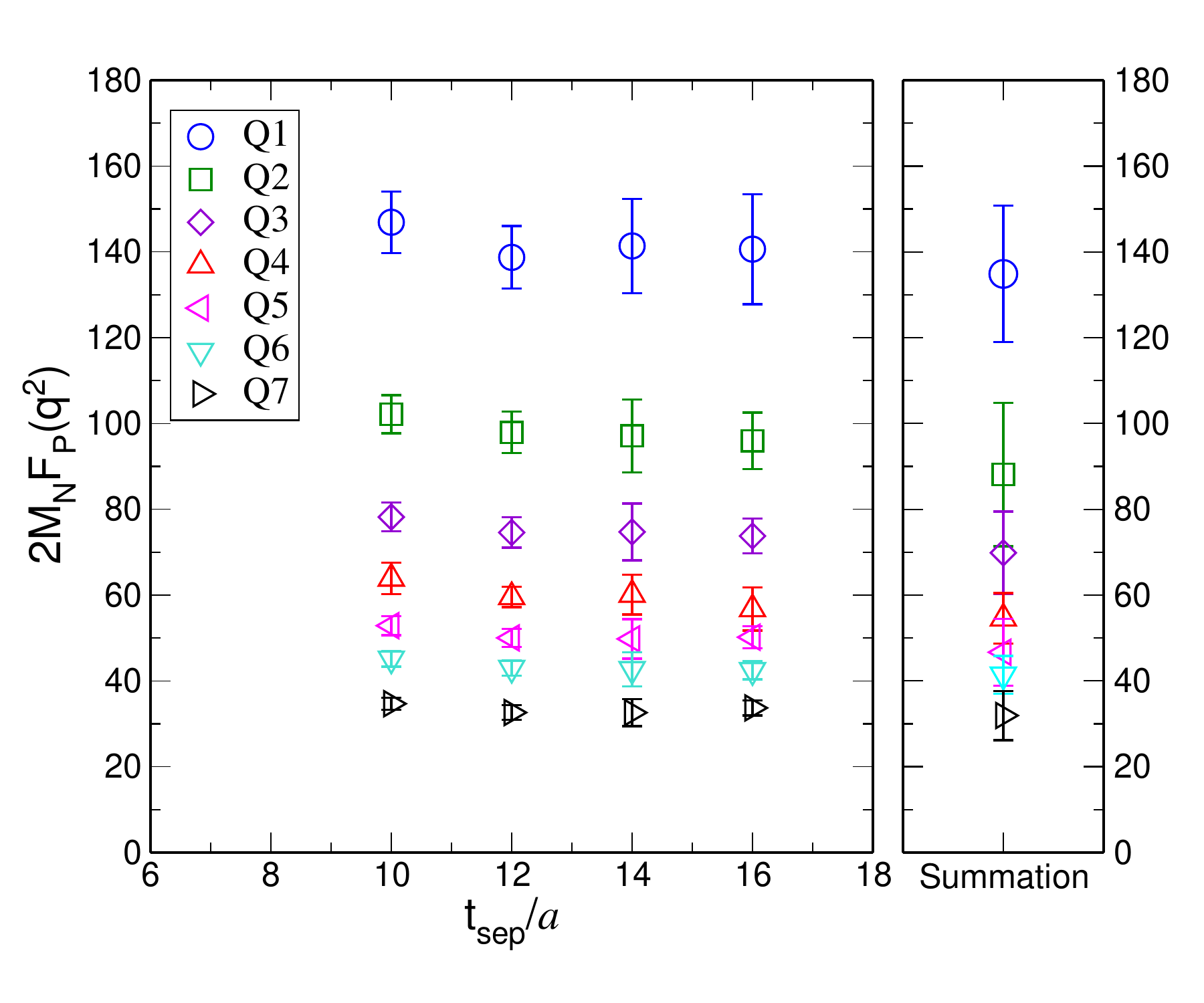}
\includegraphics[width=0.48\textwidth,bb=0 0 864 720,clip]{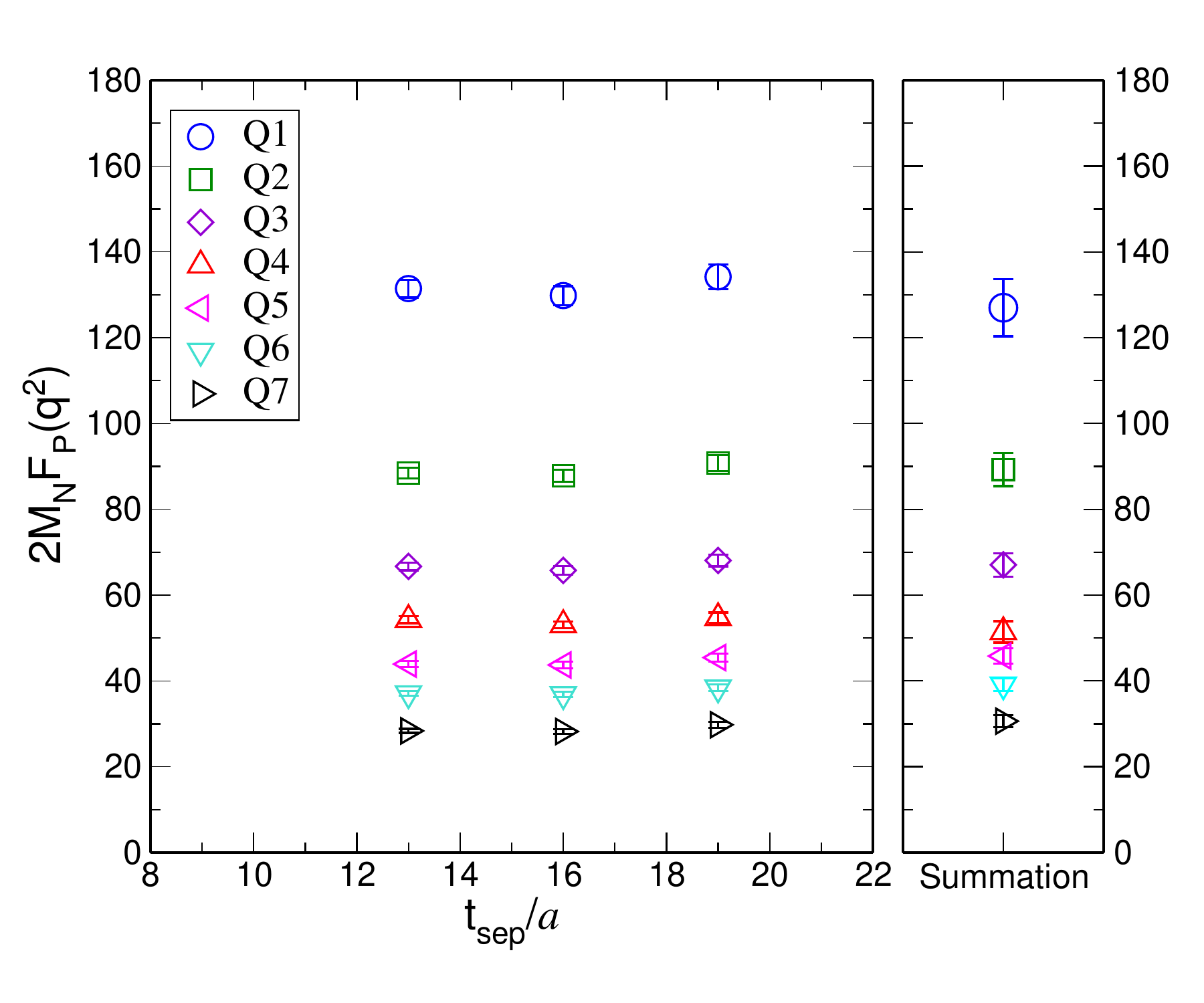}
\caption{
$t_{\mathrm{sep}}$ dependence on the $F_P$ form factor
is shown for seven lowest momentum transfers, along with a comparison of the summation method.
Results are obtained from the coarse $128^4$ lattice (left) and fine $160^4$ lattice (right).
}
\label{fig:F_P_tsep_dep}
\end{figure}

The value of $F_P(q^2)$ can be determined by the uncorrelated constant fit
in the wide range between the source and sink points.
Figure~\ref{fig:F_P_tsep_dep} shows the $t_{\mathrm{sep}}$ dependence of
the extracted values of $2M_N F_P(q^2)$ for the $128^4$ lattice ensemble
(left) and the $160^4$ lattice ensemble (right).
The results given with the
different choices of $t_{\mathrm{sep}}$ are mutually consistent with
each other within the statistical uncertainties for all seven variations
of $q^2$. There is no significant $t_{\mathrm{sep}}$ dependence at the same level as was already obtained for the $F_A$ form factor~\cite{{Shintani:2018ozy},{Tsuji:2023llh}}.
We also compare them to the results obtained using the summation method~\cite{Maiani:1987by} to assess residual excited-state contamination after removing the leading $\pi N$ contribution.
The results obtained by the two methods are consistent with each other within the statistical uncertainties.
This indicates that the systematic uncertainties stemming from the excited-state contamination are negligible within the present statistical precision and {\it well under control by the leading $\pi N$ subtraction method together with
the optimal choice of the smearing parameter for the nucleon interpolating operator.} See, Appendix~\ref{app:assesment_of_residual_excited_state_contamination} for details of the summation analysis.

\begin{figure}[h]
\centering
\includegraphics[width=0.80\textwidth,bb=0 0 792 612,clip]{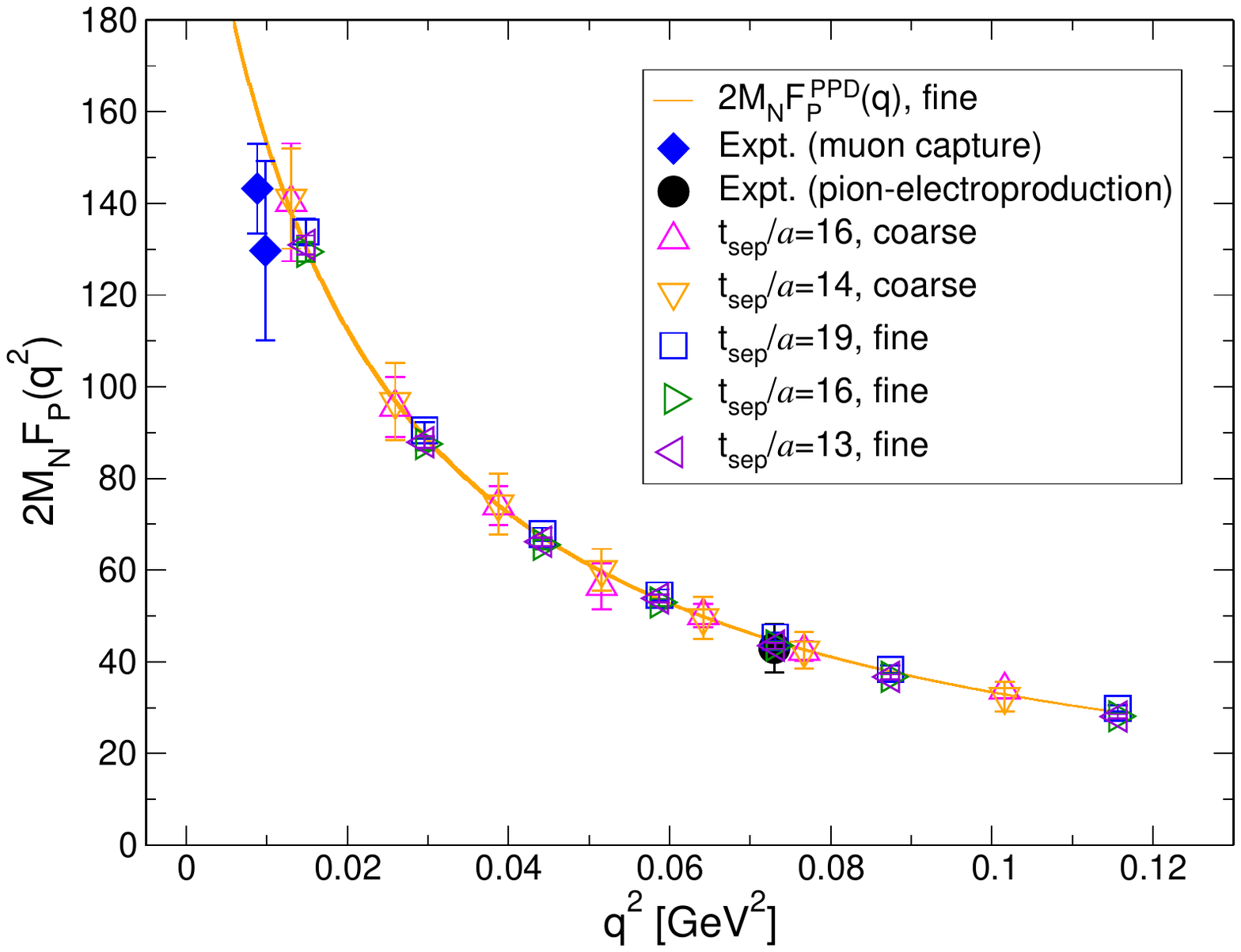}
\caption{
Results of $2M_N F_P(q^2)$ obtained by the leading $\pi N$ subtraction method as a function
of four-momentum squared $q^2$ for each dataset of 
$t_{\mathrm{sep}}=16$ (up-triangle symbols) and $t_{\mathrm{sep}}=14$ (down-triangle symbols)
from the coarse ($128^4$) lattice, and
$t_{\mathrm{sep}}=13$ (squared symbols), $t_{\mathrm{sep}}=16$ (right-triangle symbols), and $t_{\mathrm{sep}}=19$ (left-triangle symbols) from the fine ($160^4$) lattice. The solid curve is given by the PPD model. 
}
\label{fig:F_P_New_qsqr}
\end{figure}

Figure~\ref{fig:F_P_New_qsqr} shows the $q^2$ dependence of the $F_P$
form factor multiplied by $2M_N$ for the $128^4$ lattice ensemble
and the $160^4$ lattice ensemble. Two experimental results of the
muon capture~\cite{{Gorringe:2002xx},{MuCap:2007tkq},{MuCap:2012lei}} and the pion electro-production~\cite{Choi:1993vt} are marked as blue diamonds and a black filled circle. Both of our results from the coarse ($128^4$) and
fine ($160^4$) lattices are consistent with each other and 
agree well with two experimental results. 
Furthermore, the $q^2$ dependence of our results is in good agreement with the prediction of the PPD model given in Eq.~(\ref{eq:PPD_FP}),
using the simulated values of $M_\pi$, $M_N$, and $F_A(q^2)$.
In addition, our subtraction method can get more accurate results of $F_P(q^2)$, than those from the multi-state fit approach that was used in other groups~\cite{Gupta:2024qip}. See Appendix~\ref{app:Sim_Fits} for more details.

%
%

%
%

\subsection{Pseudoscalar form factor $\widetilde{G}_{P}(q^2)$}

\begin{figure}
\centering
\includegraphics[width=0.80\textwidth,bb=0 0 864 720,clip]{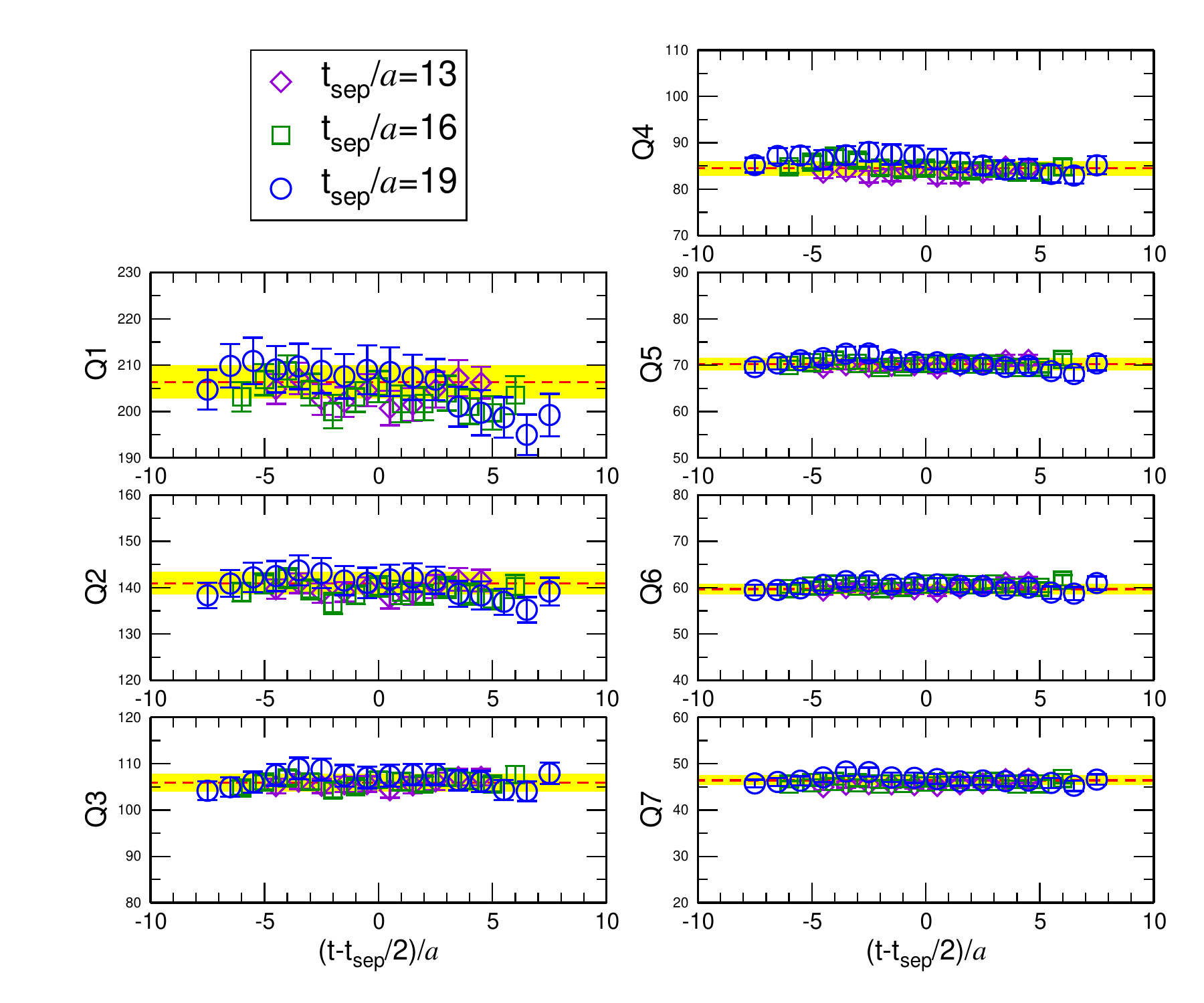}
\caption{
Using the leading $\pi N$ subtraction method as defined in Eq.~(\ref{eq:new_GP}),
the values of $\widetilde{G}_P$ calculated from 
PACS10/L160 with $t_{\mathrm{sep}}/a=13$ (diamonds), $16$ (squares) and $19$ (circles) are shown for seven nonzero momentum transfers (labeled from Q1 to Q7) as functions of the current insertion time slice $t$. The horizontal dashed line together with yellow band in each panel is calculated from the PPD model ($\widetilde{G}_P^{\rm PPD}(q^2)$) given in Eq.~(\ref{eq:PPD_GP}).
}
\label{fig:G_P_New_plateau}
\end{figure}

\begin{figure}
\centering
\includegraphics[width=0.48\textwidth,bb=0 0 864 720,clip]{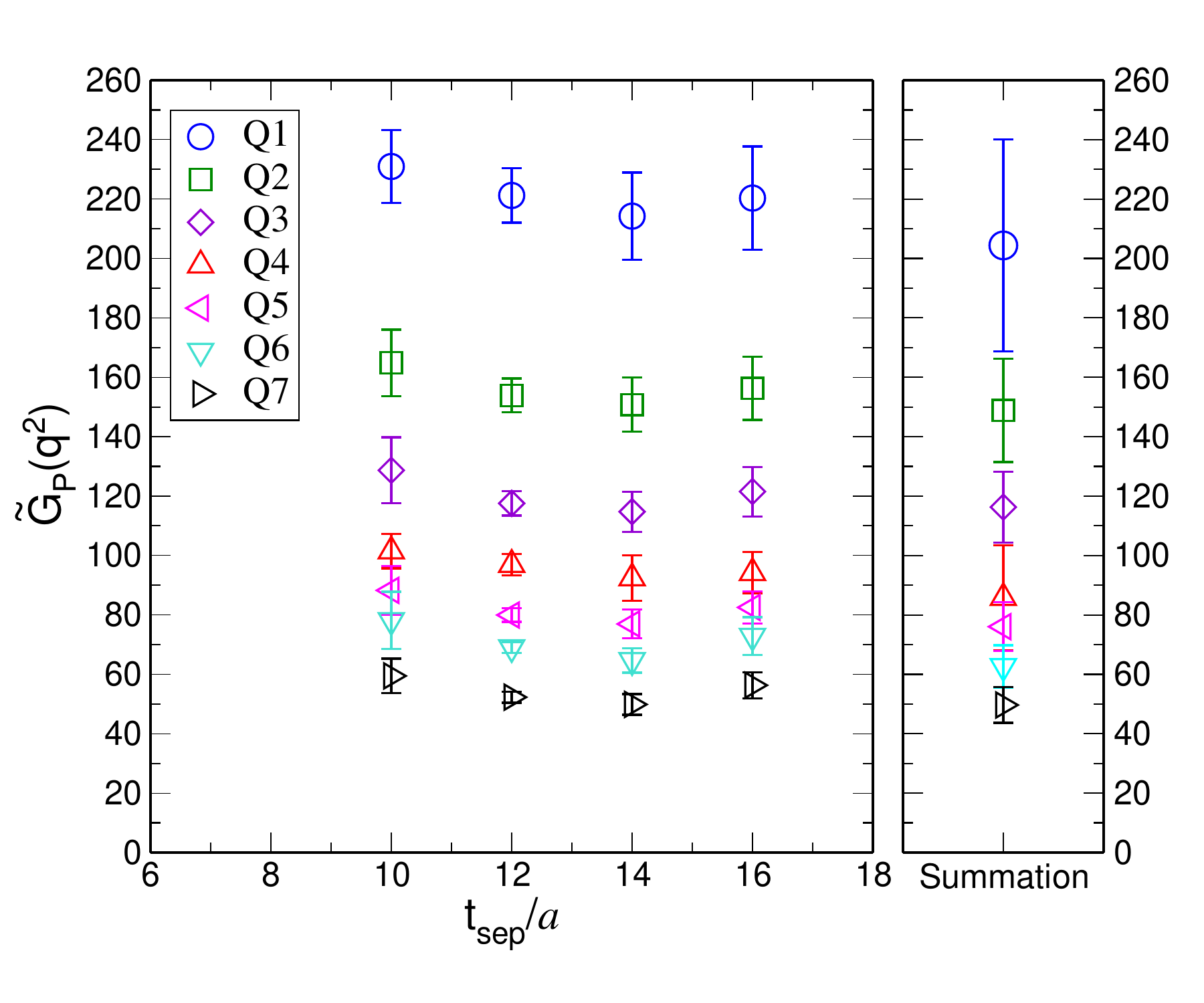}
\includegraphics[width=0.48\textwidth,bb=0 0 864 720,clip]{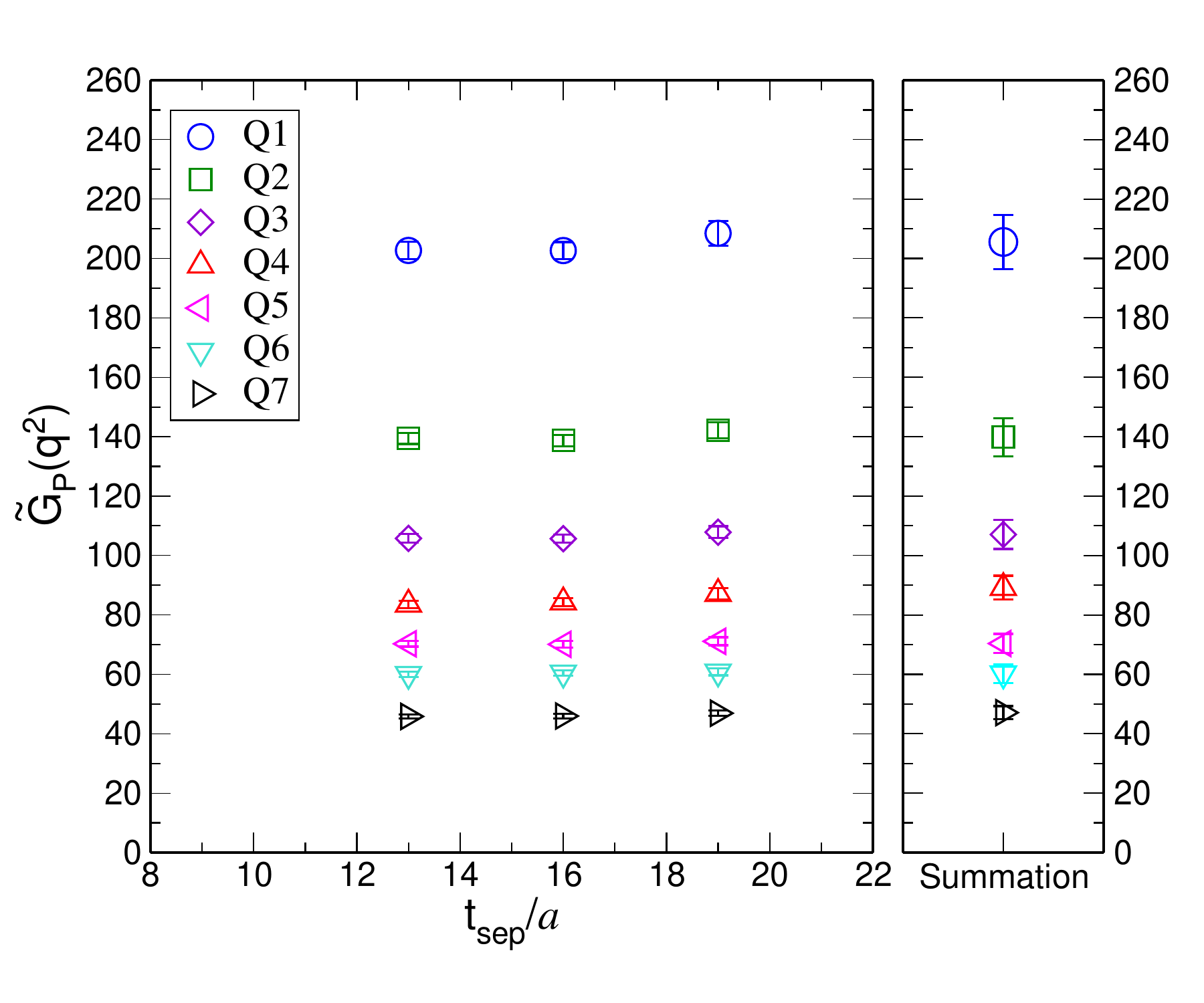}
\caption{
Same as Fig.~\ref{fig:F_P_tsep_dep} for the $\widetilde{G}_P$ form factor.
}
\label{fig:G_P_tsep_dep}
\end{figure}

The $\widetilde{G}_P$ form factor is extracted from Eq.~(\ref{eq:new_GP}) as a function of the current insertion time $t$.
As shown in Fig.\ref{fig:G_P_New_plateau}, similar to the $F_P$ form factor, the leading $\pi N$ subtraction method eliminates the slight convex shape associated with the excited-state contribution and yields a plateau behavior consistent with the PPD model for each $t_{\mathrm{sep}}$. 
As in the case of $F_P(q^2)$, the values of $\widetilde{G}_P(q^2)$ are easily determined in the standard manner using the uncorrelated
constant fit. Figure~\ref{fig:G_P_tsep_dep} shows that there is no $t_{\mathrm{sep}}$ dependence of $\widetilde{G}_P$ for all $q^2$ in either ensemble. Furthermore, there is good agreement with the results of the summation method.

\begin{figure}
\centering
\includegraphics[width=0.80\textwidth,bb=0 0 792 612,clip]{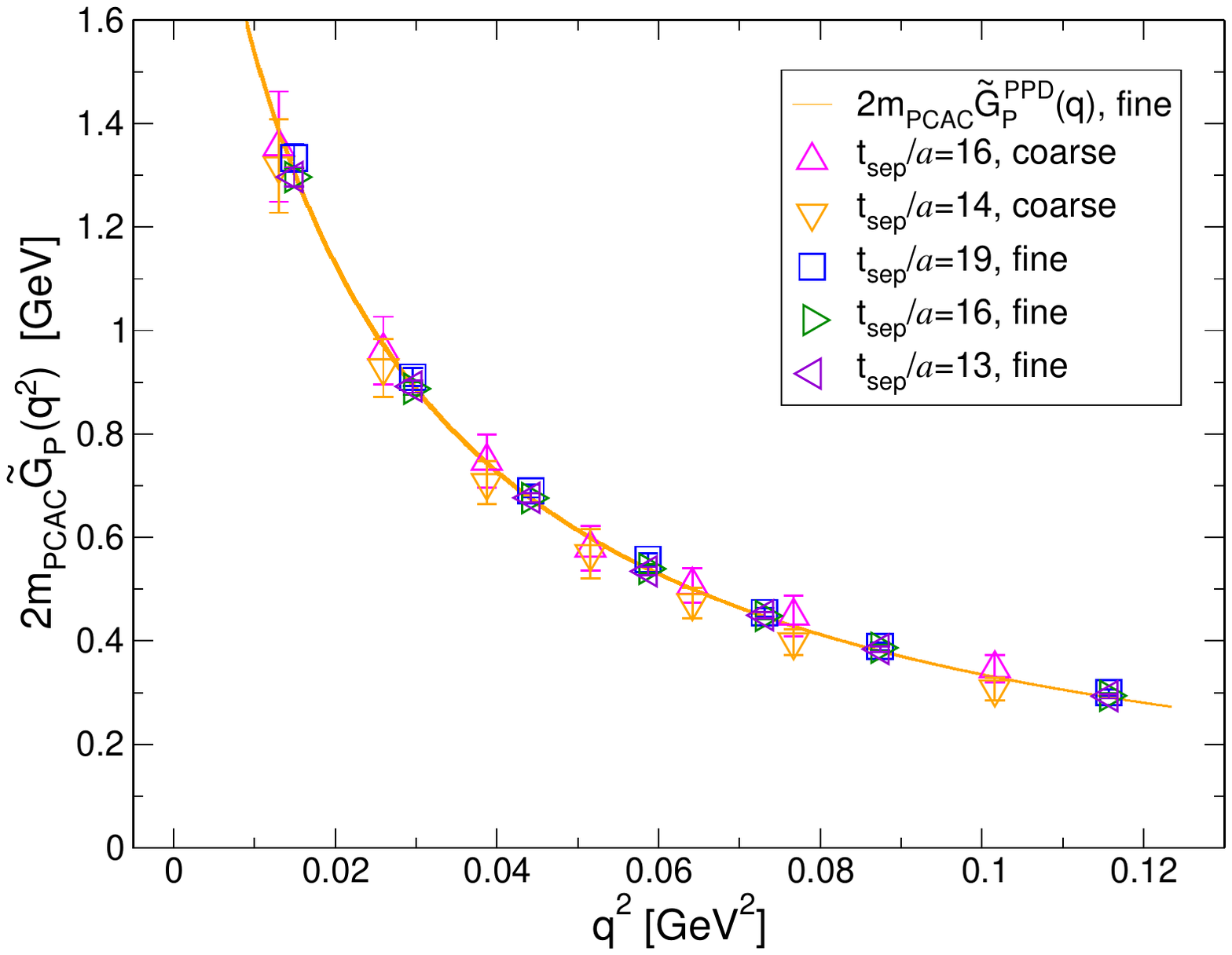}
\caption{
Same as Fig.~\ref{fig:F_P_New_qsqr} for $2m_{\rm PCAC} \widetilde{G}_P(q^2)$ obtained by the leading $\pi N$ subtraction method.
}
\label{fig:G_P_New_qsqr}
\end{figure}

In Fig.~\ref{fig:G_P_New_qsqr}, we show the $q^2$ dependence of the $\widetilde{G}_P$ form factor multiplied by $2m_{\rm PCAC}$.
The solid curve is obtained by the prediction of the PPD model given in Eq.(\ref{eq:PPD_GP}). 
All results obtained from the coarse ($128^4$) and fine ($160^4$) lattices
are fairly consistent with the PPD model
without significant $t_{\mathrm{sep}}$ dependence. 

%
%

%
%

%
%

\subsection{Induced pseudoscalar coupling $g_P^{*}$ and the pion-nucleon 
coupling $g_{\pi NN}$}

Experimentally, the main source of information about $F_P(q^2)$ comes from muon capture by the proton at rest. The induced pseudoscalar coupling
%
%
\begin{align}
\label{eq:ps_charge}
g_P^{\ast}
&=m_\mu F_P(q_0^2),
\end{align}
which is evaluated at the specific momentum transfer, $q_0^2=0.88m_\mu^2$, 
is measured by ordinary muon capture (OMC), $\mu^- + p \rightarrow \nu_\mu + n$ or radiative muon capture (RMC), $\mu^- + p \rightarrow \gamma + \nu_\mu + n$. Although there were conflicting results between those two experiments and theories, the new OMC results from the MuCap Collaboration~\cite{{MuCap:2007tkq},{MuCap:2012lei}} have settled a long-standing issue. Comprehensive overviews of the history of $g_P^{\ast}$ are given in Refs.~\cite{{Gorringe:2002xx},{Bernard:2001rs},{Sasaki:2007gw}}.

Since we do not have data point of $F_P(q^2)$ at $q^2=0$ for kinematical reason, the $q^2$ extrapolation is necessary to evaluate $g_P^{\ast}$ from our results of $F_P(q^2)$. Fortunately, thanks to our large spatial extent of more than 10 fm, the smallest momentum transfer, $q_{\rm min}^2\approx 1.2 m_\mu^2$ for PACS10/L128 and $q_{\rm min}^2\approx1.3 m_\mu^2$ for PACS10/L160, is very close to $q_0^2$. We also confirm that our results are consistent with the PPD model as shown in Fig.~\ref{fig:F_P_New_qsqr}. 
Therefore, the pion-pole structure of the $F_P(q^2)$ form factor is certainly present. (See Appendix~\ref{app:Test_PP} for a more detailed analysis.) 

In order to reduce systematic errors associated with the extrapolation
of the $F_P$ form factor in momentum transfer, we would like to use the model independent $z$-expansion method~\cite{{Boyd:1995cf},{Hill:2010yb}}. 
Unlike the $F_A$ form factor, the $F_P$ form factor seems to have the pion-pole structure. Therefore, the $z$-expansion method is applied to $G(q^2)=(q^2+M_\pi^2)F_P(q^2)$, where the pion-pole singularity can be compensated.
In the $z$-expansion method, the given $G(q^2)$
can be described by a convergent Taylor series in 
a new variable $z$ as
%
%
\begin{align}
\label{eq:zexpansion}
G(q^2)&=\sum_{k=0}^{k_{\mathrm{max}}}c_k z(q^2)^k,
\end{align}
where an infinite series expansion is truncated
at the $k_{\mathrm{max}}$-th order and 
the variable $z$ is defined by
a conformal mapping from $q^2$,
%
%
\begin{align}
\label{eq:zexpansion_z}
z(q^2)=\frac{\sqrt{t_{\mathrm{cut}}+q^2}-\sqrt{t_{\mathrm{cut}}-t_0}}{\sqrt{t_{\mathrm{cut}}+q^2}+\sqrt{t_{\mathrm{cut}}-t_0}}
\end{align}
with $t_{\mathrm{cut}}=9m_\pi^2$. A parameter $t_0$ can be
taken arbitrarily within the range of $t_{\mathrm{cut}}>t_0$.
For simplicity, $t_0=0$ is chosen in this study.
To achieve a model-independent fit, $k_{\mathrm{max}}$ should be chosen to ensure that terms $c_k z(q)^k$ become numerically negligible for $k>k_{\mathrm{max}}$.

In this study, we confirm that the value of $k_{\mathrm{max}}=3$ for the $128^4$ and $160^4$ lattices ($k_{\mathrm{max}}=4$ for the $64^4$ 
lattice) is large enough to guarantee that the $z$-expansion analysis makes a model-independent fit.~\footnote{
The model independence is verified by the convergence of the $z$-expansion under the variation of $k_{\rm max}$ for $G(q^2)$. This procedure is analogous to the one described in Ref.~\cite{Tsuji:2023llh}.} 
We then extrapolate $F_P(q^2)$ at $q^2=q_0^2$ to evaluate the value of $g_P^{\ast}$.
The results of $g_P^{\ast}$ obtained from 
the coarse ($128^4$) and fine ($160^4$) lattices are summarized in Table~\ref{tab:gp_gpinn_results}.
We also determine the value of $g_P^{\ast}$ using
the smaller $64^4$ lattice ensemble at the coarse lattice spacing. 

%
%
\begin{table}[h]
\caption{
Results for the induced pseudoscalar coupling $g_P^{\ast}$ and the pion-nucleon coupling $g_{\pi NN}$.
\label{tab:gp_gpinn_results}}
\begin{ruledtabular}
\begin{tabular}{ccccc}
 $\beta$ & $L^3\times T$ & $t_{\mathrm{sep}}/a$ & $g_P^\ast$ & $g_{\pi NN}$ \cr
 \hline
 1.82 & $128^3\times 128$ & 10 & 8.88(33) & 13.15(54) \cr
 && 12 & 8.54(32) & 12.67(60) \cr
 && 14 & 8.65(53) & 12.92(83) \cr
 && 16 & 8.34(67) & 12.18(1.24)\cr
\cline{2-5}
 & $64^3\times 64$ & 12 & 8.66(18) & 13.13(35)\cr
 &  & 14 & 8.87(19) & 13.62(37) \cr
 &  & 16 & 8.27(38) & 12.50(63) \cr
 \hline
 2.00 & $160^3\times 160$ & 13 & 8.62(11) & 13.18(20) \cr
& & 16 & 8.52(13) & 12.97(23) \cr
& & 19 & 8.74(16) & 13.24(28) \cr
\end{tabular}
\end{ruledtabular}
\end{table}
%

%
%
\begin{table}[h]
\centering
\caption{
Results of the pseudoscalar coupling $g_P^{\ast}$, the pion-nucleon coupling $g_{\pi NN}$ from the combined analysis with different data
selections for $t_{\mathrm{sep}}$ and lattice volumes.
\label{tab:summary_combined_gp_gpnn}
}
\begin{ruledtabular}
\begin{tabular}{cccccc}
$\beta$ & Ensemble & $t_{\mathrm{sep}}/a$ & \# of data & $g_P^{\ast}$ & $g_{\pi NN}$ \cr
\hline
1.82 & $64^4$ & \{14, 16\}    & 2 & 8.71(19) & 13.25(35)\cr
     &        & \{12, 14, 16\}& 3 & 8.48(24) & 12.49(68)\cr
     \cline{2-6}
     & $128^4$& \{14, 16\}    & 2 & 8.51(42) & 12.57(76)\cr
     &        & \{12, 14, 16\}& 3 & 8.52(31) & 12.56(74)\cr
      \cline{2-6}
     & $64^4$+$128^4$ & \{14, 16\}     & 4 & 8.53(23) & 12.80(41)\cr
     &                & \{12, 14, 16\} & 6 & 8.57(17) & 12.84(30)\cr
\hline
2.00 & $160^4$&  19   & 1 & 8.74(16) & 13.24(28) \cr
& & \{16, 19\} & 2 & 8.61(11) & 13.10(19)\cr     
\end{tabular}
\end{ruledtabular}
\end{table}

Since the accessible $q^2_{\mathrm{min}}$ moves away from $q^2=q_0^2$ or $q^2=0$ as the spatial volume decreases, the convergence of the $z$-expansion for the $64^4$ lattice results was expected to be much worse than for the larger volume calculation. However, if $k_{\mathrm{max}}=4$ is chosen, the convergence of the $z$-expansion is well controlled. As a result,  
the obtained values of $g_P^{\ast}$ agree
within their statistical errors
with those of the $128^4$ lattice ensemble
as listed in Table~\ref{tab:gp_gpinn_results}. 

\begin{figure}[h]
\centering
\includegraphics[width=0.80\textwidth,bb=0 0 792 612,clip]{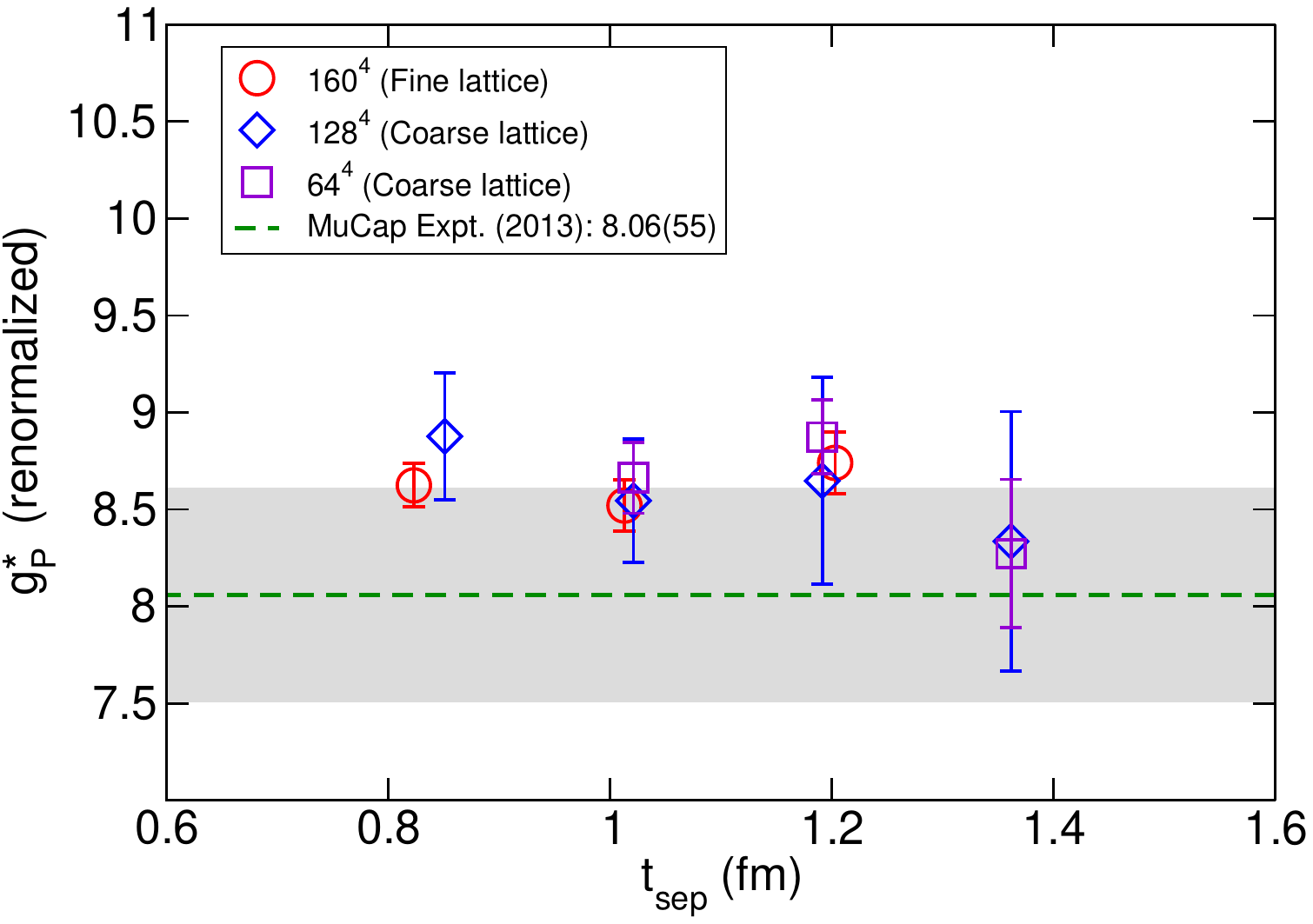}
\caption{
Source-sink separation ($t_{\mathrm{sep}}$) dependence of the renormalized values of the
induced pseudoscalar coupling $g_P^{*}$. The horizontal axis gives $t_{\mathrm{sep}}$ in physical units. The diamond and 
circle symbols are results for the $128^4$ (coarse) and $160^4$ (fine) lattices. 
The horizontal dashed line together with gray band denotes the experimental result~\cite{{MuCap:2007tkq},{MuCap:2012lei}}.
\label{fig:tsep_dep_gp}}
\end{figure}

Figure~\ref{fig:tsep_dep_gp} shows the $t_{\mathrm{sep}}$ dependence of the evaluated values of the (renormalized) induced pseudoscalar coupling $g_P^{\ast}$. As can be easily seen, the results of $g_P^{\ast}$ obtained from all the coarse ($128^4$ and $64^4$) and fine ($160^4$) lattices, show no significant $t_{\mathrm{sep}}$ dependence and agree well with each other, suggesting that the three main systematic uncertainties due to excited-state contamination,
finite volume effect and discretization effect are sufficiently small compared to the present statistical accuracies as is the axial-vector coupling $g_A$ as reported in Ref.~\cite{Tsuji:2023llh}. 
The horizontal dashed line together with gray band
indicates the result of the MuCap experiment~\cite{{MuCap:2007tkq},{MuCap:2012lei}}.
Our results are consistent with the experimental result, while the statistical uncertainties in the $160^4$ lattice results are much smaller than the experimental error. 

Next, we would like to evaluate another quantity
obtained from the $F_P$ form factor. From the phenomenological
point of view, the residue of the pion pole in the $F_P$ form factor is related to the pion-nucleon coupling $g_{\pi NN}$~\cite{Nambu:1960xd}.
The $F_P$ form factor should be expressed as
%
%
\begin{align}
\label{eq:F_P_PPD}
F_P(q^2)\simeq\frac{2F_\pi g_{\pi NN}}{q^2+M_\pi^2}
\end{align}
near the pion pole~\cite{Nambu:1960xd} with the renormalized pion decay constant $F_\pi$, which
is defined as $Z_A\langle 0|\partial_\alpha A_\alpha(x)|\pi(q)\rangle=M_\pi^2F_\pi e^{iq\cdot x}$.
Therefore, the pion-nucleon coupling $g_{\pi NN}$
can be evaluated by 
%
%
\begin{align}
\label{eq:g_piNN_from_F_P}
g_{\pi NN}=\lim_{q^2\rightarrow -M_\pi^2}(q^2+M_\pi^2)\frac{F_P(q^2)}{2F_\pi}
\end{align}
with a help of the $z$-expansion analysis on $G(q^2)=(q^2+M_\pi^2)F_P(q^2)$.

\begin{figure}[h]
\centering
\includegraphics[width=0.80\textwidth,bb=0 0 792 612,clip]{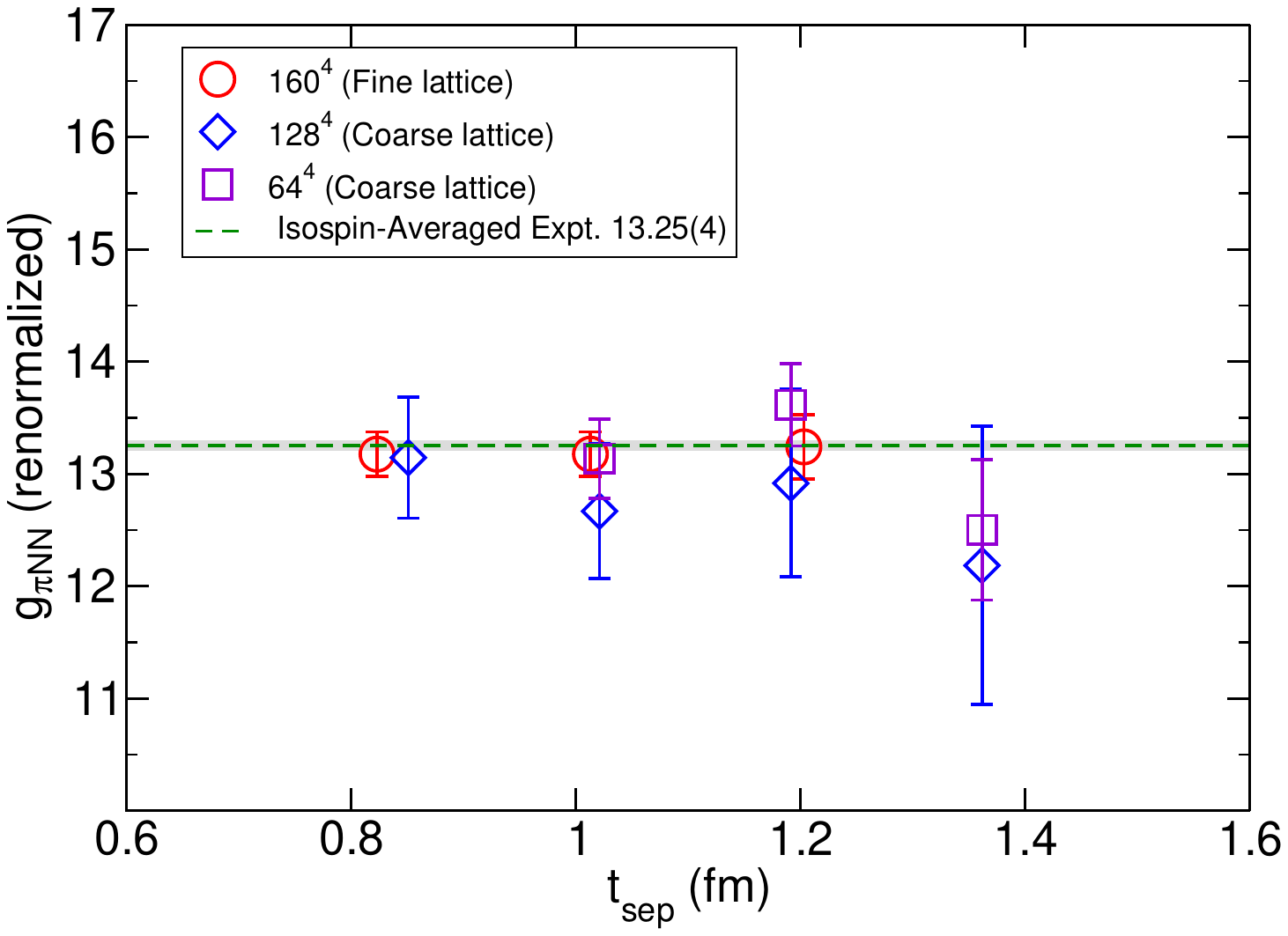}
\caption{
Same as Fig.~\ref{fig:tsep_dep_gp} for the pion-nucleon coupling $g_{\pi NN}$.
The isospin average is used for the experimental result as $g^2_{\pi NN}=\frac{1}{3}\left(
g_{\pi^0 NN}^2+2 g_{\pi^\pm NN}^2
\right)$~\cite{{Babenko:2016idp},{Limkaisang:2001yz}}. 
\label{fig:tsep_dep_gpnn}}
\end{figure}

Figure~\ref{fig:tsep_dep_gpnn} shows the $t_{\mathrm{sep}}$ dependence of the evaluated values of the (renormalized) pion-nucleon coupling $g_{\pi NN}$. As is in the case with $g_A$ and $g_P^{\ast}$, the systematic uncertainties due to excited-state contamination are well controlled, while both finite volume and lattice discretization effects do not manifest beyond statistical uncertainties. The horizontal dashed line together with gray band indicates the isospin average value of the experimental results~\cite{{Babenko:2016idp},{Limkaisang:2001yz},{Reinert:2020mcu}} as
$g_{\pi NN}^2=\frac{1}{3}\left(
g_{\pi^0 NN}^2+2g_{\pi^{\pm}NN}^2
\right)$. 
Our results for $g_{\pi NN}$ again agree well with the experimental results~\cite{{Babenko:2016idp},{Limkaisang:2001yz},{Reinert:2020mcu}}, with good accuracy.

Given the above observations, we determine the mean value of $g_P^{\ast}$ and $g_{\pi NN}$ with several combinations of ensembles and $t_{\mathrm{sep}}$ as summarized in Table~\ref{tab:summary_combined_gp_gpnn}.
It has been confirmed that systematic uncertainties due to the
finite-size effects are not statistically significant in two large spatial extents of approximately 10 and 5 fm under either selection condition of $t_{{\mathrm{sep}}}\gtrsim1.2\ \mathrm{fm}$ or $t_{{\mathrm{sep}}}\gtrsim1.0\ \mathrm{fm}$. Therefore, we prefer to
take an average of two combined results obtained from both $128^4$
and $64^4$ lattices with respect to their statistical uncertainties, which are evaluated by the super jackknife method~\cite{LHPC:2010jcs}.

Similar to the case of $g_A$ cited in Refs.~\cite{{Tsuji:2023llh},{Tsuji:2022ric}}, we quote our best estimates
of $g_P^{\ast}$ and $g_{\pi NN}$ under the condition of $t_{{\mathrm{sep}}}\gtrsim1.2\ \mathrm{fm}$.
We then evaluate the combined values of $g_P^{\ast}$ and $g_{\pi NN}$ with $t_{\mathrm{sep}}/a=\{14, 16\}$ from both $128^4$ and $64^4$ calculations for the coarse lattice result, while the results obtained with a single data set of $t_{\mathrm{sep}}/a=19$ are chosen for the fine lattice as our best estimates. Our final results for $g_A$, $g_P^{\ast}$ and $g_{\pi NN}$ are summarized in Table~\ref{tab:summary_ga_gp_gpinn_mn} together with the measured nucleon masses.
The first error is a statistical one, while the others are systematic ones. The second error is a systematic error stemming from the error of $Z_A^{\mathrm{SF}}$ for the renormalization.
The third error is evaluated by the difference between two analyses using either the $t_{{\mathrm{sep}}}\gtrsim1.2\ \mathrm{fm}$ or $t_{{\mathrm{sep}}}\gtrsim1.0\ \mathrm{fm}$ condition for the selection of datasets in a combined analysis. Two systematic errors are smaller than the statistical errors of either quantity at two lattice spacings.
Recall that the systematic error
due to contamination from excited states
is quite well controlled.

%
%

%
%
%

%
%
\begin{table*}[t]
\centering
\caption{
Summary of the axial-vector coupling $g_A$, the induced pseudoscalar coupling $g_P^{\ast}$, the pion-nucleon coupling $g_{\pi NN}$, 
and nucleon mass $M_N$ obtained at two lattice spacings
\label{tab:summary_ga_gp_gpinn_mn}
}
\begin{ruledtabular}
\begin{tabular}{ccccccccc}
$a$ [fm] & $g_A$ & $g_P^{\ast}$ & $g_{\pi NN}$ & $M_N$ [GeV]\\
\hline
0.085 & $1.288(14)_{\mathrm{stat}}(9)_{Z_A}(8)_{t_{\mathrm{sep}}}$ 
& $8.53(23)_{\mathrm{stat}}(6)_{Z_A}(4)_{t_{\mathrm{sep}}}$ 
& $12.80(41)_{\mathrm{stat}}(9)_{Z_A}(4)_{t_{\mathrm{sep}}}$ & 0.936(11)\\
0.063 & $1.264(14)_{\mathrm{stat}}(3)_{Z_A}(1)_{t_{\mathrm{sep}}}$ 
& $8.74(16)_{\mathrm{stat}}(2)_{Z_A}(13)_{t_{\mathrm{sep}}}$ 
& $13.24(28)_{\mathrm{stat}}(3)_{Z_A}(14)_{t_{\mathrm{sep}}}$ & 0.947(3)\\
\end{tabular}
\end{ruledtabular}
\end{table*}
%

%
%
\begin{figure*}[t]
\centering
\includegraphics[width=0.48\textwidth,bb=0 0 612 612,clip]{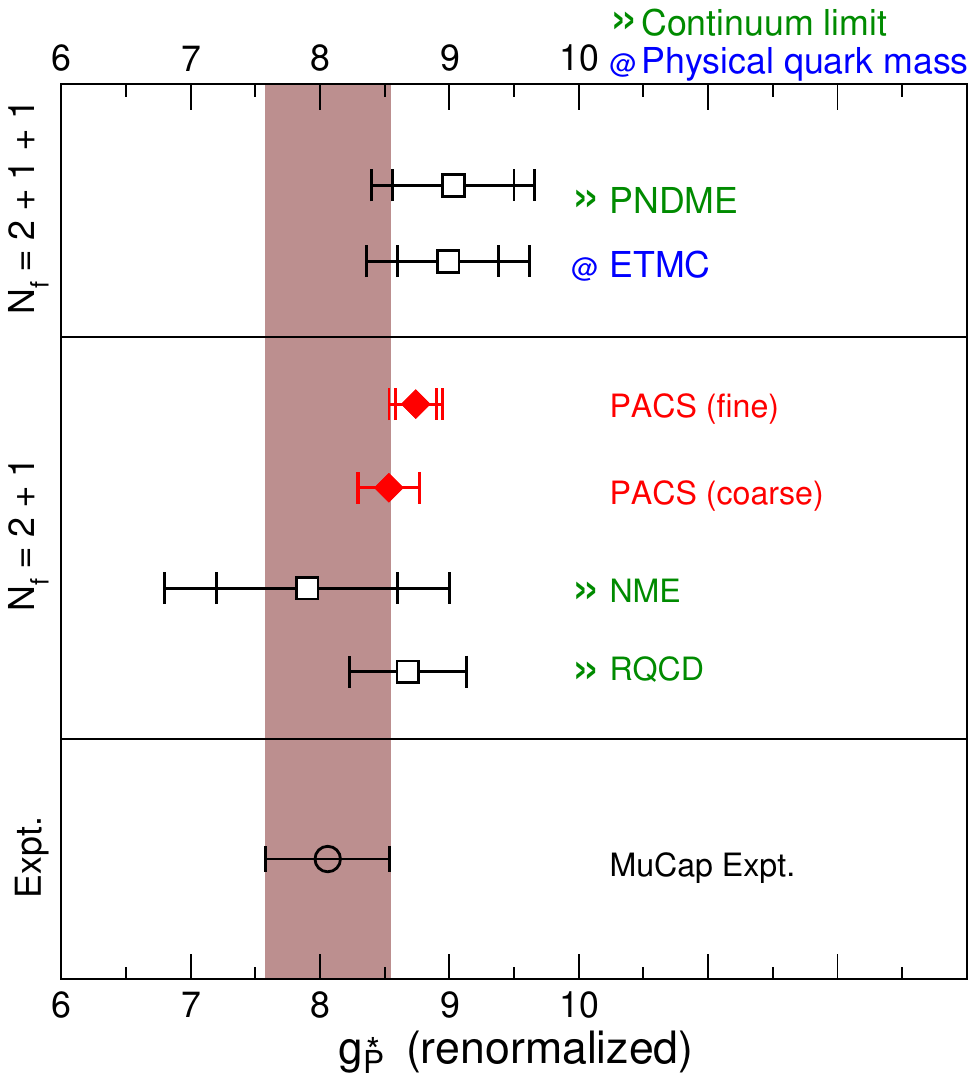}
\includegraphics[width=0.48\textwidth,bb=0 0 612 612,clip]{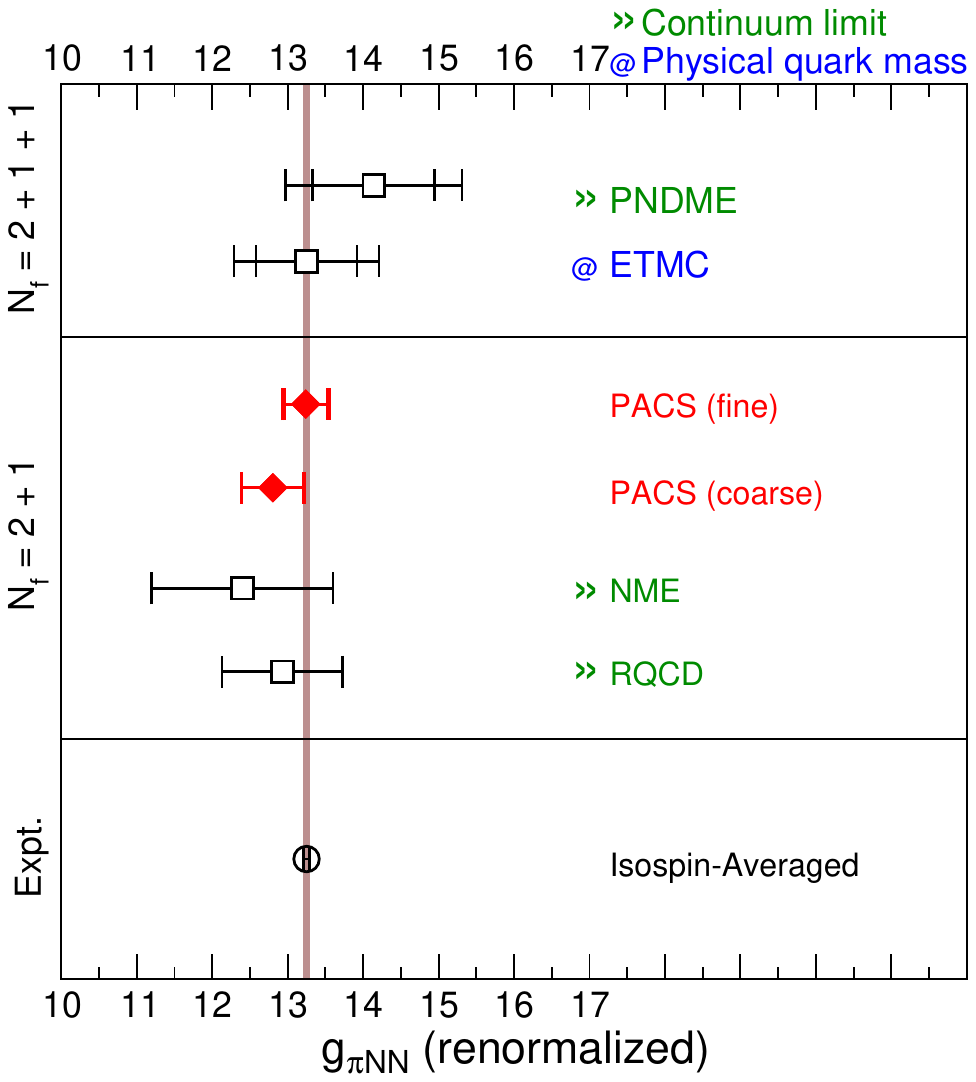}
\caption{
Comparison of our results (red diamonds) with the other lattice results (black squares)
and experimental value (black circles) for $g_P^{*}$ (left panel) and $g_{\pi NN}$ (right panel). 
The inner error bars represent the statistical uncertainties, while the outer ones represent 
the total uncertainties given by adding the statistical and systematic errors in quadrature.
Blue labels indicate that the analysis includes the data from lattice QCD simulation near
the physical point, while green labels indicate that the continuum extrapolation is achieved. 
\label{fig:comp_gp_gpNN}}
\end{figure*}

We finally compare our results of the renormalized value of $g_P^{\ast}$ (left) and $g_{\pi NN}$ (right) together with results
form the recent lattice QCD simulations~\cite{{RQCD:2019jai},{Gupta:2024qip},{Park:2021ypf},{Jang:2023zts},{Alexandrou:2023qbg}} and
experimental results~\cite{{MuCap:2007tkq},{MuCap:2012lei},{Babenko:2016idp},{Limkaisang:2001yz}} in Fig.~\ref{fig:comp_gp_gpNN}. The inner error bars represent the statistical uncertainties, while the outer ones represent 
the total uncertainties given by adding the statistical and systematic errors in quadrature.
Since our simulations are accomplished at only two lattice spacings, we have not yet performed the continuum limit extrapolation. 
Therefore, the two results obtained at different lattice spacings are plotted individually in Fig.~\ref{fig:comp_gp_gpNN}.
Our results show that the systematic errors are so well controlled that the inner and outer errors are not visible in Fig.~\ref{fig:comp_gp_gpNN}.
It is important to note that our two results are derived solely from the physical point simulations, which would suffer from large statistical fluctuations. In contrast, except for the Extended Twisted Mass Collaboration (ETMC) result~\cite{Alexandrou:2023qbg}, the other lattice results are given by the chiral-continuum extrapolation, which incorporates data obtained from simulations at heavier pion masses~\cite{{RQCD:2019jai},{Park:2021ypf},{Jang:2023zts}}.

\section{Summary}
\label{sec:summary}

We have studied nucleon form factors in the axial-vector and pseudoscalar channels in 2+1 flavor QCD using two sets of the PACS10 configurations at coarse and fine lattice spacings. Our simulations were carried out in very large spatial volumes, which allow us to access the low $q^2$ region, at the physical point essential for low-energy chiral behavior. The nucleon interpolating operator has been
adopted with well-tuned smearing parameters that guarantee ground-state dominance in $F_A(q^2)$, although the two types of pseudoscalar form factors, $F_P(q^2)$ and $G_P(q^2)$, still suffer from the excited-state contamination in our previous studies~\cite{{Shintani:2018ozy},{Tsuji:2023llh}}.

In this study, we thus propose a simple subtraction method for removing the so-called leading $\pi N$-state contamination induced by the pion-pole structure appears in $F_P(q^2)$ and $G_P(q^2)$. The leading $\pi N$-state contribution in the three-point function of the spatial axial-vector current $A_i$ was found to be eliminated by a specific linear combination of the two time derivatives of the correlator ratios
composed of the $A_4$ and $A_i$ currents.
The leading $\pi N$ subtraction method can be easily applied to existing data without the need for new calculations, provided that the excited-state contribution, which appears in the standard ratio method, is sufficiently hindered for $F_A(q^2)$ by applying the appropriate smearing to the nucleon operator.

In fact, the new correlator ratio, which includes the time derivatives of the $A_4$ and $A_i$ correlators, shows good plateau behavior regardless of the current insertion time, eliminating the slightly convex shape associated with the excited-state contribution in the standard ratio method. It is important to note that this novel method utilizes three-point functions of both spatial and temporal axial-vector currents to calculate $F_P(q^2)$.

For $G_P(q^2)$, the ``PCAC relation'' for the leading $\pi N$ contribution, which can be verified by the fact that the axial
Ward-Takahashi identity is well satisfied in terms of the nucleon three-point functions, can help to eliminate the $\pi N$ contribution in the 
three-point function of the pseudoscalar current in a similar way in the case of $F_P(q^2)$.
The same subtraction prescription was also effective for $G_P(q^2)$, completely eliminating the slightly convex shape of the time dependence that is visible in the standard ratio method, and yielding a constant value that is insensitive to the choice of $t_{\mathrm{sep}}$.

The results of both $F_P(q^2)$ and $G_P(q^2)$ obtained from the new analysis surprisingly agree well with the prediction of the PPD model in the range of $0.01 \lesssim q^2 \lesssim 0.12$ $[{\rm GeV}^2]$ measured in this paper. There is experimental information coming from two experiments for $F_P(q^2)$. One is the experiment 
of the muon capture by the proton, the other is the pion electro-production experiment. Our results also agree well with these experimental results.

Finally, we determine the values of $g_P^{\ast}$ and $g_{\pi NN}$ from $F_P(q^2)$ using the $z$-expansion method
with the pion-pole singularity taken into account. Our final results
are
%
%
\begin{align}
g_P^{\ast}&=\left\{
\begin{array}{lr}
\makebox[8em][l]{8.53(23)(6)(4)} & \beta=1.82 \\
\makebox[8em][l]{8.74(16)(2)(13)}& \beta=2.00\\
\end{array}
\right.\\
g_{\pi NN}&=\left\{
\begin{array}{lcr}
\makebox[8em][l]{12.80(41)(9)(4)} & \beta=1.82 \cr
\makebox[8em][l]{13.24(28)(3)(14)}& \beta=2.00\cr
\end{array}
\right.
\end{align}
where the first error is statistical, while
the others are systematic ones. The second
error is evaluated stemming from the error
of $Z_A^{\mathrm{SF}}$ for the renormalization. 
The third error is evaluated by the difference between two analyses using either the $t_{{\mathrm{sep}}}\gtrsim1.2\ \mathrm{fm}$ or $t_{{\mathrm{sep}}}\gtrsim1.0\ \mathrm{fm}$ condition for the selection of datasets in a combined analysis. 

Since our lattice simulations are currently running at the third lattice spacing to perform the continuum limit extrapolation~\cite{Tsuji:2024scy}, 
we may estimate the size of the systematic uncertainties associated with the finite lattice spacing for $g_P^{\ast}$ and $g_{\pi NN}$ by comparing the current results obtained at two different lattice spacings.
Similar to the case of $g_A$ as reported in Ref~\cite{Tsuji:2023llh}, this analysis reveals that the discrepancy between the two results obtained at different lattice spacings is smaller than their statistical errors for both values of $g_P^{\ast}$ and $g_{\pi NN}$.
Furthermore, taking into account the preliminary results obtained at the third lattice spacing~\cite{Tsuji:2024scy}, we expect that the lattice discretization error will be maintained at a level of a few percent for both $g_P^{\ast}$ and $g_{\pi NN}$.

A comprehensive study that employs the current results for $F_P(q^2)$ and $G_P(q^2)$ along with $F_A(q^2)$ to verify
the generalized GT relation and the PPD model is outlined in a separate paper~\cite{Tsuji:2025quu} and not included in this paper.
Although the axial radius $r_A$ is most relevant for current and future neutrino experiments, there is a possible discretization error of about 10\% in this quantity as pointed out in Ref.~\cite{Tsuji:2023llh}.
Therefore, it is necessary to continue our lattice QCD calculations for the axial structure of the nucleon with the third lattice spacing toward the continuum limit. Such a study is underway~\cite{Tsuji:2024scy}.

\begin{acknowledgments}

We would like to thank members of the PACS collaboration for useful discussions. K.-I.~I. is supported in part by MEXT as ``Feasibility studies for the next-generation computing infrastructure".
K.~S. is supported by JST, The Establishment of University Fellowships towards the creation of Science Technology Innovation, Grant Number JPMJFS2106.
We also thank Y. Namekawa for his careful reading of the manuscript.
Numerical calculations in this work were performed on Oakforest-PACS in Joint Center for Advanced High Performance Computing (JCAHPC) and Cygnus  and Pegasus in Center for Computational Sciences at University of Tsukuba under Multidisciplinary Cooperative Research Program of Center for Computational Sciences, University of Tsukuba, and Wisteria/BDEC-01 in the Information Technology Center, the University of Tokyo. 
This research also used computational resources of the K computer (Project ID: hp180126) and the Supercomputer Fugaku (Project ID: hp20018, hp210088, hp230007, hp230199, hp240028, hp240207, hp250037) provided by RIKEN Center for Computational Science (R-CCS), as well as Oakforest-PACS (Project ID: hp170022, hp180051, hp180072, hp190025, hp190081, hp200062),  Wisteria/BDEC-01 Odyssey (Project ID: hp220050) provided by the Information Technology Center of the University of Tokyo/JCAHPC.
The  calculation employed OpenQCD system~\cite{OpenQCD}. 
This work is supported by the Japan Lattice Data Grid (JLDG) constructed over the SINET6 of NII.
This work was also supported in part by Grants-in-Aid for Scientific Research from the Ministry of Education, Culture, Sports, Science and Technology (Nos.~22K03612, 23H01195, 23K03428, 25KJ0404) and MEXT as ``Program for Promoting Researches on the Supercomputer Fugaku'' (Search for physics beyond the standard model using large-scale lattice QCD simulation and development of AI technology toward next-generation lattice QCD; Grant Number JPMXP1020230409). 

\end{acknowledgments}
\appendix

\section{Nucleon ground-state contribution in the nucleon three-point function}
\label{app:NN_nucleon_3pt_function}

In Sec.~\ref{Sec:Method}, we define the nucleon three-point function $C_{J}^{5z}(t; \bm{p}^{\prime}, \bm{p})$
in Eq.~(\ref{eq:3ptC}), where the initial and final states carry fixed momenta $\bm{p}$ and $\bm{p}^\prime$, respectively.
The leading contribution of the nucleon three-point function can be described by
the ground-state contribution $C_{J}^{5z}(t; \bm{p}^{\prime}, \bm{p})_{NN}$ as
%
%
\begin{align}
C_{J}^{5z}(t; \bm{p}^{\prime}, \bm{p})=
C_{J}^{5z}(t; \bm{p}^{\prime}, \bm{p})_{NN}
+\cdot\cdot\cdot,
\end{align}
where the ellipsis denotes excited-state contributions. The explicit formula of 
the ground-state contribution can be expressed by
%
%
\begin{align}
\label{eq:NN_contribution}
C_{J}^{5z}(t; \bm{p}^{\prime}, \bm{p})_{NN}&=
\frac{Z_N(\bm{p})}{2E_N(\bm{p})}e^{-E_N(\bm{p}) t}
\times
\widetilde{\Lambda}^{NN}_{J}(\bm{p}^\prime, \bm{p})\cr
&\times
\frac{Z_N^{\ast}(\bm{p}^\prime)}{2E_N(\bm{p}^\prime)}e^{-E_N(\bm{p}^\prime)(t_{\mathrm{sep}}-t)},
\end{align}
where $Z_N(\bm{p})$ ($Z_N^{\ast}(\bm{p}^\prime)$) corresponds to the spectral weight 
between the nucleon interpolating operator and the initial (final) state of the nucleon, 
$\widetilde{\Lambda}^{NN}_{J}(\bm{p}^\prime, \bm{p})$ is defined as
%
%
\begin{align}
\label{eq:Trace_3pt}
&\widetilde{\Lambda}^{NN}_{J}(\bm{p}^\prime, \bm{p}) \cr
&={\mathrm{Tr}}\left[
{\mathcal{P}^{5z}}
(-i\gamma\cdot p^\prime+M_N){\mathcal{O}_{\alpha}}(q)
(-i\gamma\cdot p+M_N)
\right]
\end{align}
with ${\mathcal{O}_{\alpha}}(q)$ containing the respective nucleon form factors.
For the axial-vector ($A_\alpha$) and pseudoscalar ($P$) currents, 
${\mathcal{O}_{\alpha}}(q)$ have the following relativistically covariant decomposition
in terms of the nucleon form factors as
%
%
\begin{align}
{\mathcal{O}_{A_{\alpha}}}(q)&=\gamma_\alpha 
\gamma_5\widetilde{F}_A(q^2)+i(p-p^\prime)_\alpha \gamma_5 \widetilde{F}_P(q^2), \\
{\mathcal{O}_{P}}(q)&=\gamma_5 \widetilde{G}_P(q^2).
\end{align}
After the trace operation is performed in the right-hand side of Eq.~(\ref{eq:Trace_3pt}),  
the general expression of $\widetilde{\Lambda}^{NN}_{J}(\bm{p}^\prime, \bm{p})$ 
for both the axial-vector and pseudoscalar cases are given by
%
%
\begin{align}
\widetilde{\Lambda}^{NN}_{A_\alpha}(\bm{p}^\prime, \bm{p})&=
\widetilde{F}_A(q^2){\Lambda}^{NN}_{5, \alpha}(\bm{p}^\prime, \bm{p})\cr
&+i(p-p^\prime)_\alpha\widetilde{F}_P(q^2) {\Lambda}^{NN}_{5}(\bm{p}^\prime, \bm{p}),
\\
\widetilde{\Lambda}^{NN}_{P}(\bm{p}^\prime, \bm{p})&=\widetilde{G}_P(q^2) {\Lambda}^{NN}_{5}(\bm{p}^\prime, \bm{p}),
\end{align}
where
%
%
\begin{align}
{\Lambda}^{NN}_{5, \alpha}(\bm{p}^\prime, \bm{p})&=2M_N\left[\left(M_N - i(p_4^\prime + p_4)
\right)\delta_{3\alpha}\right.\cr
&\left.+i(p_3^\prime+p_3)\delta_{4\alpha}\right]\cr
&+2\left[(p_3^\prime p_\alpha+ p_\alpha^\prime p_3) -(p^\prime \cdot p) \delta_{3\alpha} \right],
\\
{\Lambda}^{NN}_{5}(\bm{p}^\prime, \bm{p}) &=2\left[ i(p_3 - p_3^\prime) M_N +
(p_4^\prime p_3 - p_3^\prime p_4) 
\right].
\end{align}

In this study, we consider only the case at the rest frame of the final state ($\bm{p}^\prime =\bm{0}$), which
leads to $\bm{p} =\bm{q}$. For this specific kinematics, we get
%
%
\begin{align}
{\Lambda}^{NN}_{5, \alpha}(\bm{0}, \bm{q})
&=2(-ip_4^\prime+M_N)(-ip_4+M_N)\delta_{3\alpha}\cr
&+2iq_3(-ip_4^{\prime}+M_N)\delta_{4\alpha},
\\
{\Lambda}^{NN}_{5}(\bm{0}, \bm{q}) &=i2M_N(-ip_4^\prime+M_N) q_3,
\end{align}
which lead to
%
%
\begin{align}
\widetilde{\Lambda}^{NN}_{A_{\alpha}}(\bm{0}, \bm{q})&=
2(-ip_4^\prime+M_N)
\left[\big((-ip_4+M_N)\delta_{3\alpha}+
iq_3\delta_{4\alpha}\big)\right.\cr
&\times \widetilde{F}_A(q^2)\left.-q_{3}(p-p^\prime)_\alpha\widetilde{F}_P(q^2)\right], \\
\widetilde{\Lambda}^{NN}_{P}(\bm{0}, \bm{q})&=i2(-ip_4^\prime+M_N) q_3 \widetilde{G}_P(q^2).
\end{align}

In the standard ratio method described in Sec.~\ref{Sec:Method}, the quantity 
of $\widetilde{\Lambda}^{NN}_{J}(\bm{p}^\prime, \bm{p})$ is extracted from the ratio of the three-point function and two-point functions defined in Eq.~(\ref{eq:ratio_3pt_2pt}).
In our choice of the kinematics with $p_4=iE_N(\bm{q})$ and $p_4^\prime=iM_N$, $\widetilde{\Lambda}^{NN}_{J}(\bm{p}^\prime, \bm{p})$ is expressed as follows~\cite{Sasaki:2007gw},
%
%
\begin{align}
\widetilde{\Lambda}^{NN}_{A_{\alpha}}(\bm{0}, \bm{q})&=4M_N(M_N+E_N(\bm{q}))\cr
&\times\left(
\delta_{3\alpha}+\frac{iq_3}{M_N+E_N(\bm{q})}\delta_{4\alpha}\right)\widetilde{F}_A(q^2)
-4M_Nq_{3}q_\alpha\widetilde{F}_P(q^2), \cr
\widetilde{\Lambda}^{NN}_{P}(\bm{0}, \bm{q})&=i4M_N q_3 \widetilde{G}_P(q^2). \nonumber
\end{align}

\section{Spectral decomposition on the nucleon three-point function}
\label{app:MOS}

Following an argument given by Meyer-Ottnad-Schulz~\cite{Meyer:2018twz}, let us consider the
spectral decomposition on the nucleon ground-state contribution of the nucleon three-point function in the continuum as follows:
%
%
\begin{align}
\label{eq:SpectDecomp_3pt}
&\hat{C}_{J}^{5z}(t; \bm{p}^{\prime}, \bm{p})_{NN}\cr
&=
\int \frac{dp_4^\prime}{2\pi} \int \frac{dp_4}{2\pi}
\frac{Z_N(\bm{p})}{p^2+M_N^2}e^{ip_4 t}
\times
\widetilde{\Lambda}^{NN}_{J}(\bm{p}^\prime, \bm{p})\cr
&\times
\frac{Z_N^{\ast}(\bm{p}^\prime)}{{p^\prime}^2+M_N^2}e^{ip_4^\prime(t_{\mathrm{sep}}-t)}.
\end{align}
for the axial-vector current
including both of $A_4$ and $A_i$ ($i=1,2,3$), and the pseudoscalar correlator. In this expression, the ground-state contribution defined in Eq.~(\ref{eq:NN_contribution}) can be obtained
after performing the $p_4$ and $p_4^\prime$ integrals by contour integration. 
Furthermore, if the form factors contained in $\widetilde{\Lambda}^{NN}_{J}(\bm{p}^\prime, \bm{p})$ 
have singularities in the complex plane, additional contributions may appear.

To actually see such a singularity in the
axial-vector case of $\widetilde{\Lambda}^{NN}_{A_{\alpha}}(\bm{p}^\prime, \bm{p})$, we use the dispersive representation of the $F_A$ and $F_P$ form factors
%
%
\begin{align}
{F}_{A,P}(q^2)&=\int^\infty_{s_0}\frac{ds}{\pi}\frac{\mathrm{Im}{F}_{A,P}(-s)}{s+q^2}\;.
\end{align}
For the $F_A(q^2)$, $s_0$ represents a starting point of branch cut associated with a three-pion threshold as $s_0=9M_\pi^2$, while for the $F_P(q^2)$, $s_0$ represents a singularity point due to the pion-pole position as $s_0=M_\pi^2$. 
In the case of the $F_P(q^2)$, the presence of the pion pole on the real axis in the negative $q^2$ region leads to 
%
%
\begin{align}
\mathrm{Im}{F}_{P}(-s)=2\pi F_\pi g_{\pi NN} \delta(s-m_\pi^2)
+\cdot\cdot\cdot,
\end{align}
which can generate the additional contribution
from the expression of (\ref{eq:SpectDecomp_3pt}).
The additional contribution for the pseudoscalar case of
$\widetilde{\Lambda}^{NN}_{P}(\bm{p}^\prime, \bm{p})$ also
can be generated, since the $G_P(q^2)$ has the pion pole on the real axis in the negative $q^2$ region as well
as the $F_P(q^2)$.

To see it more explicitly, let us consider for our choice of the kinematics ($\bm{p}^\prime=\bm{0}$ and $\bm{p}=\bm{q}$) calculated in this study. In the axial-vector case, 
$\hat{C}_{A_\alpha}^{5z}(t; \bm{0}, \bm{q})_{NN}$
contains the following integral associated
with the pion-pole structure embedded in the $F_P(q^2)$,
%
%
\begin{widetext}
\begin{align}
\label{eq:total_integral}
I_{\alpha}(t, t_{\mathrm{sep}})=\int \frac{dp_4^\prime}{2\pi} \int \frac{dp_4}{2\pi}
\underbrace{\frac{e^{ip_4 t}}{p_4^2+E_N^2(\bm{q})}}_{\mathrm{(I)}}
\;
\underbrace{
\frac{(p-p^\prime)_{\alpha}}{(p_4 - p_4^\prime)^2+\bm{q}^2+M_\pi^2}
}_{\mathrm{(II)}}
\;
\underbrace{
\frac{(-ip_4^\prime+M_N)e^{ip_4^\prime(t_{\mathrm{sep}}-t)}}{{p_4^\prime}^2+M_N^2}
}_{\mathrm{(III)}},
\end{align}
\end{widetext}
which is classified into three sections, where section (II) is responsible for the pion-pole structure.

In considering the two pole contributions arising from section (II), it is worth transforming the equation as follows:
%
%
\begin{widetext}
\begin{align}
\frac{(p-p^\prime)_{\alpha}}{(p_4 - p_4^\prime)^2+\bm{q}^2+M_\pi^2}=
\left\{
\begin{array}{ll}
\frac{1}{2}\left(
\frac{1}{(p_4 - p_4^\prime)-iE_\pi(\bm{q})}
+\frac{1}{(p_4 - p_4^\prime)+iE_\pi(\bm{q})}
\right) & \mbox{for}\;\alpha=4 \cr
-\frac{iq_i}{2E_\pi(\bm{q})} \left(
\frac{1}{(p_4 - p_4^\prime)-iE_\pi(\bm{q})}
-\frac{1}{(p_4 - p_4^\prime)+iE_\pi(\bm{q})}
\right) & \mbox{for}\;\alpha=i\neq 4 \cr
\end{array}
\right.
\end{align}
\end{widetext}
where the relative signs of the two contributions differ in the case of the temporal and spatial components.~\footnote{
In Ref.~\cite{Meyer:2018twz}, the original discussion addressed only the temporal component of the axial-vector current and omitted the spatial one.
} By the standard way of evaluating residues, the integral $I_\alpha$ given by Eq.~(\ref{eq:total_integral}) can yield three contributions:
%
%
\begin{align}
I_{\alpha}(t, t_{\mathrm{sep}})=I_{\alpha}^{\mathrm{A}}(t, t_{\mathrm{sep}})+I_{\alpha}^{\mathrm{B}}(t, t_{\mathrm{sep}})+I_{\alpha}^{\mathrm{C}}(t, t_{\mathrm{sep}}).
\end{align}
The evaluation of each contribution is based on the following residues in the lower half of the complex plane:
%
%
\begin{itemize}
\item{Case A:} (I) and (III) contain poles: $p_4=iE_N(\bm{q})$ and $p_4^\prime = iM_N$
\item{Case B:} (II) and (III) contain poles: $p_4=i(M_N+E_\pi(\bm{q}))$ and $p_4^\prime = iM_N$
\item{Case C:} (I) and (II) contain poles: $p_4=iE_N(\bm{q})$ and $p_4^\prime = i(E_N(\bm{q})+E_\pi(\bm{q}))$
\end{itemize}
%
and results in the following form
\begin{widetext}
\begin{align}
I_{\alpha}^{\mathrm{A}}(t, t_{\mathrm{sep}})&=\frac{1}{2E_N(\bm{q})}\frac{q_\alpha}{q^2+M_\pi^2}e^{-E_N(\bm{q})t}e^{-M_N(t_{\mathrm{sep}}-t)},  
\label{eq:Aterm}
\\
I_{\alpha}^{\mathrm{B}}(t, t_{\mathrm{sep}})&=
-\frac{1}{2E_\pi(\bm{q})}\frac{r^{+}_{\alpha}}{(M_N+E_\pi(\bm{q}))^2-E_N^2(\bm{q})}e^{-(M_N+E_\pi(\bm{q}))t}e^{-M_N(t_{\mathrm{sep}}-t)}, 
\label{eq:Bterm}
\\
I_{\alpha}^{\mathrm{C}}(t, t_{\mathrm{sep}})&=
\frac{1}{4E_N(\bm{q})E_\pi(\bm{q})}\frac{r^{-}_\alpha}{E_N(\bm{q})+E_\pi(\bm{q})-M_N}e^{-E_N(\bm{q})t}e^{-(E_N(\bm{q})+E_\pi(\bm{q}))(t_{\mathrm{sep}}-t)},
\label{eq:Cterm}
\end{align}
\end{widetext}
where $q=(i(E_N(\bm{q})-M_N), \bm{q} )$ and $r^{\pm}=(iE_\pi(\bm{q}),\pm \bm{q})$.
The contribution of $I_{\alpha}^{\mathrm{A}}$ provides 
the usual ground-state contribution, while the other two
contributions $I_{\alpha}^{\mathrm{B}}$ and $I_{\alpha}^{\mathrm{C}}$ are nothing but  
the pole-enhanced $\pi N$ contributions, which are represented in terms of $\Delta_{+}(t, t_{\mathrm{sep}}; {\bm q})$ 
and $\Delta_{-}(t, t_{\mathrm{sep}}; {\bm q})$
in Eqs.~(\ref{Eq:A_space}) and (\ref{Eq:A_time}), respectively. 

Furthermore, we observe that these coefficients explicitly contain the pion-pole contributions as a function of $q^2$, indicating that their derivation is based on the pion-pole structure of $F_P(q^2)$.
It is important to recall that $q^2=2M_N(E_N(\bm{q})-M_N)$ under the kinematics selected for the nucleon ground-state contribution. Consequently, the coefficient of Eq.~(\ref{eq:Bterm}) contains 
the pion-pole contribution as
%
%
\begin{align}
&\frac{1}{(M_N+E_\pi(\bm{q}))^2-E_N^2(\bm{q})}\cr
&=
\frac{1}{2E_N(\bm{q})}
\left(
\frac{E_\pi(\bm{q})+E_N(\bm{q})-M_N}{q^2+M_\pi^2}
-\frac{1}{
M_N+E_\pi(\bm{q})+E_N(\bm{q})}
\right),
\end{align}
while the coefficient of Eq.~(\ref{eq:Bterm}) can be expressed as
%
%
\begin{align}
\frac{1}{E_N(\bm{q})+E_\pi(\bm{q})-M_N}
=\frac{E_\pi(\bm{q})-(E_N(\bm{q})-M_N)}{q^2+M_\pi^2}.
\end{align}
Thus, in the heavy nucleon mass limit,
the coefficients of Eqs.~(\ref{eq:Bterm}) and (\ref{eq:Cterm}) become the following expressions
%
%
\begin{align}
I_{\alpha}^{\mathrm{B}}(t, t_{\mathrm{sep}})&\xrightarrow[M_N\rightarrow \infty]{}
-\frac{1}{4E_N(\bm{q})}\frac{r_\alpha^{+}}{q^2+M_\pi^2}e^{-(M_N+E_\pi(\bm{q}))t}e^{-M_N(t_{\mathrm{sep}}-t)}, 
\\
I_{\alpha}^{\mathrm{C}}(t, t_{\mathrm{sep}})&\xrightarrow[M_N\rightarrow \infty]{}
\frac{1}{4E_N(\bm{q})}\frac{r_\alpha^{-}}{q^2+M_\pi^2}e^{-(M_N+E_\pi(\bm{q}))t}
e^{-E_N(\bm{q})t}e^{-(E_N(\bm{q})+E_\pi(\bm{q}))(t_{\mathrm{sep}}-t)},
\end{align}
which can reproduce the leading $\pi N$
contribution discussed in the ChPT-based analysis~\cite{Bar:2018xyi}.

It is not difficult to verify that the pseudoscalar case can be
calculated similarly to the spatial component of the axial-vector to ensure that the leading $\pi N$ contributions are represented in terms of $\Delta_+(t,t_{\mathrm{sep}};\bm{q})$ in Eq.~(\ref{Eq:P_space}).

\section{Derivation of Eq.~(\ref{eq:piN_AWTI})}
\label{app:derivation_eq_25}

First, we introduce the following correlator ratio
%
%
\begin{align}
\mathcal{D}_{J}^{5z}(t; \bm{q})\equiv \frac{\partial_4 C_{J}^{5z}(t; \bm{q})}{C_{2}(t, t_{\mathrm{sep}}; \bm{q})},
\end{align}
where the numerator is the time-derivative of the three-point functions. We then consider the relation 
between $\mathcal{D}_{J}^{5z}(t; \bm{q})$ and $\mathcal{R}_{J}^{5z}(t; \bm{q})$. Assuming that the time derivative can be treated like a continuous derivative with respect to time, we get 
%
%
\begin{align}
\partial_4 {\cal R}_{J}^{5z}(t; \bm{q})
&=\frac{\partial_4 C_{J}^{5z}(t; \bm{q})}{C_2(t, t_{\mathrm{sep}}; \bm{q})}-{\cal R}_{J}^{5z}(t; \bm{q})\frac{\partial_4 C_2(t, t_{\mathrm{sep}}; \bm{q})}{C_2(t, t_{\mathrm{sep}}; \bm{q})} \cr
&={\cal D}_{J}^{5z}(t; \bm{q})+(E_N(\bm{q})-M_N){\cal R}_{J}^{5z}(t; \bm{q}),
\label{eq:derivative_rule1}
\end{align}
where the second line is obtained using the following relation
%
%
\begin{align}
\partial_4 C_2(t, t_{\mathrm{sep}};\bm{q})=-(E_N(\bm{q})-M_N) C_2(t, t_{\mathrm{sep}};\bm{q}).
\end{align}

In Eq.~(\ref{eq:AWTI}), the derivative of the nucleon three-point function with respect to the coordinate are evaluated by
%
%
\begin{align}
\partial_{i}C_{A_i}^{5z}(t; \bm{q})=iq_iC_{A_i}^{5z}(t;\bm{q})
\end{align}
for the spatial component ($i=1,2,3)$.
Therefore, Eq.~(\ref{eq:AWTI}) can be 
rewritten as
%
%
\begin{align}
Z_A\left[
iq_i {\cal R}_{A_i}^{5z}(t; \bm{q})+
{\cal D}_{A_4}^{5z}(t; \bm{q})
\right]=2m_{\mathrm{PCAC}}
{\cal R}_P^{5z}(t; \bm{q}).
\end{align}
Here, using Eq.~(\ref{eq:derivative_rule1}), 
${\cal R}_P^{5z}(t; \bm{q})$ can be expressed in terms of ${\cal R}_{A_\alpha}^{5z}(t; \bm{q})$ and
its time derivative as
%
%
\begin{widetext}
\begin{align}
{\cal R}_P^{5z}(t; \bm{q})=\frac{Z_A}{2m_{\mathrm{PCAC}}}\left[
iq_i{\cal R}_{A_i}^{5z}(t; \bm{q})
+\partial_4 {\cal R}_{A_4}^{5z}(t; \bm{q})-(E_N(\bm{q})-M_N){\cal R}_{A_4}^{5z}(t; \bm{q})
\right].
\label{eq:R_P_express}
\end{align}
\end{widetext}
Inserting Eqs.~(\ref{Eq:A_space}) and~(\ref{Eq:A_time}) into the right-hand side of Eq.~(\ref{eq:R_P_express}), we get
%
%
\begin{widetext}
\begin{align}
{\cal R}_P^{5z}(t; \bm{q})=&iq_3K^{-1}
\frac{Z_A}{2m_{\mathrm{PCAC}}}\left[2M_N\widetilde{F}_A(q^2)-2M_N(E_N(\bm{q})-M_N)\widetilde{F}_P(q^2)-q_i^2\Delta_{+}(t,t_{\mathrm{sep}};\bm{q})\right.\nonumber\\
&\left.+E_\pi(\bm{q})
\left\{\partial_4\Delta_{-}(t,t_{\mathrm{sep}};\bm{q})-(E_N(\bm{q})-M_N)\Delta_{-}(t,t_{\mathrm{sep}};\bm{q})\right\}
\right]\nonumber\\
=&iq_3K^{-1}
\frac{Z_A}{2m_{\mathrm{PCAC}}}\left[
2M_N\widetilde{F}_A(q^2)-q^2\widetilde{F}_P(q^2)-M_\pi^2\Delta_{+}(t,t_{\mathrm{sep}};\bm{q})
\right],
\label{eq:extend_GGT}
\end{align}
\end{widetext}
where we used $q^2=2M_N(E_N(\bm{q})-M_N)$ and Eq.~(\ref{eq:Time_Derivative_Delta}) at the stage of going from the first line to the second line. 
By comparing the time-independent and time-dependent parts of both sides of Eq.~(\ref{eq:extend_GGT}), 
two equalities are obtained, respectively. 
The generalized Goldberger-Treiman (GT) relation
%
%
\begin{align}
Z_A\left[
2M_N \widetilde{F}_A(q^2) - q^2 \widetilde{F}_P(q^2)
\right]=2m_{\rm PCAC} \widetilde{G}_P(q^2)
\end{align}
is obtained from the time-independent part of Eq.~(\ref{eq:extend_GGT}), while the time-dependent part of Eq.~(\ref{eq:extend_GGT}) provides Eq.~(\ref{eq:piN_AWTI}) as
%
%
\begin{align}
\Delta_P(t,t_{\mathrm{sep}};\bm{q}) =Z_A\frac{ M_\pi^2}{2m_{\mathrm{PCAC}}}\Delta_{+}(t,t_{\mathrm{sep}};\bm{q}).
\end{align}

It is worth noting that the elimination of the leading $\pi N$ contribution from ${\cal R}_P^{5z}(t; \bm{q})$ using Eq.~(\ref{eq:piN_AWTI}) ensures that the generalized GT relation holds for the ground-state contribution among three form factors, namely, $F_A(q^2)$, $F_P(q^2)$, and $G_P(q^2)$, as long as Eq.~(\ref{eq:AWTI}) is satisfied.

\section{Simultaneous fit approach}
\label{app:Sim_Fits}

In Sec.~\ref{sec:New_method}, we proposed the leading $\pi N$ subtraction method to determine $\widetilde{F}_P(q^2)$ and $\widetilde{G}_P(q^2)$ as defined in Eqs.~(\ref{eq:new_FP}) and (\ref{eq:new_GP}), where the leading $\pi N$ contribution in the three-point functions of the axial-vector and pseudoscalar currents are subtracted by adding the specific linear combination of the time derivatives of the $A_4$ and $A_i$ correlators. Here, instead, we employ
a simultaneous fit method to multiple datasets with different values of $t_{\mathrm{sep}}$ for the residual time dependence of the correlator ratios, including the leading $\pi N$ contribution, given by the terms of $\Delta_{\pm}(t, t_{\mathrm{sep}}; \bm{q})$, as defined in Eq.~(\ref{Eq:Delta_ex}). 

This is almost similar to the multi-state analysis, used in the other group, if the energy of excited states 
is treated as a free parameter. In this study, we fix the energies of the excited states appearing in the time dependence in $\Delta_{\pm}(t, t_{\mathrm{sep}}; \bm{q})$ by the non-interacting estimates
as
$\Delta E(\bm{q},-\bm{q})=E_\pi(\bm{q})+M_N-E_N(\bm{q})$ and
$\Delta E(\bm{0},\bm{q})=E_\pi(\bm{q})+E_N(\bm{q})
-M_N$, while the $t$-independent coefficients $B$ and $C$ are treated as free parameters. 

We apply the simultaneous fits to each of the following four quantities:
%
%
\begin{align}
W_{A_i}(t)&= -K\frac{\overline{\cal R}^{5z}_{A_i}(t, {\bm q})}{q_iq_3}, \\
W_{A_4}(t)&= K \frac{{\cal R}^{5z}_{A_4}(t, {\bm q})}{iq_3}, \\
W_{P}(t)&= K \frac{{\cal R}^{5z}_{P}(t, {\bm q})}{iq_3}, \\
W_{\Delta_+}(t)&=
\frac{K}{(\Delta E_N)^2-E^2_\pi}\cr
&\times
\left[
\Delta E_N\frac{\partial_4 \overline{\cal R}^{5z}_{A_i}(t, {\bm q})}{q_iq_3}
+\frac{\partial_4 {\cal R}^{5z}_{A_4}(t, {\bm q})}{iq_3}
\right],
\end{align}
where $\Delta E_N=E_N(\bm{q})-M_N$ and $E_\pi=E_\pi(\bm{q})$. $W_{A_i}(t)$ and
$W_{P}(t)$ correspond to $\widetilde{F}^{\mathrm{std}}_P(q^2)$
and 
$\widetilde{G}^{\mathrm{std}}_P(q^2)$ in the standard ratio method. 

\begin{figure*}[t]
\centering
\includegraphics[width=0.48\textwidth,bb=0 0 864 720,clip]{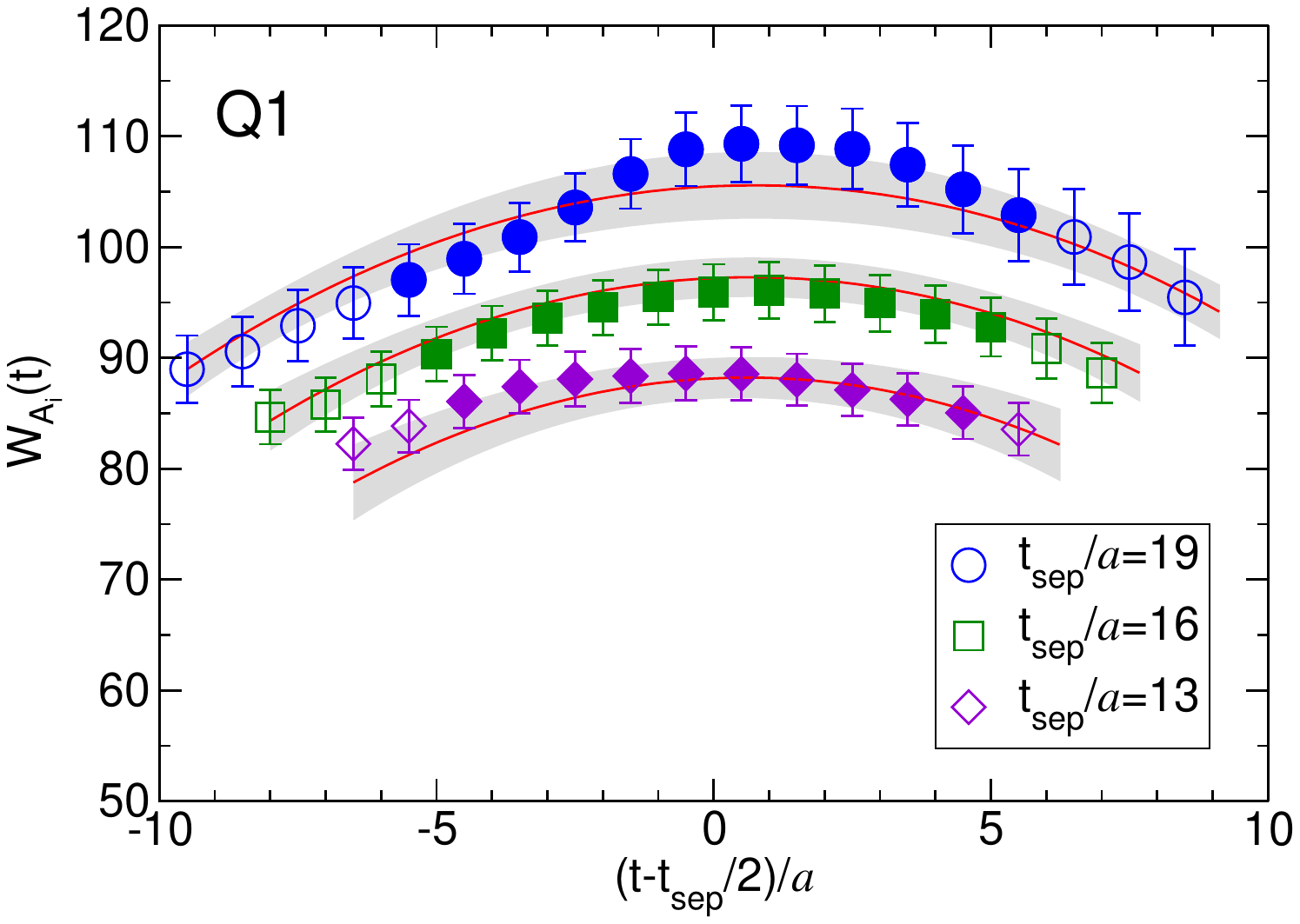}
\includegraphics[width=0.48\textwidth,bb=0 0 864 720,clip]{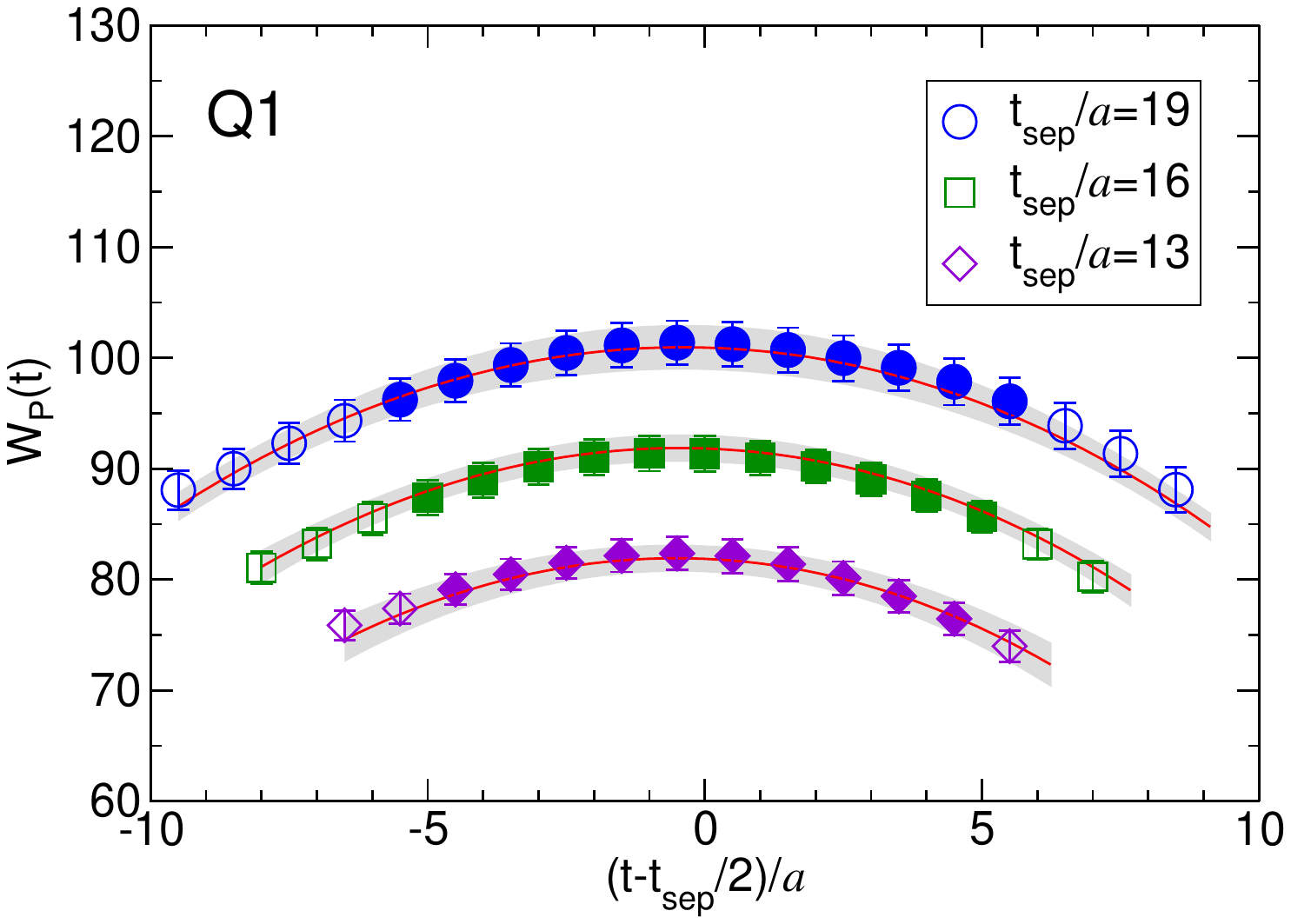}
\includegraphics[width=0.48\textwidth,bb=0 0 864 720,clip]{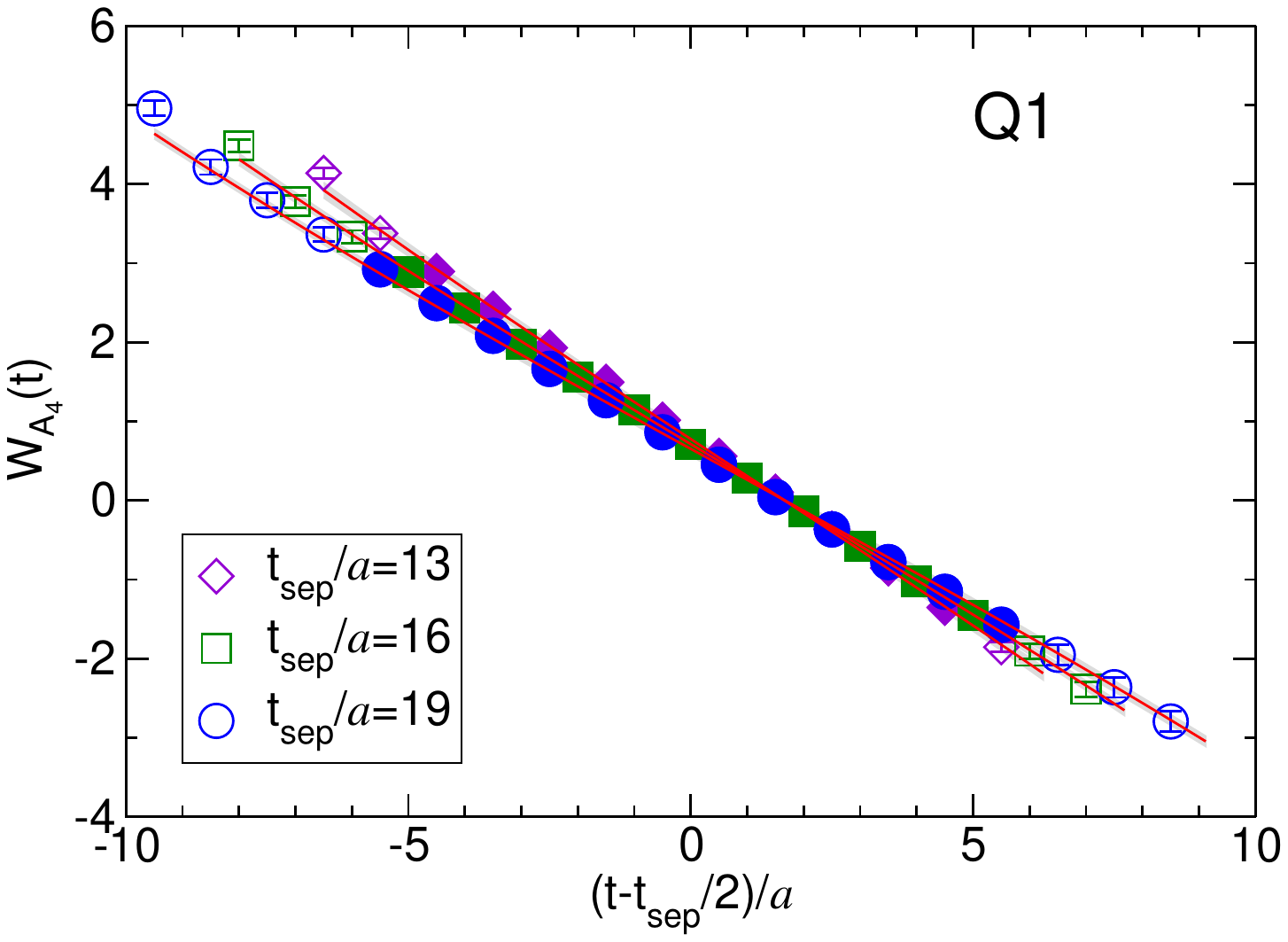}
\includegraphics[width=0.48\textwidth,bb=0 0 864 720,clip]{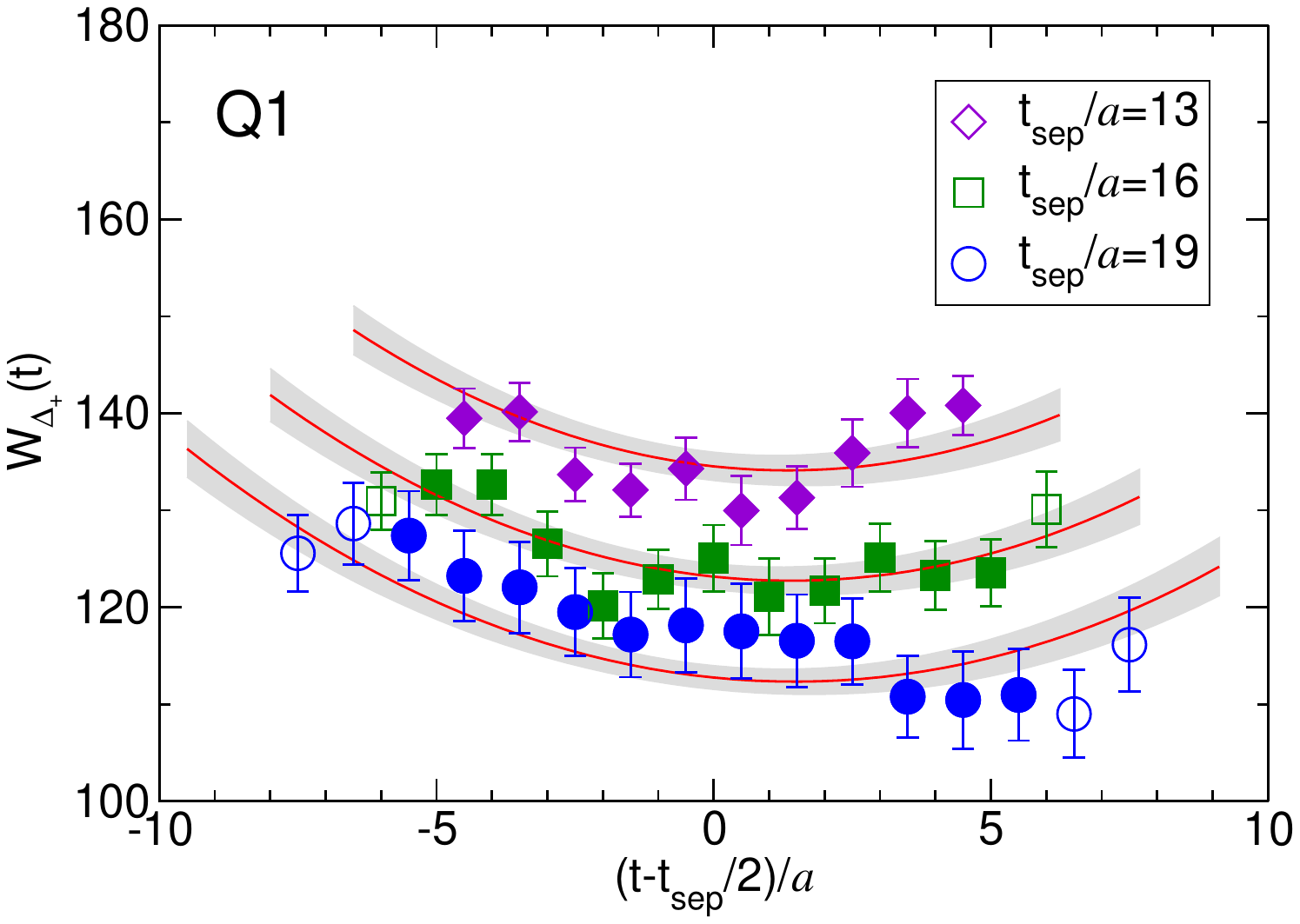}
\caption{
Typical results of the simultaneous fits on $W_{A_i}(t)$ (upper-left), $W_{P}(t)$ (upper-right), $W_{A_4}(t)$ (lower-left) and
$W_{\Delta_{+}}(t)$ (lower-right) 
at the lowest momentum transfer for the $160^4$ lattice ensemble. Each panel contains three datasets obtained with $t_{\mathrm{sep}}/a=13$ (diamonds), $16$ (squares) and $19$ (circles).
The red curves and the gray bands represent the resulting fitted curves, while the data points used to fit are marked as filled symbols.
}
\label{fig:sim_fits_curves}
\end{figure*}

The resulting fitted curve and the data points (marked as filled symbols) used to fit are shown in Fig.~\ref{fig:sim_fits_curves} for the lowest momentum transfer (Q1) as typical examples.
The fit results successfully reproduce the slight
convex shapes associated with the excited-state contamination and their $t_{\mathrm{sep}}$ dependence that appear in $W_{A_i}(t)$ and $W_{P}(t)$, as well as the peculiar behavior that depends linearly on the current insertion time $t$ in $W_{A_4}(t)$. Similarly, the corresponding $\pi N$ contribution extracted in $W_{\Delta_{+}}(t)$ is also well described by the fit results. 

\begin{figure*}[t]
\centering
\includegraphics[width=0.48\textwidth,bb=0 0 864 720,clip]{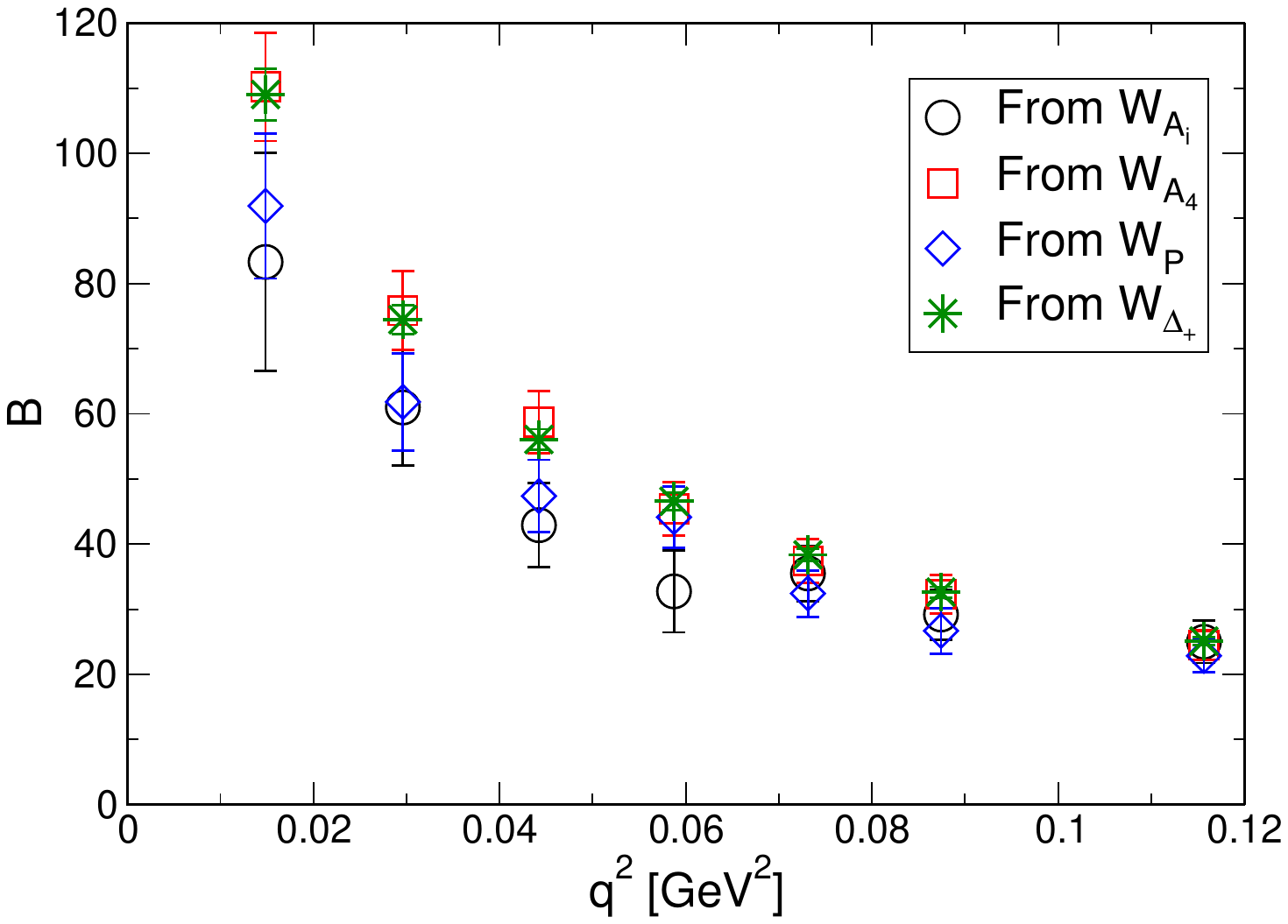}
\includegraphics[width=0.48\textwidth,bb=0 0 864 720,clip]{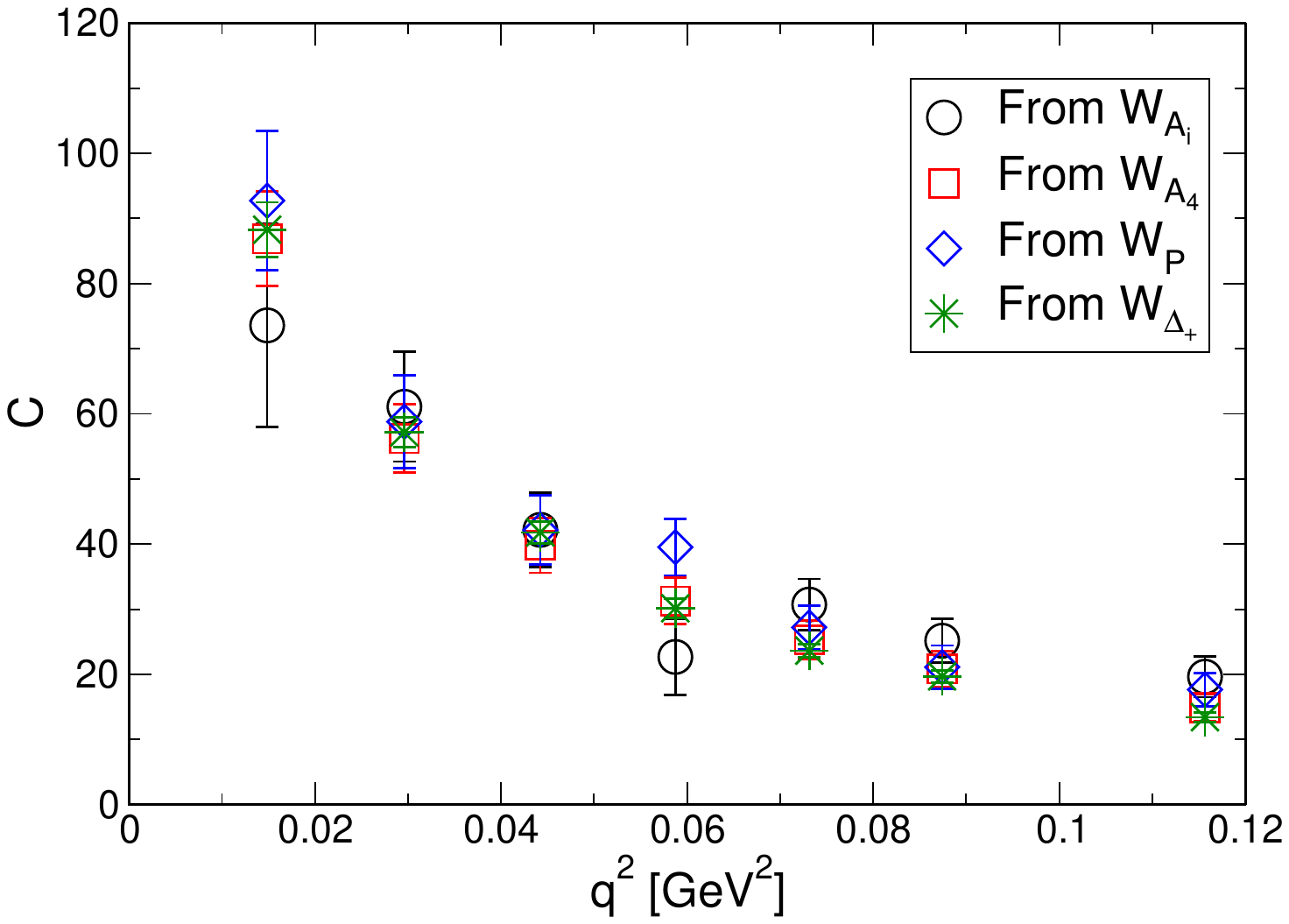}
\caption{
Two amplitudes of $B(\bm{q})$ (left) and $C(\bm{q})$ (right) appearing in the form of $\Delta_\pm(t,t_{\mathrm{sep}};\bm{q})$ as defined in Eq~(\ref{Eq:Delta_ex}) are evaluated in four different ways as a function of $q^2$. 
}
\label{fig:Ampl_B_and_C}
\end{figure*}

\begin{figure*}[t]
\centering
\includegraphics[width=0.48\textwidth,bb=0 0 864 720,clip]{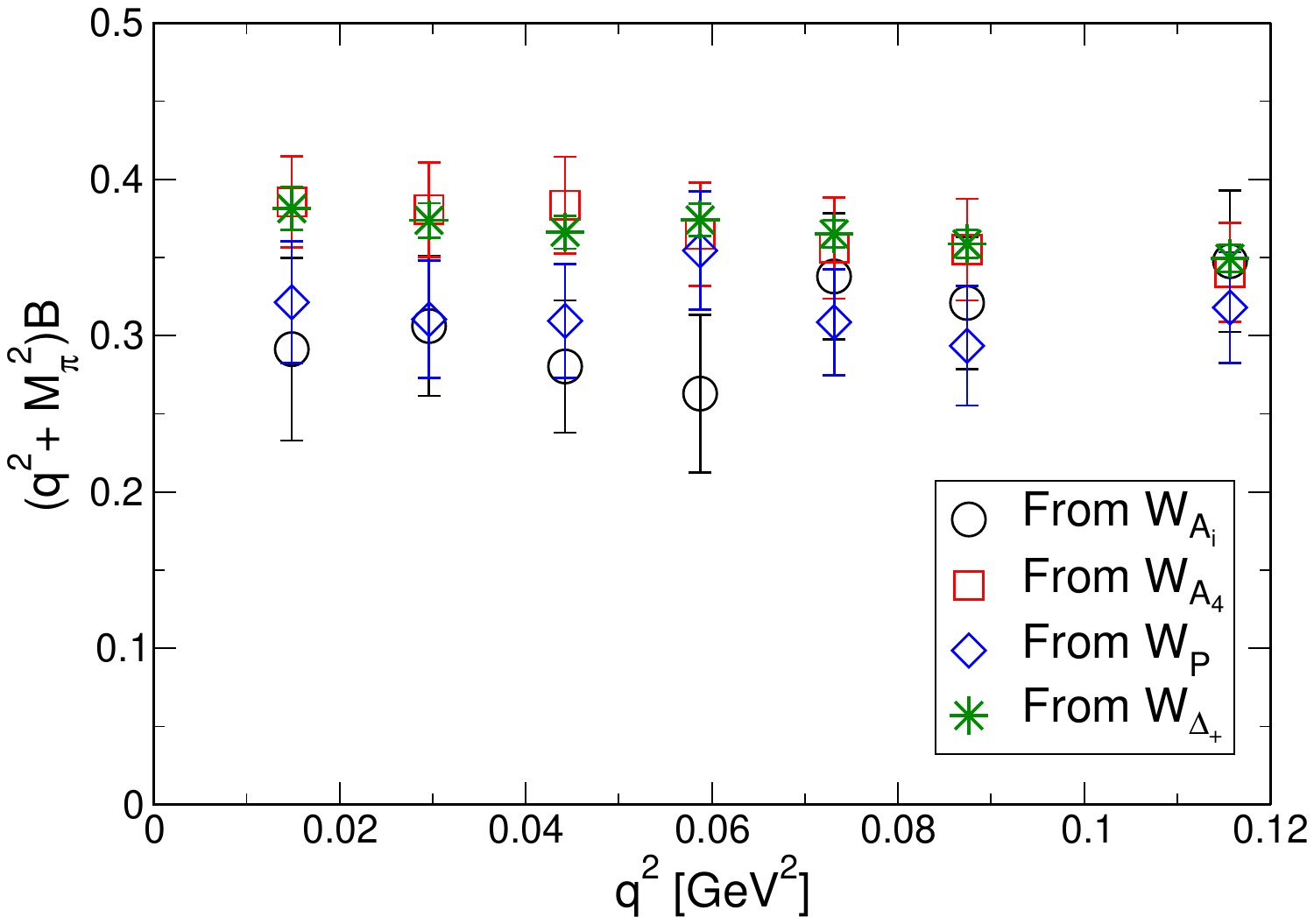}
\includegraphics[width=0.48\textwidth,bb=0 0 864 720,clip]{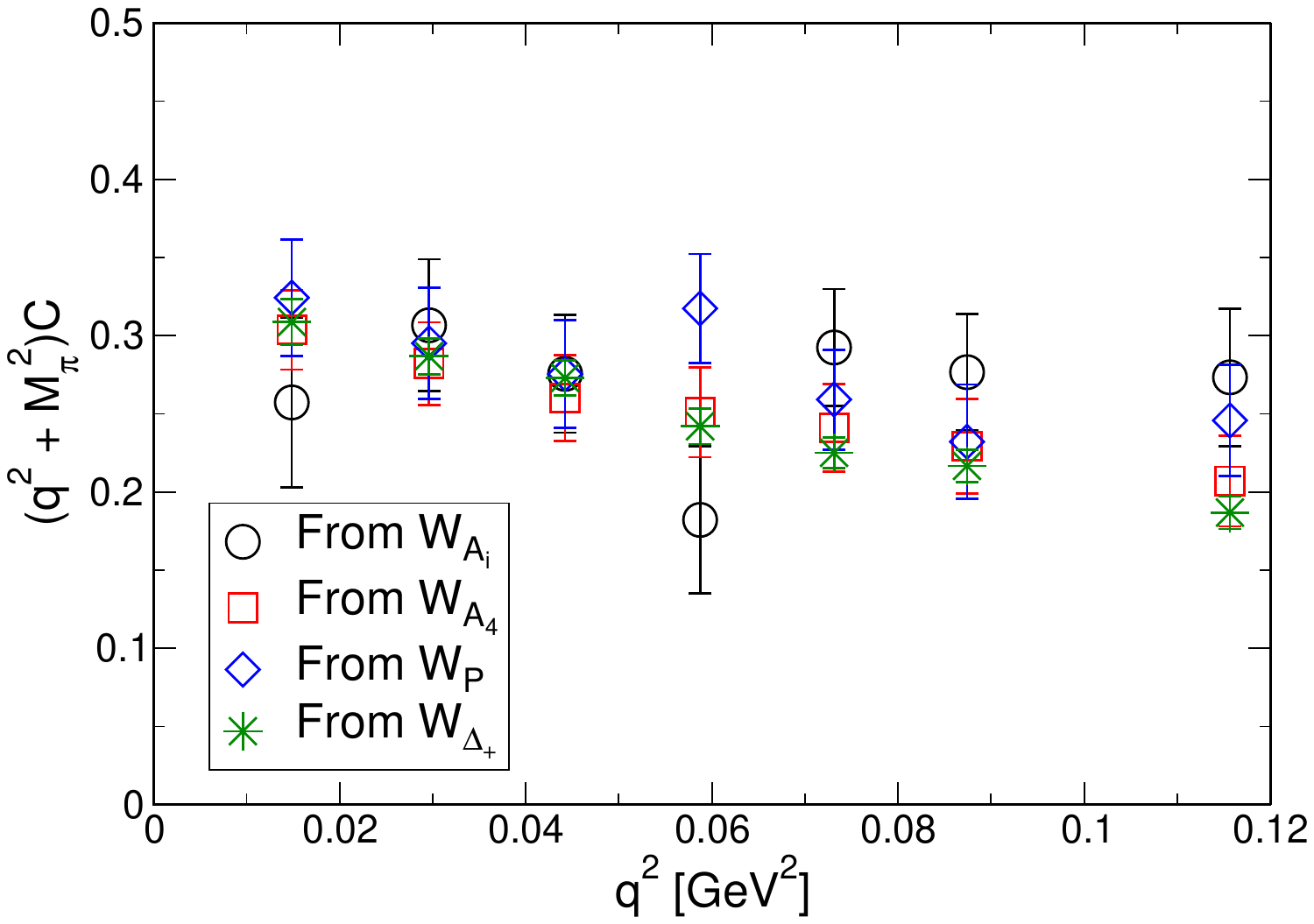}
\caption{
Two amplitudes of $B(\bm{q})$ (left) and $C(\bm{q})$ (right) seem to contain the pion-pole contribution as a function of $q^2$. 
}
\label{fig:Ampl_B_and_C_pion_pole}
\end{figure*}

In Fig.~\ref{fig:Ampl_B_and_C}, we next show
two amplitudes of $B(\bm{q})$ and $C(\bm{q})$ appearing in the form of $\Delta_\pm(t,t_{\mathrm{sep}};\bm{q})$
are evaluated in four different ways.
All four results are consistent with each other in both amplitudes $B(\bm{q})$ (left) and $C(\bm{q})$ (right) as a function of the four-momentum squared $q^2$, which is given by $2M_N(E_N(\bm{q})-M_N)$.
More interestingly, their $q^2$ dependence is very similar, and in both cases they grow rapidly when $q^2$ is small, as in the $F_P$ and $G_P$ form factors. In Fig.~\ref{fig:Ampl_B_and_C_pion_pole}, we plot
the values of $(q^2+M_\pi^2)B(\bm{q})$ and $(q^2+M_\pi^2)C(\bm{q})$ as a function of $q^2$. 
It has been demonstrated that the data points exhibit a flat tendency in the region of $q^2$ that has been utilized in this study. This observation is consistent with the analytical results obtained in Appendix~\ref{app:MOS}, which show that the corresponding $B(\bm{q})$ and $C(\bm{q})$ possess the pion-pole nature as a function of $q^2$.

Intuitively, if the nucleon form factor has a pion-pole structure such as $F_P(q^2)$ and $G_P(q^2)$, the contribution from the excited state should be relatively suppressed in comparison with the ground-state contribution in the low $q^2$ region.
This is simply because the magnitude of the form factor increases rapidly with lower $q^2$. 
In fact, however, the $\pi N$-state contamination is much more pronounced in both cases of $F_P(q^2)$ and $G_P(q^2)$. 
The underlying reason for this reversal trend is now straightforward: the amplitudes of the terms contributed by the leading $\pi N$ excited states are also enhanced by the pion-pole nature, in conjunction with the ground-state contributions, $F_P(q^2)$ and $G_P(q^2)$. In other words, the ground-state contribution and the leading $\pi N$ contribution are both enhanced by the 
pion-pole effect, causing their magnitudes to compete with each other at lower $q^2$.

\begin{figure*}[t]
\centering
\includegraphics[width=0.49\textwidth,bb=0 0 864 720,clip]{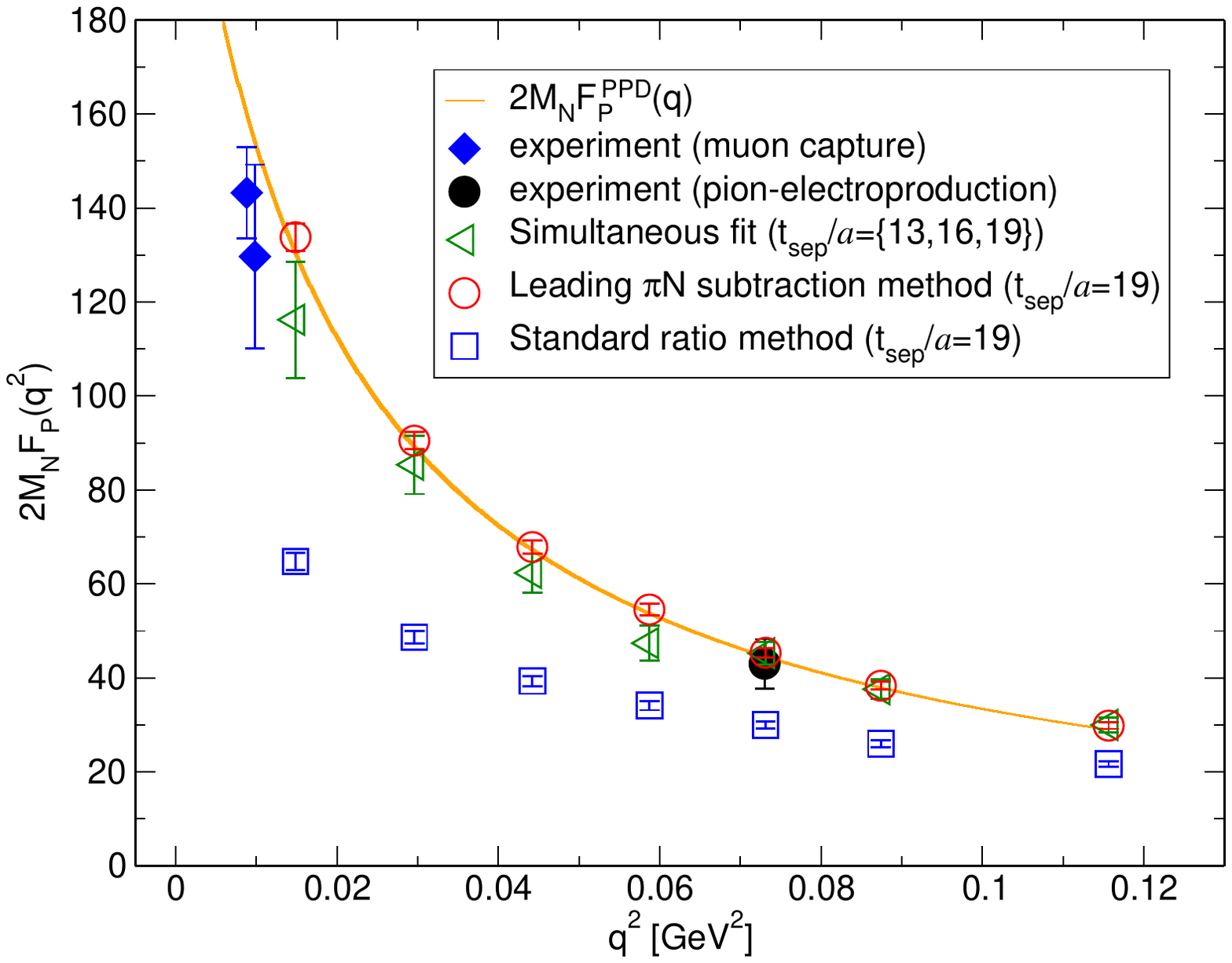}
\includegraphics[width=0.49\textwidth,bb=0 0 864 720,clip]{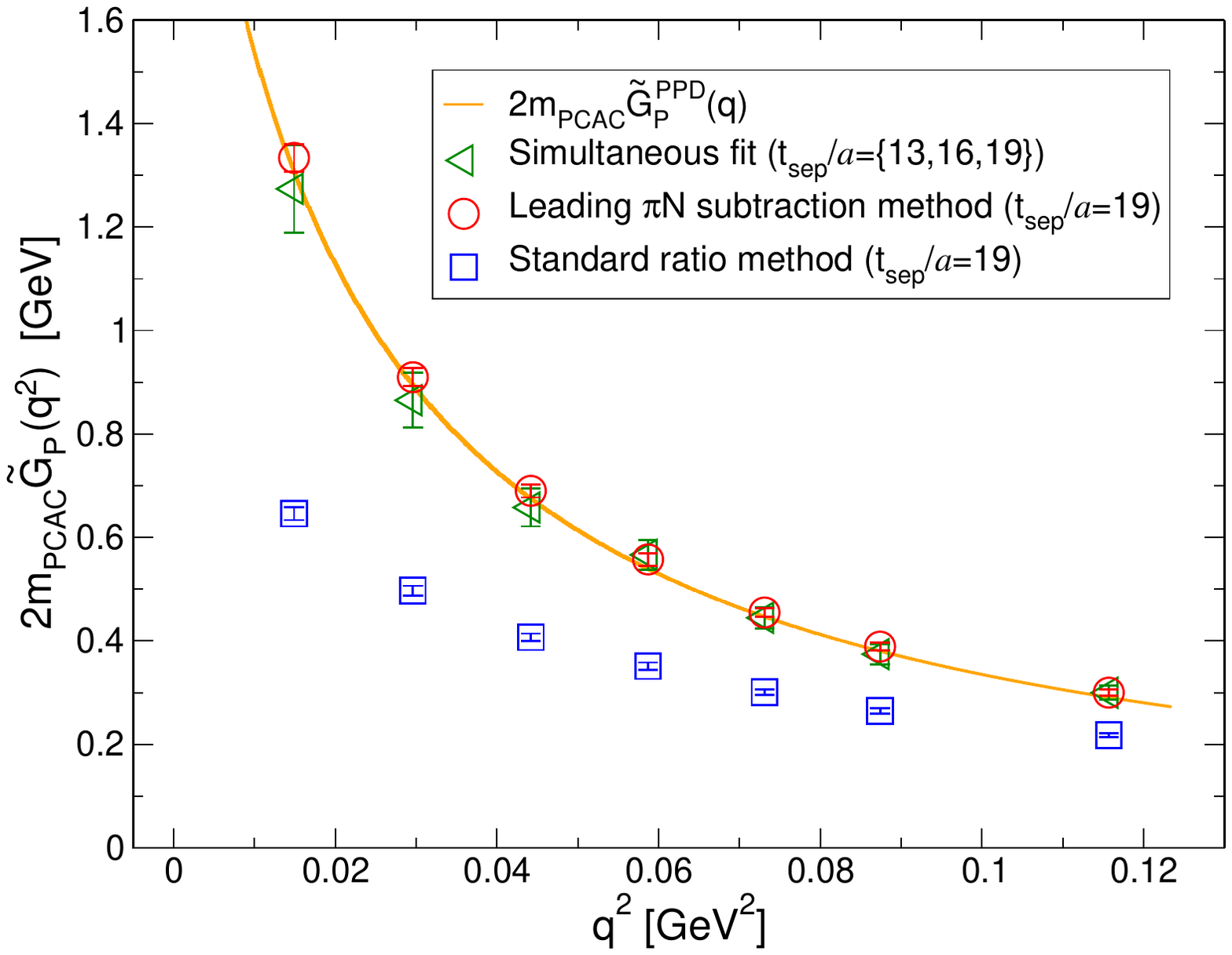}
\caption{
Comparison of values of $2M_N F_P(q^2)$ (left)
and $2m_{\mathrm{PCAC}}\widetilde{G}_P(q^2)$ (right) computed by three methods including the standard ratio method (blue squares), the leading $\pi N$ subtraction method (red circles) and the simultaneous fit procedure (green left-triangles)
using the $160^4$ lattice ensemble (PACS10/L160).}
\label{fig:three_methods_comp}
\end{figure*}

Finally, for $F_P(q^2)$ and $\widetilde{G}_P(q^2)$, the results obtained here are compared with those obtained by the leading $\pi N$ subtraction method as well as the standard ratio method, as shown in Fig.~\ref{fig:three_methods_comp}. For this comparison, we use the $160^4$ ensemble (PACS10/L160) with $t_{\mathrm{sep}}/a=19$. 
The standard ratio method (marked as blue squares) underestimates the values of $F_P(q^2)$. However, when the leading $\pi N$ contributions identified in the form of 
$\Delta_\pm(t,t_{\mathrm{sep}};\bm{q})$
are removed by the two approaches — the subtraction approach (marked as red circles) and the simultaneous fit approach (marked as green left-triangles) — the resulting values of $F_P(q^2)$ reproduce well the experimental data points and the PPD model. Furthermore, it should be emphasized here that the leading $\pi N$ subtraction method can produce smaller uncertainties than the simultaneous fit approach, especially in the low $q^2$ region.

%
%

%
%

%
%

%
%

\section{Test for pion-pole structure}
\label{app:Test_PP}
As discussed in Ref~\cite{Ishikawa:2018rew}, both of $F_P^{\rm std}(q^2)$ and $\widetilde{G}^{\rm std}_P(q^2)$ obtained from the standard ratio method are significantly underestimated in the low-$q^2$ region compared to the PPD model, while the ratio of $\widetilde{G}^{\rm std}_P(q^2)/F_P^{\rm std}(q^2)$ shows no appreciable $q^2$ dependence and
gives a nearly constant value, which is barely consistent with the bare low-energy constant $B_0$ evaluated by the ratio of $M_\pi^2/(2m_{\rm PCAC})$. 
We have therefore concluded that this is simply because the expected pion-pole structure in both cases of $F_P^{\rm std}(q^2)$ and $\widetilde{G}^{\rm std}_P(q^2)$ is
simultaneously distorted due to strong $\pi N$ excited-state contamination~\cite{Ishikawa:2018rew}.

\begin{figure*}[t]
\centering
\includegraphics[width=0.48\textwidth,bb=0 0 792 612,clip]{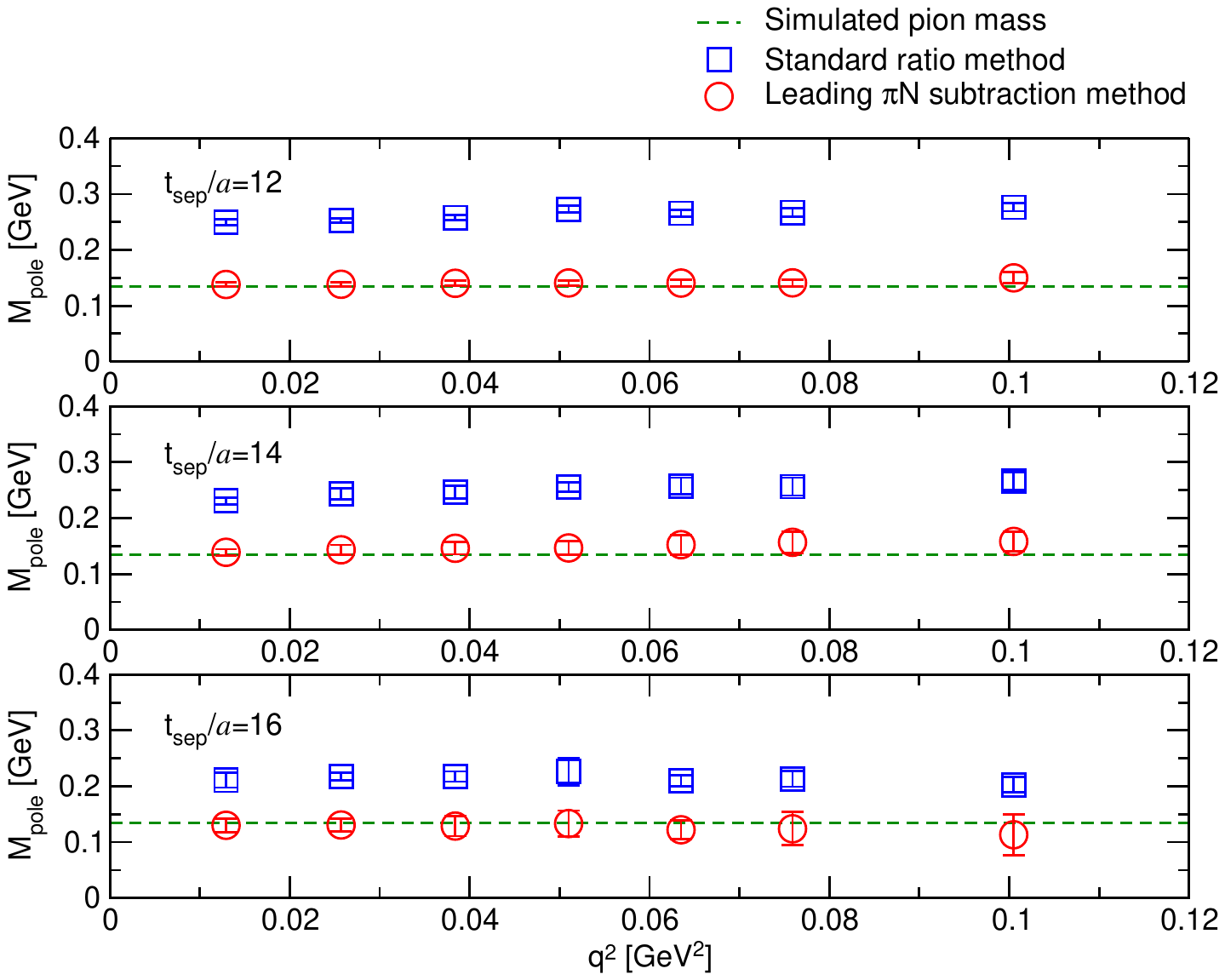}
\includegraphics[width=0.48\textwidth,bb=0 0 792 612,clip]{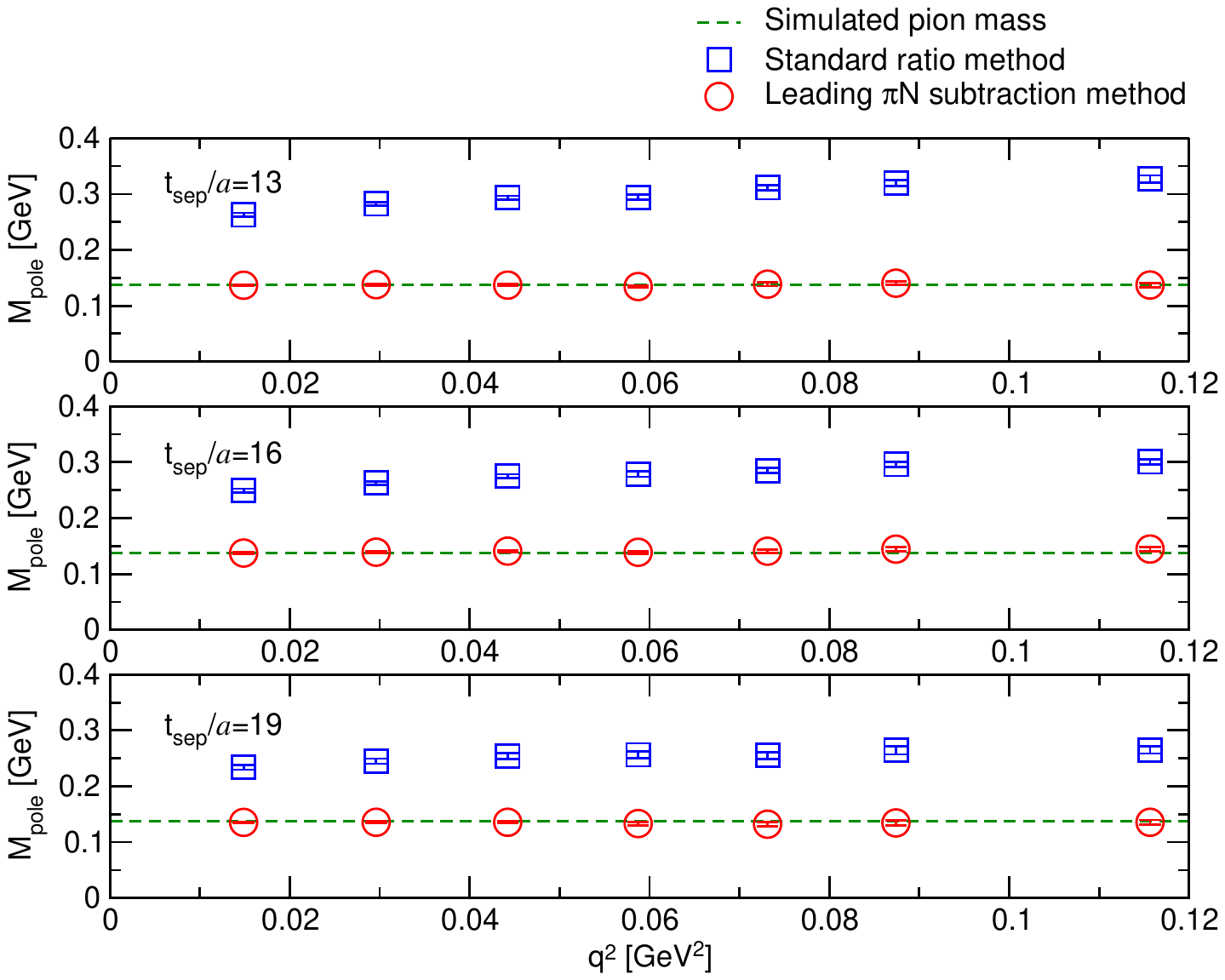}
\caption{
Check for the ``pion-pole'' structure in the $F_P$ form factor with $128^4$ (left) and $160^4$ (right) results. The blue square symbols denote the effective pole masses obtained from $F_P^{\mathrm{std}}(q^2)$ evaluated by the standard ratio method, while the
red circle symbols denote those of $F_P(q^2)$ evaluated by the leading $\pi N$ subtraction method. 
}
\label{fig:mpole_plateau}
\end{figure*}

On the other hand, as described in the text, the new analysis to remove the $\pi N$-state contamination from both $F_P$ and $G_P$ form factors makes results compatible with the PPD model. To see this fact, we define the effective
``pion-pole'' mass from $F_P(q^2)$~\cite{Yamazaki:2009zq} as
%
%
\begin{align}
M_{\rm pole}=\sqrt{\frac{2M_N F_A(q^2)}{F_P(q^2)}-q^2},
\end{align}
which would exhibit a flat $q^2$ dependence if $F_P(q^2)$ possesses the pole structure.
In Fig.\ref{fig:mpole_plateau}, we plot the effective pole mass as a function
of $q^2$. The horizontal dashed line at each panel represents the value of the
simulated pion mass $M_\pi$ in lattice units. 
The blue symbols are obtained from $F_P^{\rm std}(q^2)$ given by the standard ratio analysis,
while the red symbols are obtained for the new analysis. 
Although there is no significant $q^2$ dependence of the effective pole mass $M_{\mathrm{pole}}$ in either case, the new analysis gives fairly consistent values with the simulated pion mass. 

Why does the strong $\pi N$ excited-state contamination seem to affect the pion-pole structure only as a distortion of the pole mass in the PPD model?
In our observation, $M_{\mathrm{pole}}>M_\pi$ for $F_P^{\mathrm{std}}(q^2)$, while 
$F_P(q^2)$ certainly has the pion pole as shown in Fig.~\ref{fig:mpole_plateau}. Therefore
we see
%
%
\begin{align}
M_{\mathrm{pole}}^2-M_\pi^2=2M_NF_A(q^2)\left(\frac{1}{F_{P}^{\mathrm{std}}(q^2)}
-\frac{1}{F_{P}(q^2)}
\right)>0,
\end{align}
because of $F_P(q^2)>F_P^{\mathrm{std}}(q^2)$.

First of all,  the axial Ward-Takahashi identity for the nucleon three-point function, where both the ground-state and
excited-state contributions are entangled, is verified in Ref.~\cite{Tsuji:2023llh}. This observation suggests that the ground-state contribution and the excited-state contribution must separately satisfy the axial Ward-Takahashi identity.
For the ground state, this is nothing but
the generalized GT relation. Therefore, once the leading $\pi N$ contribution on the axial-vector three-point function is identified as $\Delta_\pm(t)$ term, 
the leading $\pi N$ contribution on the pseudoscalar three-point function can be described by $\Delta_P(t)=Z_A B_0 \Delta_+(t)$, where
the factor $Z_AB_0$ is determined to hold the axial Ward-Takahashi identity for the leading $\pi N$ contribution. 
Using the terms of $\Delta_{+}(t)$ and $\Delta_{P}(t)$, the true ground-state contribution, $\widetilde{F}_P(q^2)$ ($\widetilde{G}_P(q^2)$) and  $\widetilde{F}^{\mathrm{std}}_P(q^2)$ 
($\widetilde{G}^{\mathrm{std}}_P(q^2)$)
evaluated by the standard ratio method are related as follows:
%
%
\begin{align}
\widetilde{F}_P(q^2)-\widetilde{F}^{\mathrm{std}}_P(q^2)&=\Delta_{+}(t,t_{\mathrm{sep}};\bm{q}),\\
\widetilde{G}_P(q^2)-\widetilde{G}^{\mathrm{std}}_P(q^2)&=\Delta_P(t,t_{\mathrm{sep}};\bm{q}).
\end{align}
Therefore, we get the following relation
\begin{align}
B_0 &=\frac{\widetilde{G}_P(q^2)-\widetilde{G}^{\mathrm{std}}_P(q^2)}{F_P(q^2)-F^{\mathrm{std}}_P(q^2)}.
\end{align}
Here, we easily see that 
\begin{align}
\frac{\widetilde{G}^{\mathrm{std}}_P(q^2)}{{F}^{\mathrm{std}}_P(q^2)}
\approx\frac{\widetilde{G}_P(q^2)}{{F}_P(q^2)}
\approx B_0
\end{align}
is fulfilled, 
if both ${F}_P(q^2)$ and $\widetilde{G}_P(q^2)$ are well described by the PPD model that leads to
$\widetilde{G}^{\mathrm{PPD}}_P(q^2)/{F}^{\mathrm{PPD}}_P(q^2)=B_0$.

%
%

\section{Numerical data of nucleon form factors}
\label{app:data_for_the_form_factors_versus_q^2}

The results of the two nucleon form factors
$F_P(q^2)$ and $\widetilde{G}_P(q^2)$ obtained
with the standard ratio method
are summarized in Tables~\ref{tab:FP_qdep_128},
\ref{tab:FP_qdep_160}, \ref{tab:GP_qdep_128}
and \ref{tab:GP_qdep_160}.

%
%
\begin{table}[t]
\caption{Results of the $F_P(q^2)$ calculated at the coarse ($128^4$) lattice.}
\label{tab:FP_qdep_128}
\centering
\begin{ruledtabular}
\begin{tabular}{crrrr}
\hline
\multicolumn{1}{c}{\bf{PACS10/L128}} & 
\multicolumn{1}{c}{$t_{\rm sep}/a=10$} & \multicolumn{1}{c}{$t_{\rm sep}/a=12$} & \multicolumn{1}{c}{$t_{\rm sep}/a=14$} & \multicolumn{1}{c}{$t_{\rm sep}/a=16$} \\
\cline{2-5}
\multicolumn{1}{c}{$q^2\ [\mathrm{GeV}^2]$} & \multicolumn{1}{c}{$2M_N F_P(q^2)$} & \multicolumn{1}{c}{$2M_N F_P(q^2)$} & \multicolumn{1}{c}{$2M_N F_P(q^2)$} & \multicolumn{1}{c}{$2M_N F_P(q^2)$} \\
\hline
0.0129 & 146.9(7.1)& 138.7(7.3) & 141.4(11.0) & 140.6(12.8)\\
0.0257 & 102.1(4.5) & 97.9(4.8) & 97.1(8.5) & 96.0(6.6)\\
0.0384 & 78.2(3.4) &
74.6(3.5)  &
74.7(6.6)  &
73.8(4.1)  \\
0.0510 & 63.9(3.7) &
59.6(2.4) &
60.1(4.6) &
56.8(5.0) \\
0.0635 & 52.9(2.2) &
50.0(2.1) &
49.8(4.6) &
50.2(2.6) \\
0.0759 & 45.2(1.8) &
43.0(1.8) &
42.7(4.0) &
42.5(2.1) \\
0.1005 & 34.7(1.4) &
32.6(1.7) &
32.6(3.2) &
33.7(1.8) \\
\hline
\end{tabular}
\end{ruledtabular}
\end{table}
%

%
%
\begin{table}[t]
\caption{Results of the $F_P(q^2)$ calculated at the fine ($160^4$) lattice.}
\label{tab:FP_qdep_160}
\centering
\begin{ruledtabular}
\begin{tabular}{crrr}
\hline
\multicolumn{1}{c}{\bf{PACS10/L160}}& 
\multicolumn{1}{c}{$t_{\rm sep}/a=13$} & \multicolumn{1}{c}{$t_{\rm sep}/a=16$} & \multicolumn{1}{c}{$t_{\rm sep}/a=19$}\\
\cline{2-4}
\multicolumn{1}{c}{$q^2\ [\mathrm{GeV}^2]$} & \multicolumn{1}{c}{$2M_N F_P(q^2)$} & \multicolumn{1}{c}{$2M_N F_P(q^2)$} & \multicolumn{1}{c}{$2M_N F_P(q^2)$} \\
\hline
0.0149 & 131.4(2.1) &
129.8(2.2) &
134.2(2.9) \\
0.0296 & 88.4(1.3)&
87.8(1.4) &
90.8(1.8) \\
0.0442 & 66.7(0.9) &
65.8(1.0) &
68.1(1.4) \\
0.0587 & 54.3(0.8) &
53.0(0.8) &
54.7(1.3) \\
0.0731 & 43.9(0.7) &
43.7(0.8) &
45.4(0.9) \\
0.0874 & 37.1(0.5)  &
36.9(0.6) &
38.4(0.8) \\
0.1156 & 28.4(0.5) &
28.2(0.5)  &
29.8(0.7)  \\
\hline
\end{tabular}
\end{ruledtabular}
\end{table}
%

%
%
\begin{table}[t]
\caption{Results of the $\widetilde{G}_P(q^2)$ calculated at the coarse ($128^4$) lattice.}
\label{tab:GP_qdep_128}
\centering
\begin{ruledtabular}
\begin{tabular}{crrrr}
\hline
\multicolumn{1}{c}{\bf{PACS10/L128}}& 
 \multicolumn{1}{c}{$t_{\rm sep}/a=10$} & \multicolumn{1}{c}{$t_{\rm sep}/a=12$} & \multicolumn{1}{c}{$t_{\rm sep}/a=14$} & \multicolumn{1}{c}{$t_{\rm sep}/a=16$} \\\cline{2-5}
\multicolumn{1}{c}{$q^2\ [\mathrm{GeV}^2]$} & \multicolumn{1}{c}{$\widetilde{G}_P(q^2)$} & \multicolumn{1}{c}{$\widetilde{G}_P(q^2)$} & \multicolumn{1}{c}{$\widetilde{G}_P(q^2)$} &\multicolumn{1}{c}{ $\widetilde{G}_P(q^2)$} \\
\hline
0.0129 & 231.0(12.3) &
221.2(9.2) &
214.2(14.7) &
220.3(17.3) \\
0.0257 & 164.9(11.2) &
153.9(5.7) &
150.8(9.1) &
156.3(10.6) \\
0.0384 & 128.7(11.1) &
117.6(4.1) &
114.7(6.7) &
121.5(8.3) \\
0.0510 & 101.5(5.8) &
96.9(3.6) &
92.4(7.7) &
94.2(7.0) \\
0.0635 & 88.3(8.1) &
80.0(2.3) &
77.0(4.9) &
82.5(5.4) \\
0.0759 & 78.2(9.6) &
69.0(1.8) &
64.7(4.1) &
72.9(6.3) \\
0.1005 & 59.5(5.8) &
52.3(1.9) &
49.9(3.5) &
56.3(4.3) \\
\hline
\end{tabular}
\end{ruledtabular}
\end{table}
%

%
%
\begin{table}[t]
\caption{Results of the $\widetilde{G}_P(q^2)$ calculated at the fine ($160^4$) lattice.}
\label{tab:GP_qdep_160}
\centering
\begin{ruledtabular}
\begin{tabular}{crrr}
\hline
\multicolumn{1}{c}{\bf{PACS10/L160}}& 
\multicolumn{1}{c}{$t_{\rm sep}/a=13$}  & \multicolumn{1}{c}{$t_{\rm sep}/a=16$}  & \multicolumn{1}{c}{$t_{\rm sep}/a=19$}\\
\cline{2-4}
\multicolumn{1}{c}{$q^2\ [\mathrm{GeV}^2]$} & \multicolumn{1}{c}{$\widetilde{G}_P(q^2)$} & \multicolumn{1}{c}{$\widetilde{G}_P(q^2)$} & \multicolumn{1}{c}{$\widetilde{G}_P(q^2)$} \\
\hline
0.0149 & 202.7(2.9) &
202.7(2.9) &
208.5(4.1) \\
0.0296 & 139.4(1.9) &
138.7(1.9) &
142.2(2.7) \\
0.0442 & 105.8(1.4) &
105.7(1.4) &
107.9(2.0) \\
0.0587 & 83.6(1.3) &
84.3(1.3) &
87.1(1.9) \\
0.0731 & 70.3(0.9) &
70.1(1.1) &
71.1(1.3) \\
0.0874 &  60.0(0.9) &
60.5(0.9) &
60.8(1.2) \\
0.1156 & 45.8(0.7) &
45.9(0.8) &
46.9(0.9) \\
\hline
\end{tabular}
\end{ruledtabular}
\end{table}

\clearpage

\section{Assessment of residual excited-state contamination}
\label{app:assesment_of_residual_excited_state_contamination}
In this study, we mainly adopted the standard ratio method 
to determine the $F_P(q^2)$ and $\widetilde{G}_P(q^2)$ after 
subtracting the leading $\pi N$ contribution. 
Since more than three sets of $t_{\mathrm{sep}}$ were carried out
in the standard ratio method, the summation method~\cite{Maiani:1987by} can also be used to determine the $F_P(q^2)$ and $\widetilde{G}_P(q^2)$
within the leading $\pi N$ subtraction method.
The analysis is conducted in accordance with the guidelines described in Appendix D in Ref.~\cite{Tsuji:2024scy}.
The summation method is applied to four datasets of $t_{\mathrm{sep}}=\{10,12,14,16\}$ for the $128^4$ lattice, and three datasets of
$t_{\mathrm{sep}}=\{13,16,19\}$ for the $160^4$ lattice.
The obtained values of the $F_P(q^2)$ and $\widetilde{G}_P(q^2)$ are summarized in Table~\ref{tab:summation_method}.
The results obtained by the two methods, the standard ratio and summation methods, are consistent with each other within statistical uncertainties. This finding suggests that the residual excited-state contamination, subsequent to the subtraction of the leading $\pi N$ contamination, is well controlled by the optimal choice of the smearing parameter for the nucleon interpolating operator.

%
%
\begin{table}[ht]
\caption{Results of the $F_P(q^2)$ and $\widetilde{G}_P(q^2)$
using the summation method after removing leading $\pi N$ contributions.}
\label{tab:summation_method}
\centering
\begin{ruledtabular}
\begin{tabular}{crrcrr}
\hline
\multicolumn{3}{c}{\bf{PACS10/L128}}  & \multicolumn{3}{c}{\bf{PACS10/L160}}  \\
\cline{1-3}\cline{4-6}
\multicolumn{1}{c}{$q^2\ [\mathrm{GeV}^2]$} & \multicolumn{1}{c}{$2M_N F_P(q^2)$} & \multicolumn{1}{c}{$\widetilde{G}_P(q^2)$} &
\multicolumn{1}{c}{$q^2\ [\mathrm{GeV}^2]$}
& \multicolumn{1}{c}{$2M_N F_P(q^2)$} & \multicolumn{1}{c}{$\widetilde{G}_P(q^2)$} \\
\hline
0.0129 & 134.9(15.9) & 204.4(35.6)& 
0.0149 & 127.0(6.7) & 205.6(9.1)\\
0.0257 & 88.1(16.7) & 148.8(17.3)&
0.0296 & 89.3(3.8)  & 139.9(6.4)\\
0.0384 & 69.9(9.6)  & 116.3(11.9)&
0.0442 & 67.0(2.7)  & 107.1(4.9)\\
0.0510 & 54.6(5.9)  & 85.8(17.7) &
0.0587 & 51.4(2.5)  & 89.1(4.0)\\
0.0635 & 46.7(7.8)  & 76.1(8.2)&
0.0731 & 45.8(1.8)  & 70.4(3.2)\\
0.0759 & 41.4(4.4)  & 62.7(7.1)&
0.0874 & 39.1(1.5)  & 60.2(3.1)\\
0.1005 & 31.9(5.7)  & 49.7(6.0)&
0.1156 & 30.6(1.4)  & 47.1(2.2)\\
\hline
\end{tabular}
\end{ruledtabular}
\end{table}

\clearpage

\end{document}